# Cold Physics and Chemistry: Collisions, Ionization and Reactions inside Helium Nanodroplets Close to Zero K.

A. Mauracher[1], O. Echt[2], A.M. Ellis[3], S. Yang[3], D.K. Bohme[4], J. Postler[1], A. Kaiser[1], S. Denifl[1] and P. Scheier[1,*]

[1]Institut für Ionenphysik und Angewandte Physik, Leopold-Franzens-Universität Innsbruck, Technikerstr. 25, A-6020 Innsbruck, Austria

[2]Department of Physics, University of New Hampshire, Durham, USA

[3]Department of Chemistry, University of Leicester, University Road, Leicester LE1 7RH, United Kingdom

[4]Department of Chemistry, York University, Toronto, Ontario M3J 1P3, Canada

*Corresponding author: paul.scheier@uibk.ac.at

# Contents

# 1   Introduction

Superfluid helium is an extraordinary illustration of the consequences of quantum mechanics on the macroscopic scale. The discovery of superfluid behavior in liquid helium was first made in independent studies by Kapitza [1] and Allen and Misener [2]. $^4$He becomes a superfluid below its so-called λ-point, which occurs at 2.17 K. At these low temperatures it is no longer appropriate to think of the liquid as a collection of distinct helium atoms colliding and moving randomly with different energies, as one would with a classical liquid. Instead, quantum properties come to the fore and the behavior of the liquid is best described by collapse of the system into a ground state macroscopic wavefunction extending across the entire liquid [3-5]. This collective wavefunction, in which the $^4$He atoms become indistinguishable, is a form of Bose-Einstein condensation, although the atoms in superfluid helium interact far more strongly than in a gaseous Bose-Einstein condensate. Many of the extreme properties of superfluids derive from this macroscopic quantum mechanical behavior, including the near-zero viscosity, exceptionally high thermal conductivity, and formation of quantized vortices.

The possibility of adding atoms and molecules (dopants) to superfluid helium has been considered many times. If this could be successfully achieved, then one could use selected dopants as optical probes of superfluid phenomena. On the other hand, the superfluid offers the opportunity to explore how atoms and molecules behave at low temperatures in a chemically inert environment. For example, one might wish to use this matrix as a means of simplifying the spectrum of a molecule, or perhaps to form atomic and molecular clusters which are difficult or impossible to make by any other means. Unfortunately, bulk superfluid helium is not conducive to such studies. Liquid helium is the least effective of any known solvents in terms of providing stabilization to a solute. Consequently, when an atomic or molecular solute is added, it rapidly migrates to the walls of the container, since this delivers a lower net energy. This rapid phase separation has severely limited the study of molecules in bulk liquid helium. Work that has been done has been largely limited to small metal clusters, delivered into the liquid by laser ablation techniques (see for example [6-10]).

Nanoscale helium droplets (also referred to as helium nanodroplets - HNDs), formed *in vacuo*, offer an alternative to bulk superfluid helium as a matrix. Although first made back in 1961 [11], there was a long wait until the early 1990s before gaseous dopants were successfully added to these droplets [12]. Since HNDs are not in contact with any container walls, phase separation is avoided and so the addition of atoms and molecules offers the opportunity to place dopants inside the droplets, where they will remain indefinitely if not perturbed, *e.g.* by a collision of the droplet with the vacuum vessel. A significant proportion of the early studies of molecules in HNDs were concerned with using the molecules as optical probes to characterize the helium environment and to try and prove superfluid behavior [13-17]. However, the study of atoms and molecules in HNDs has branched out into other areas of investigation and HNDs are now being exploited by many groups across the world to explore a wide variety of chemical and physical phenomena.

This is by no means the first review focusing on HNDs. Several reviews have appeared which provide a general overview of the subject [18-21]. Furthermore, there are other reviews focusing on more specialized aspects of HND research, such as their use in spectroscopy and dynamics [22-27], photoionization and ultrafast processes [28-30], chemical reaction dynamics [31] and the study of metal clusters [32]. In addition, there have been several reviews that discuss theoretical aspects of pure and doped HNDs [33-35].

This review has two principal aims. The first of these is to provide a comprehensive overview of the applications of HNDs in the study of collections of atoms and molecules, *i.e.* clusters and complexes. These clusters and complexes must form through collisions inside HNDs, hence the title of this review. A second aim is to provide a particularly detailed overview of the many studies of ions, both positive and negative, that have been carried out in HNDs.

## 2 Formation and size distributions of helium nanodroplets

HNDs can be produced by expansion of precooled helium through a small nozzle into vacuum. Depending on the stagnation condition of the expansion (pressure, temperature and nozzle diameter) various average droplet sizes <$N$> and distribution widths can be achieved. Subcritical expansion, i.e. the helium is still gaseous when it passes the nozzle, leads to an average droplet size in the order of $10^3$ to $10^4$ helium atoms [36]. In the critical expansion regime, the expansion conditions are close to the liquid-vapor critical point of helium ($p_c$ = 2.3 bar, $T_c$ = 5.2 K). Large fluctuations in the average cluster size and mean velocity are observed [37]. Supercritical expansion leads to an average size of <$N$> = $10^5$ to <$N$> = $10^6$ and is described in detail below [38]. The largest cluster sizes are achieved by Rayleigh breakup. The expansion conditions are chosen in such a way, that helium is already in its superfluid state before expansion. During the expansion the liquid jet breaks up and leads to clusters with average size between <$N$> = $10^9$ and <$N$> = $10^{12}$ [39].

The most common way to produce HNDs is via supersonic expansion [38] of pressurized and precooled He through a small nozzle into vacuum. The first beam of HNDs formed by means of this technique was achieved by Becker and coworkers [11, 40]. They measured the velocity distribution of He cluster beams via a chopped beam method and determined the total mass flux via a micro membrane pressure gauge. The average droplet size <$N$> and the droplet size distribution $p(N)$ were first measured for small HNDs (< $10^4$ He atoms) by deflecting a continuous beam of HNDs using heavy rare gas atoms (Ar, Kr, Xe) or $SF_6$ from a secondary molecular beam source [41, 42]. For slightly larger HNDs with mean sizes between $10^5$ and $10^7$ helium atoms the deflection of charged droplets (ionized via electron impact) in an electric field was used [43, 44]. All these early studies obtained a log-normal distribution (see Equation (1)) for small droplets and a linear-logarithmic distribution for larger droplets.

$$lnP(N; \mu, \sigma) = \frac{1}{N\sigma\sqrt{2\pi}} e^{-\frac{(\ln(N) - \mu)^2}{2\sigma^2}}, N > 0 \qquad (1)$$

A log-normal distribution generally leads to a better description of original data across different fields of science compared to a normal (Gaussian) distribution [45] and this is particularly true for growth processes such as cluster formation [46]. For expansion conditions that, according to Toennies and coworkers [19, 44], lead to an average droplet size between $10^5$ and $10^7$, we have recently measured cluster size distributions of charged droplets by deflection in a 90° cylindrical electrostatic sector field. As in the earlier studies [41-44], the method depends on the fact that the velocity distribution of particles in a supersonic expansion is extremely narrow and independent of size and thus the kinetic energy is proportional to the mass of a particle in such a molecular beam [47]. Figure 1 shows the experimental setup and in Figure 2 measured size distributions are depicted for various source conditions. However, one has to take into consideration that the geometric cross section for a HND containing 12 million He atoms is 28 times larger than for a HND containing 80000 He atoms, assuming spherical droplets with fixed densities. Thus for the same electron current, there is a much higher probability of collisions between a large HND and more than one electron which may underestimate the high-mass range of the droplet size distribution. Both, the formation of multiply charged HNDs and Coulomb explosion will lead to a reduced mass per charge value of an initially heavy HND.

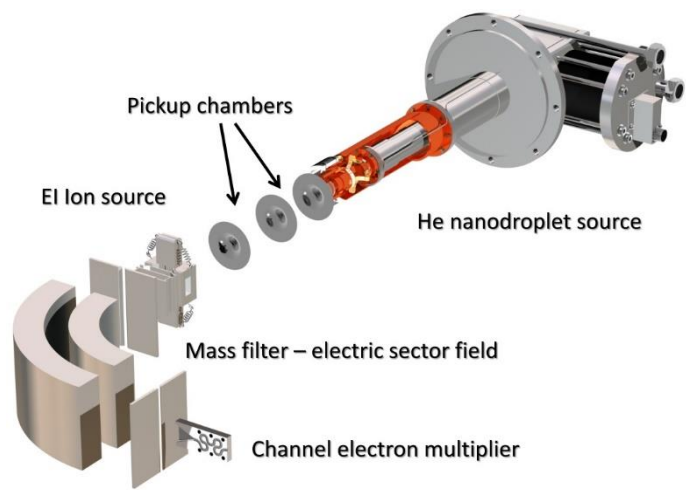

*Figure 1 Schematic diagram of the experimental setup to determine the size distribution of charged HNDs. As all clusters from a supersonic expansion have the same velocity, an electrostatic sector field analyzer is suitable to determine the mass (distribution).*

Figure 2 shows four droplet size distributions, measured via expansion of 20 bar through a pinhole nozzle with a diameter of 5 µm at four different nozzle temperatures. All data can be fitted perfectly with a log-normal distribution yielding average droplet sizes of 11.6 (5.31 [48], 9 [44]), 4.4 (2.4 [48], 3 [44]) and 1.1 (1.82 [48], 1.3 [44]) million for nozzle temperatures of 8.1 K, 8.7 K and 9.0 K, respectively, and 69000 (325000 [48], 140000 [44]) for 9.5 K. The values in the brackets are average droplet sizes for the same expansion conditions published in the literature. For these expansion conditions, we measured up to a droplet size of about seven million He atoms. This value is determined by the maximum voltage that can be applied to the cylindrical plates of the 90° sector field and corresponds to a kinetic energy of more than 7 keV (the velocity of the beam expanded at a temperature of 8.1 K is 207 m/s [48]). Thus, this method will inevitably fail for very heavy droplets formed via Rayleigh breakup [19], *e.g.* the kinetic energy will be 6 GeV for a droplet containing $10^{13}$ He atoms with a speed of 165 m/s [11]. The voltage required to deflect such a singly charged HND through our 90° sector would be 2.6 GV, which is not feasible.

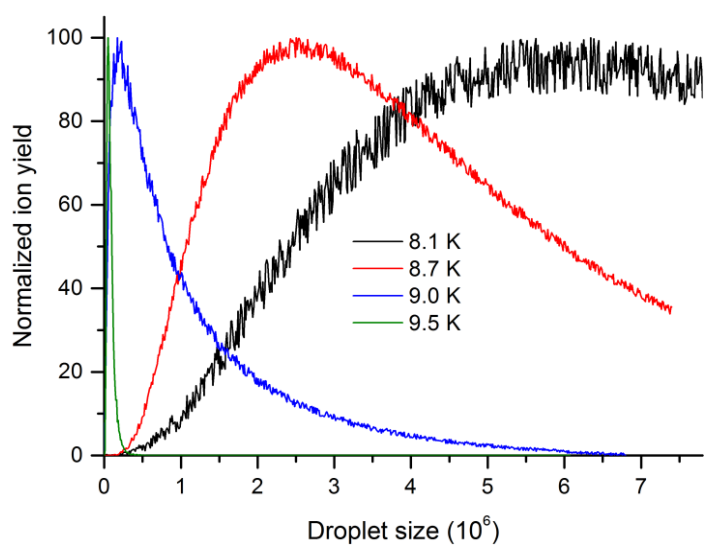

*Figure 2 Size distribution of neutral HNDs measured with the setup shown in Figure 1 for different temperatures of the He stagnation chamber and corrected by a factor $N^{-2/3}$ to account for the geometric (ionization) cross section of different droplet sizes. Parameters: $p_{He}$ = 20 bar, nozzle diameter = 5 µm, electron energy = 30 eV, electron current = 5 µA. The average droplet size is 11.6, 4.4 and 1.1 million for nozzle temperatures of 8.1 K, 8.7 K and 9.0 K, respectively, and 69000 for 9.5 K.*

To determine the size distributions for larger droplets, another approach is required and the attenuation of the droplet beam through collisions with argon and helium gas at room temperature has been employed for this purpose [48]. The results obtained are in good agreement with previous measurements in the size range $<N>$ = $10^5$–$10^7$, which provides confidence in the method. This method has been used to determine sizes up to $10^{10}$ atoms.

Mass spectra obtained via electron ionization show a large increase in the intensity of the $He_4^+$ signal upon increase of the droplet size. Vilesov and co-workers have shown that this can be related to the size of the HNDs. In particular, they have suggested using the ratio of the ion yields of $He_4^+$ to either $He_2^+$ or $He_3^+$, to provide a secondary size standard in the droplet size range of $<N>$ = $10^4$–$10^9$ atoms [48]. A different method for the determination of the size of large He droplets (about $10^{10}$ atoms) was also introduced recently by the same research group [49]. From the pulse height distribution of the ion bursts obtained by multiple electron ionization of a single large HND, both the size distribution and droplet flux in a beam of large He droplets of about $10^{10}$ atoms were measured. Based on the assumption that the amplitudes of the resulting pulses are proportional to the droplet cross section, the results show that the size distribution for smaller droplets ($<N>$ between $10^6$–$10^7$ atoms) can deviate considerably from a log-normal distribution.

Very recently, the emergence of powerful X-ray light sources at free electron laser facilities has enabled the characterization of the shape of large HNDs ($N_{He} = 10^9 - 10^{11}$, which corresponds to droplets with radii $R = 300 - 1000 nm$) [50, 51]. Most of the droplets exhibit spherical to oblate shapes with aspect ratios up to 1.5. In addition, approximately 1% of the diffraction images exhibit pronounced intensity anomalies radiating away from the image center. These images indicate oblate metastable superfluid droplets with angular momenta beyond the stability range of classical droplets. The high angular momentum that is responsible for the deformation of the droplets also gives rise to the formation of quantum vortices. X-ray scattering has also been used to image these vortices. In this case, Xe atoms were added to the helium droplets: the Xe atoms bind to the vortices and act as strong X-ray scatterers. The formation of quantum vortices has also been studied theoretically in pristine HNDs via density functional theory [52].

Very large droplets of helium in the size range of several µm to cm have been formed by various techniques. Among these techniques are rapid cooling, pumping of a cell filled with gaseous or liquid He [53], and formation of large droplets via coalescence [54]. The size of these droplets were large enough to be determined by optical means. Large droplets have also been produced by means of a piezoelectric transducer immersed under liquid helium [55, 56]. Charged helium surfaces in a strong electric field can become unstable and emit multiply charged droplets, as observed by Boyle and Dahm [57]. Niemela reported a method of producing a fountain of charged droplets by operating a field emission point just below the liquid surface [58] . Because of the much smaller surface to volume ratio compared to HNDs, and thus limited rate of evaporative cooling, these large drops have temperatures in the kelvin range as opposed to the mK range of smaller droplets.

## 2.1 Mass spectra of pristine helium nanodroplets

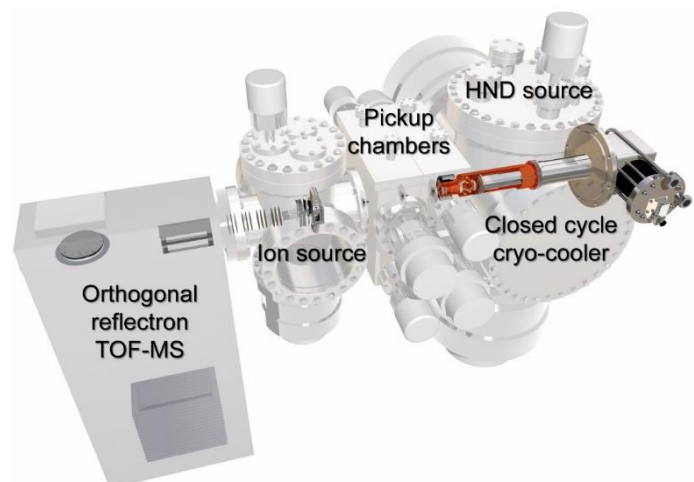

*Figure 3 Schematic drawing of the experimental apparatus in Innsbruck.*

Owing to the high ionization energy of helium, 24.59 eV [59, 60], electron ionization was the first method utilized to form ions from pristine [36, 61-63] and doped HNDs [12, 64-67]. In combination with mass spectrometry, cluster size distributions of foreign molecules were measured. Early studies often utilized quadrupole mass filters with rather poor mass resolution. This makes it very difficult to extract the individual contributions of isobaric ions such as $He_7^+$ and $N_2^+$ at m/z = 28 or $He_8^+$ and $O_2^+$ at m/z = 32. Nitrogen and oxygen originate from the residual gas in the ion source. Sector field mass spectrometer instruments provide higher resolving power, but the rather low ion yield of $He_n^+$ ions requires slow scan speeds for a reasonable signal-to-noise ratio. The temperature of HND sources cooled by a closed cycle cryocooler oscillates with a periodicity of typically seconds by about ± 0.05 K. Furthermore, mechanical oscillations of the closed cycle cryostat lead to small displacements of the neutral HND beam with respect to the skimmer and electron beam, which is located several tens of cm downstream. These effects result in instabilities of the yield of $He_n^+$ ions and make a comparison of the peak heights of neighboring peaks difficult for mass spectrometer systems that slowly scan the mass range. Time-of-flight mass spectrometers measure the whole mass range in every extraction pulse and therefore temporal fluctuations of the HND beam and ion source contribute to every mass peak in the same way, and so will be averaged out. Long-term drifts of voltage supplies or thermal expansion/shrinkage of the length of the flight tube can be corrected as mass spectra are typically saved every few seconds while the entire sequence of measurements may last for many hours.

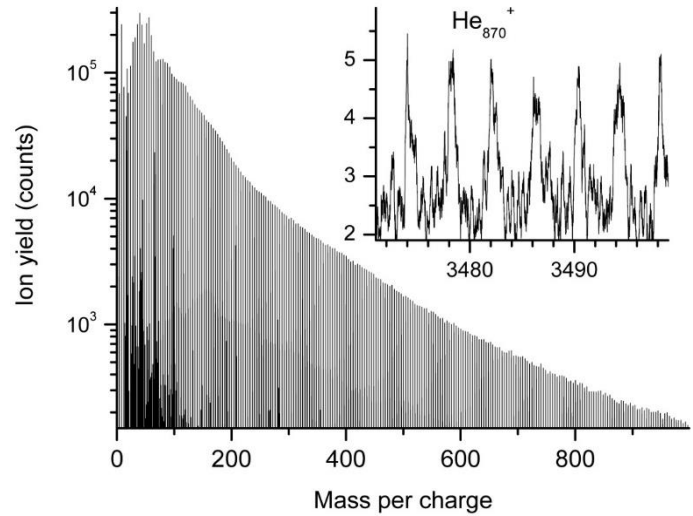

*Figure 4 Mass spectrum obtained via electron ionization of pristine HNDs. The operating conditions were $p_{He}$ = 2.5 MPa, $T_{He}$ = 9 K, $<N>$ = $10^6$ [48], $E_{el}$ = 78 eV, $I_{el}$ = 54 µA.*

Figure 4 shows a mass spectrum measured with the TOF-MS shown in Figure 3. The mass spectrum was accumulated over 800 s with a total ion yield of 67 kHz and a TOF operating frequency of 10 kHz. Except for a peak at m/z = 18, which results from residual water in the instrument, the mass spectrum is entirely dominated by $He_n^+$ ions. At lower masses up to about m/z = 120, some pronounced intensity anomalies are present. For larger masses the size distribution smoothly decreases with increasing cluster size. The inset shows a section of the same mass spectrum near the end of the mass range (which is determined by the TOF pulse repetition frequency). Even here, where m/z is well in excess of 3000, one can still easily see mass peaks that are spaced by 4 mass units that can be assigned to $He_n^+$ ions. In order to determine the contribution of every $He_n^+$ ion and correct for possible overlapping isobaric ions, we evaluated this mass spectrum with the program IsotopeFit, which is a versatile tool to analyze complex mass spectra that contain one or more cluster ion series [68]. Figure 5 shows the result of this analysis; the yield of $He_n^+$ ions is plotted as a function of the cluster size $N$. The error bars indicate possible uncertainties during the fitting routine due to the presence of isobaric ions. For instance, the error for $He_7^+$ is unusually large and the inset shows an expanded view of the original mass spectrum. At m/z = 28 one can identify four ions (indicated by the four colored peaks) in the measured mass spectrum (solid black line). The shoulder on the high-mass side of the $He_7^+$ peak is common for all $He_n^+$ ions and results from metastable decay of the helium cluster ions on their way through the TOF-MS. Note that the fitting algorithm uses a custom peak shape for each helium peak, which takes into account that fragmentation of ions through collisions with residual gas and with grid wires leads to a small satellite peak that arrives a few ps after the main peak, whereas a Gaussian is used for all other peaks originating from dopants in the HNDs and residual gas ionized in the ion source. For example, the full width at half maximum for the peak from $He_{130}^+$ at m/z = 520.338 is $\Delta$m/z = 0.074, which gives a mass resolution of m/$\Delta$m = 7031. It is interesting to note that the average cluster size distribution determined via the ratio of the ion yields of $He_4^+$ and $He_2^+$ does not agree with the values reported in [48], most likely due to a smaller pinhole diameter of the nozzle used. The accuracy stated by the manufacturer (Günther Frey GmbH) is ± 1 µm for a 5 µm pinhole.

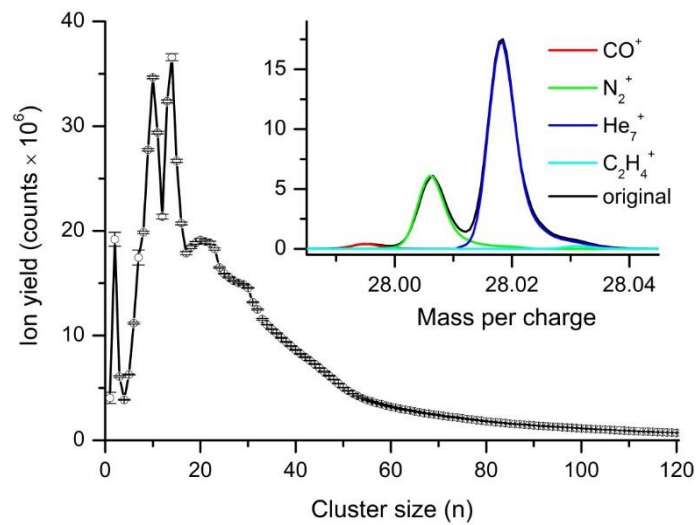

*Figure 5 Cluster size distribution determined via IsotopeFit [68] from the mass spectrum shown in Figure 4. The inset is an expanded view of the isobaric ions at m/z = 28, where four different ions could be assigned; they reproduce the measured ion yields reasonably well.*

Although the mass resolution of early studies utilizing quadrupole mass filters was insufficient to separate isobaric ions such as $N_2^+/He_7^+$ or $O_2^+/He_8^+$, the authors were able to report magic number ions [69]. Stephens and King [63] reported these magic numbers for both isotopes of helium ($^3$He and $^4$He) and measured ion intensities upon electron ionization of neutral He cluster beams utilizing a quadrupole mass filter. Helium was expanded through a 5 µm pinhole at 4.2 K and 0.53 bar for $^4$He and 3.2 K and 0.73 bar for $^3$He. Kobayashi *et al.* formed cluster ions of $^4$He by passing He$^+$ through a 10 cm long drift tube, which was cooled with liquid helium to about 4.5 K and filled with He at pressures between 2.7 and 8 Pa [70]. Two years later, Buchenau *et al.* [36] determined cluster size distributions of $He_n^+$ via electron ionization of neutral He clusters formed by expansion of helium through a 5 µm pinhole at various temperatures (ranging from 5 to 20 K) and pressures (ranging from 8 to 20 bar). The group of Janda reported electron ionization of neutral HNDs ranging from an average droplet size between 100 and 15000. Magic number helium cluster ions were seen at n = 7, 10, and 14, and very weakly for n = 22, 30, 54, 57, and 77 [66]. In 1996 Toennies and coworkers reported the first mass spectrum of $He_n^+$ ions formed via single photon ionization, which was achieved utilizing synchrotron radiation [71] at photon energies above the ionization energy of atomic He.

In Figure 6, cluster size distributions reported in the literature [36, 63, 66, 70, 71] are compared with a cluster size distribution recorded in our laboratory as determined from the mass spectrum shown in Figure 4 and measured with the time-of-flight mass spectrometer system shown in Figure 3. Except for very low cluster sizes, all mass spectra reveal intensity anomalies at the same cluster sizes as reported previously (indicated by the vertical dashed lines in Figure 6). This indicates that for $He_n^+$ (n > 4) it is the stability of the ion that determines the abundance of a particular cluster size, rather than the means of formation of the helium clusters or the method used for ionization. However, for $He_n^+$ with n < 5, the behavior is somewhat different. According to the drift tube experiment of Kobayashi *et al.* [70] and theoretical considerations [72-76], $He_3^+$ is by far the most stable cationic helium cluster in this low-mass range, and yet $He_2^+$ and $He_7^+$ are more abundant products in the electron ionization of large HNDs [48, 63, 77, 78]. Buchenau *et al.* suggested that recombination of two metastable $He_2^*$ excimers, inside or on the surface of a large HND, was the reason for this observation [77]. Von Issendorff *et al.* concluded from photodissociation experiments of $He_4^+$ and measurements of the fragment kinetic energy release that the magic tetramer is formed in a metastable electronically excited state [78]. A recent on the enhanced production of $He_4^+$ in large

HNDs by means of pulsed electron impact excitation of pure HNDs was presented by Fine *et al.* together with a qualitative model to explain their findings [79], see also section 5.6.

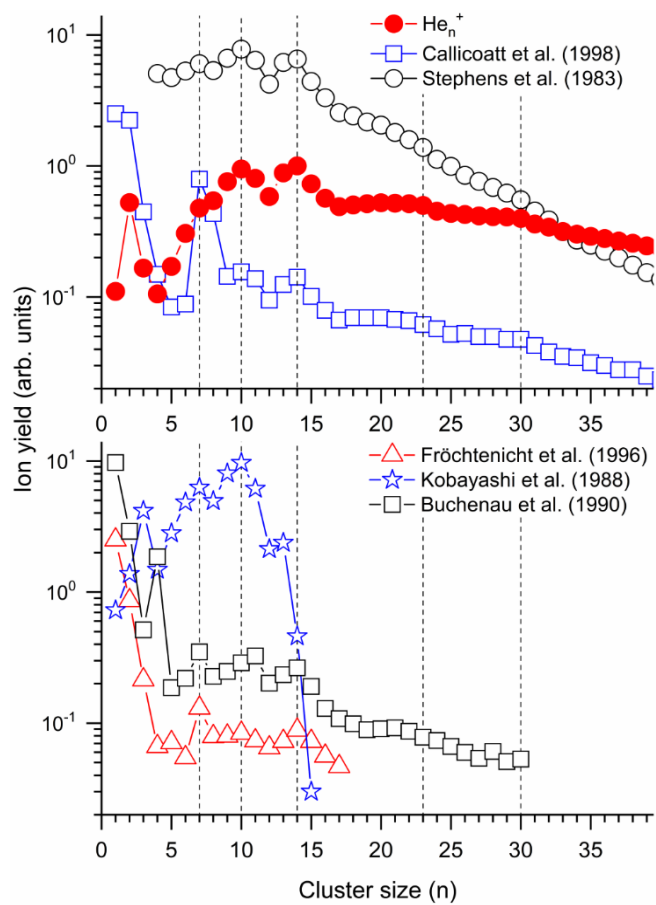

*Figure 6 Cluster size distributions for $He_n^+$ taken from the literature (open symbols) and compared to the values obtained with the TOF mass spectrometer system schematically shown in Figure 3 (red solid circles). Magic numbers are indicated with dashed vertical lines at n = 7, 10, 14, 23 and 30.*

In 2004 Brühl *et al.* deduced the size distribution of neutral He clusters from the diffraction angles of a beam of these clusters sent through a 100 nm period transmission grating [19]. Magic numbers were reported at n = 10-11, 14, 22, 26-27, and 44 atoms. These magic numbers occur at threshold sizes for which quantized excitations calculated with the diffusion Monte Carlo method are stabilized, thereby providing the first experimental confirmation for the energy levels of $^4$He clusters. Figure 7 compares the normalized cluster size distributions obtained from a transformation of the diffraction patterns from [19] and from the ion intensities shown in Figure 5. The fairly good agreement of the magic numbers of neutral and charged He clusters may be a coincidence as ionization leads to the formation of a tightly bound ionic core of two or three helium atoms [72, 73, 76] and the binding energy then released leads to the evaporation of up to several thousand monomers.

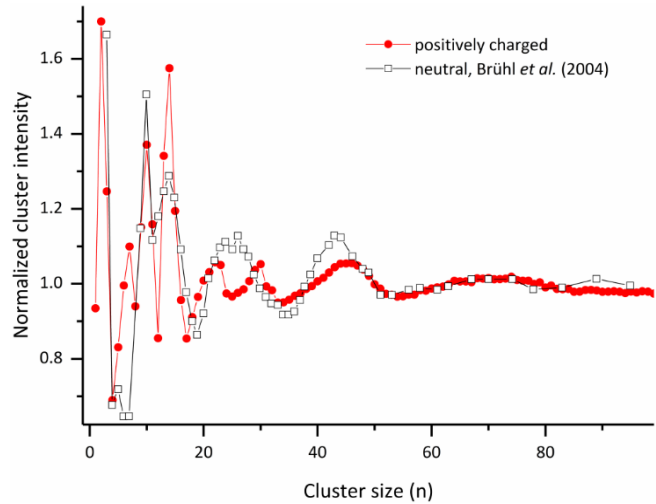

*Figure 7 Normalized cluster size distribution obtained from the transformation of diffraction patterns by passing a beam of neutral He clusters through a 100 nm period transmission grating [19] (open squares). Solid dots represent the ion cluster distribution obtained by normalization of the curve shown in Figure 5 with a multi Gaussian fit.*

## 3   Pickup of atoms and molecules (dopants)

In contrast to bulk He, most atoms and molecules are easily picked up inside HNDs. This was demonstrated first for Ne atoms by Scheidemann and Toennies in 1990 [12]. Vibrational spectra of HNDs doped with $SF_6$ molecules were recorded by Goyal *et al.* in 1992 [13].Other early studies on the capture and formation of clusters of noble gas atoms, as well as of $H_2O$ and of $SF_6$ molecules, were reported by Lewerenz *et al.* [65]. The neutral noble gas atoms were found to coagulate within the HND. This was also the case with $H_2O$ and $SF_6$ molecules. According to calculations of classical trajectories, after the first molecule is captured and thermalized by the HND, a second captured molecule migrates through the droplet and is attracted to the first molecule through van der Waals interaction ($\propto -1/r^6$) and the formation of a dimer between the two dopants occurs within 0.1–10 ns [65]. There is generally enough time between the dopant pickup and probing ($\approx$ 1 ms) by ionizing electrons or photons, for dopants picked up by the droplet to agglomerate. The heat of reaction is released through collective excitation of the surrounding helium leading to evaporation of He atoms. The formation of multiple clusters independently within a single large droplet is also possible. This concept of multi-center growth is discussed in Section 3.3. Dopants can be introduced in the HND by pick-up, where the HND collects vaporized dopants on the fly-by, by electrospray ionization followed by trapping [80] or ion deceleration [81], by laser ablation techniques [82-84], or by merged beams [85]. While dopant efficiencies for neutrals are usually high, they can go down for the pickup of ions with high kinetic energies [86].

## 3.1 Collision dynamics in helium nanodroplets

The dynamics of cluster formation in HNDs is still a matter of debate. In recent work, Hauser *et al.* studied collisions of two metal atoms in HNDs and the mass dependence of their collision times [87].

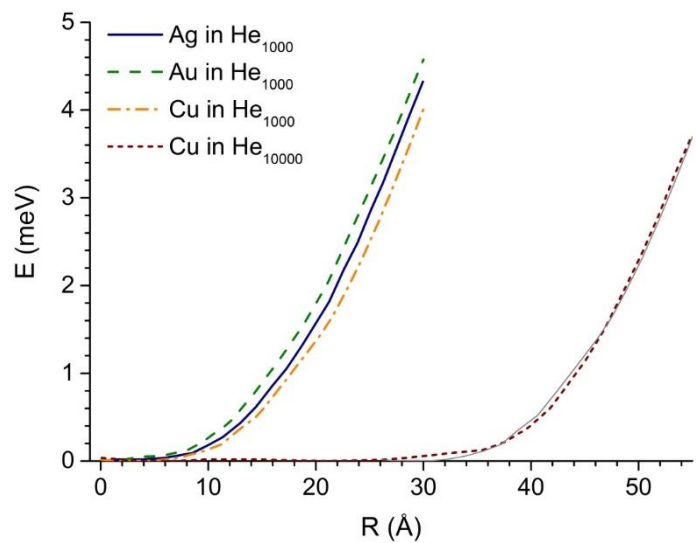

*Figure 8 Calculated confinement potentials of Ag, Au, Cu in He$_{1000}$ droplets and of Cu in ten times heavier He$_{10000}$ droplets with data taken from Ref. [87]. The thin gray line is a copy of the dot dashed line moved to the right by 26 Å; it shows that the droplet size has very little influence on the shape of the confinement potential in the edge region.*

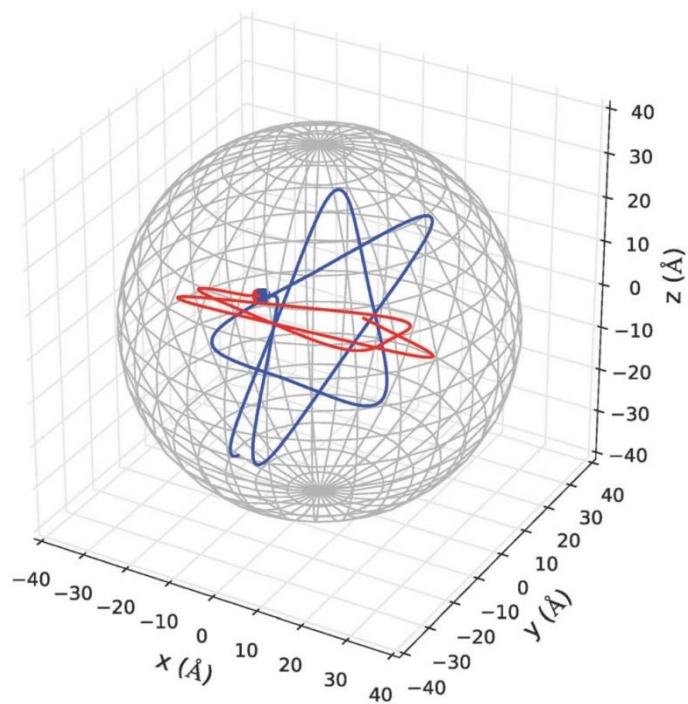

*Figure 9 Trajectories of two Cu dopants until collision inside a HND with 5000 helium. Adapted from Ref. [87] (Fig. 5b) licensed under CC BY 3.0..*

The solvation energy of single atoms embedded in HNDs and the influence of the He environment on dopant-dopant dimer interactions were treated with a He-density functional approach. Dopant-dopant interactions were corrected for the He-rearrangement energy that is needed to create density holes for the dopants. This gives very small corrections to the dimer potential energy curve of Au$_2$, for example in a range of -11.2 meV to +1.9 meV depending on the Au-Au distance. For comparison the Au-Au dimer potential well depth was given by $D_e$ = 2.199 eV. The solvation energy for single-atoms leads to a shallow, spherically symmetric confinement potential, shown in Figure 8, with a minimum at the center of the droplet. In large droplets this basin is very flat in the center and only shows a significant rise near the outer region of the droplet. This confinement potential is not very strong; the solvation energy of Au in He$_{1000}$ was reported to be 21.8 meV. However, such a shallow potential suffices to hold the dopants in the droplet because the kinetic energy of gold at the Landau velocity lies below 3.2 meV. Interestingly, the dopants can cross the estimated surface region of the HND and their reflection points can lie outside the droplet in the case of heavy atoms such as Au, whereas it lies inside for lighter atoms such as Cu.

The classical simulations of Hauser *et al.* paint a vivid picture of two atom collisions in HNDs. Particles picked up by the droplet are rapidly cooled to velocities below the Landau limit of ~ 56 m/s [88]. Inside large droplets the particles move without friction along straight lines. In the outer region of the droplet (or beyond) the dopants are reflected by the confinement potential arising from the solvation energy. The reflections continue until the particles come close enough to attract each other. Two typical trajectories that lead to collision are shown in Figure 9. The Au-Au binding energy is rapidly quenched by helium evaporation and a dimer is formed inside the HND. The average time until collision is slightly longer for heavier atoms because, as a result of their higher kinetic energies, their turning points are beyond those for lighter species. This mass-dependence of the collision time becomes smaller for larger droplets, where the collision partners spend most of their time well inside the droplet at constant velocity, and only a small fraction of the total time in the outer region. A statistical sampling yielded collision times of 515 ps for Cu and 1374 ps for Au in small He$_{2000}$ droplets, and 11031 ns for Cu and 14452 ns for Au in large droplets with a radius of 1000 Å. For large droplets the authors replaced the confinement potential with a hard wall potential at the turning points. In summary, Hauser *et al.* showed that the dimer binding energy has only a minor influence on the collision times, whereas a strong mass dependence was predicted for small HNDs.

Time dependent calculations of the quantum dynamics of the He response to classical or quantum impurities have recently become possible with the Barcelona-Toulouse $^4$He DFT code for atomic impurities with and without imprinted vortices [52, 89, 90]. For example, the code allows for simulating the pickup and capture process of atomic dopants and the response of the He density that can react rather violently to impinging impurities and may form solvation shells later on. Videos of the HND dynamics for a Rb dopant that is first excited and then ionized can be found in the Supporting Information of Ref. [91]. Here, the interplay between the heliophobic Rb* and the heliophilic Rb$^+$ and the timescales between excitations govern the overall dynamics. In a recent combined experimental and theoretical study, the formation and desorption dynamics of RbHe exciplexes initiated by laser excitation of Rb atoms attached to HNDs was investigated leading to detailed analysis of the real-time dynamics of an alkali-metal exciplex, still some uncertainties remain [92].

### 3.2 Growth of unusual structures

Cluster growth in HNDs can lead to unusual structures and in particular also to very weakly bound structures that would not form easily in the gas phase. Typical examples are linear molecules with a strong permanent electric dipole moment, such as HCN and HCCCN, that grow into chains in HNDs. An incoming molecule is oriented in the electrostatic field by the dipole from the (chain of) polar molecule(s) in the center of the HND, and vice versa [93, 94]. The two moieties move presumably along the field lines and the molecules are attached non-covalently in a head-to-tail arrangement, even if the linear structure is energetically unfavorable compared to other configurations. At elevated temperatures the rotations of the molecules will average out the dipole interactions, rendering chain formation unlikely. The low temperature of a HND environment thus allows for self-assembled cluster growth leading to structures that are neither in the energetic minimum, nor in thermodynamic equilibrium.

Another example of unusual structure formation is the cyclic water hexamer. This structure is stabilized by rapid quenching in the HND, whereas in the gas phase this structure does not form [95]. The time between capturing dopants is much longer than the time to cool molecules in the HND and the energy released by cluster formation is removed efficiently by the liquid helium. Thus clusters can be trapped in local minima and are not able to surmount barriers towards lower energy basins in the potential energy surface.

Gomez *et al.* formed elongated structures of Ag atoms in HNDs in 2012 and attributed their appearance to the existence of quantized vortices [96]. We refer to Section 4.4 for more details on quantum vortices in and nanowires from HNDs.

### 3.3 Multicenter growth of metallic nanoclusters

Multi-center growth will occur when the rate of addition of a dopant exceeds the nucleation frequency [97]. The average time between two pickup events and the average recombination time dependence is illustrated for $Ag_n$ clusters in Figure 10.

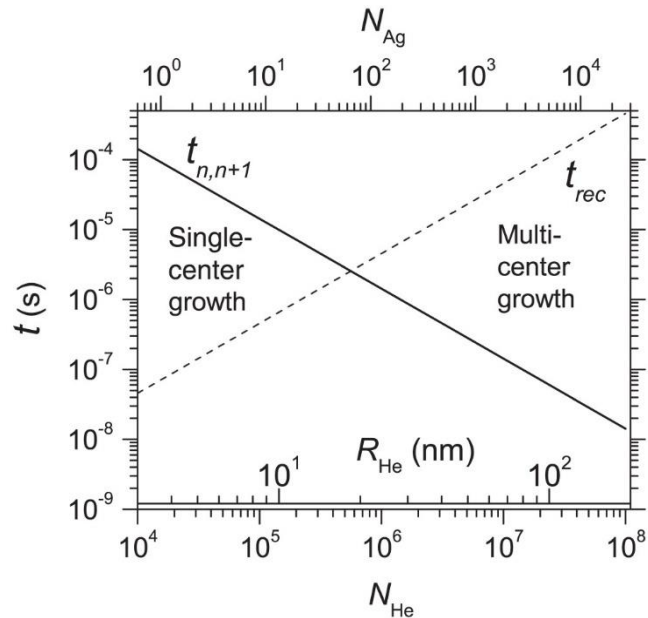

*Figure 10 With increasing HND size the time between two pick-up events $t_{n,n+1}$ decreases, whereas the recombination time $t_{rec}$ for clustering inside the HND increases. Multicenter growth is dominant in large HND, where $t_{rec} \gg t_{n,n+1}$. Reprinted figure with permission from Ref. [97]. Copyright 2011 by the American Physical Society.*

As a consequence, local nucleation centers can form at several different sites within a droplet. Given sufficient time, the individual particles grown at these separate sites will agglomerate into a granular composite but the individual grains are unlikely to merge into a single compact structure at 0.37 K. Strong evidence for exactly this process was provided by Loginov *et al.*, who recorded optical spectra of silver clusters inside HNDs with different mean sizes [97]. For relatively small droplets, which gave small $Ag_n$ clusters, the spectra were consistent with an intense plasmon-like absorption in the near-ultraviolet. However, for droplet sizes exceeding $10^6$ helium atoms, a second absorption feature appeared which spanned the red and near-infrared parts of the spectrum. Since a similar absorption feature is seen in the spectra of colloidal metal nanoparticle aggregates, this new band was attributed to a composite of smaller particles formed by multi-center growth. Silver clusters have been probed near the plasmon resonance of large spherical Ag clusters (355 nm) and at 532 nm, where multicenter grown, branched Ag clusters absorb. The saturation of the depletion signal at 532 nm was attributed to melting of the branched structure into a compact spherical cluster within the droplet [98]. The observation of Ag/Au core-shell nanoparticles (see Section 4.3) with two distinct Ag cores may also be the product of multi-center, or in this case two-center, growth of the Ag cores [99]. In large droplets it is possible that two distinct core nanoparticles, grown at very different positions within the droplet, would have insufficient time to recombine before they arrive in the second pick-up region.

## 3.4 Core-shell nanoclusters

The effect of a protective layer of the co-dopants Ar, $O_2$, $N_2$, CO, $CO_2$, NO, and $C_6D_6$ on water cluster fragmentation upon electron impact ionization was studied by Liu *et al.* in 2011 [100]. In some cases the co-dopants act as a protective layer against fragmentation; a higher yield of $(H_2O)_n^+$ clusters compared to protonated $(H_2O)_nH^+$ clusters can be achieved than without co-doping. This softening effect is efficient for $N_2$, $O_2$, $CO_2$ and $C_6D_6$, whereas for CO and NO secondary reactions can also favor fragmentation for some water cluster ions.

Silver/ethane core-shell clusters have been grown in large HNDs of up to $10^7$ atoms by Loginov *et al.* [101, 102]. Infrared spectroscopy revealed a frequency shift of a specific C-H stretching vibration for the ethane molecules that surround the Ag core that was sensitive to the location of the ethane molecules relative to the silver core. The intensity ratios of the absorptions from these 'inner' and 'outer' ethane molecules was shown to be consistent with a model of a densely packed silver core surrounded by dense ethane shells. Larger clusters of about 3000 Ag atoms showed a five times larger adsorption capacity than suggested by dense packing, which was attributed to the formation of elongated or ramified structures. Similar investigations were carried out for large silver clusters with shells of methane, ethylene, and acetylene also formed in HNDs [103].

Metallic core-shell nanoclusters created by multicenter growth are treated in Section 4.3 and core-shell Ag-Au nanowires created along vortices in HNDs are treated in Section 4.4.

## 3.5 Growth on the edge of helium nanodroplets

Submersion of a dopant atom or molecule into a HND is governed by the balance between the energy lowering provided by the interaction of the dopant with helium and the energy needed to create space for the dopant in the HND (which arises from the helium-helium interactions). These interactions differ significantly for different dopants and lead to the classification of two different species: heliophobic and heliophilic. To help distinguish between heliophobic and heliophilic species Ancilotto *et al.* [104] proposed the dimensionless parameter

$$\lambda = \frac{\rho \epsilon_d r_m}{2^{1/6} \sigma} \qquad (2)$$

where $\rho$ is the helium number density, $\epsilon_d$ and $r_m$ are the potential depth and equilibrium separation of the He-dopant diatomic interaction, and $\sigma$ is the surface tension. Dopants become submerged if $\lambda$ is higher than the value of liquid $^4$He, $\lambda_{He} = 1.9$ (and *vice versa*). Using density functional theory, Dalfovo predicted that alkali atoms are not stable in the center of HNDs [105], whereas many other species are stable. Ancilotto *et al.* calculated surface binding energies of 10-20 K for alkali atoms near the liquid-vapor phase of a HND [106]. Due to the very weak interaction between alkali-atoms and helium atoms, alkali atoms are prime examples of species that are heliophobic and prefer to locate at the HND surface. This has been confirmed by several experiments: for example, two types of transitions were observed by laser induced fluorescence spectroscopy of alkali atoms attached to HNDs, corresponding to excitation of states with orbitals aligned parallel or perpendicular to the HND surface [107, 108]. Note also that fluorescence spectra of alkali atoms and small alkali clusters formed at the HND surface are barely shifted compared to gas phase spectra [109].

Although alkali atoms and small alkali clusters remain at the surface of HNDs, larger clusters of alkali atoms have been shown to submerge into the droplet at a given size, notably n ≥ 21 for $Na_n$ clusters [110] and n ≥ 80 for $K_n$ clusters [111] (see Section 7.2 for further information).

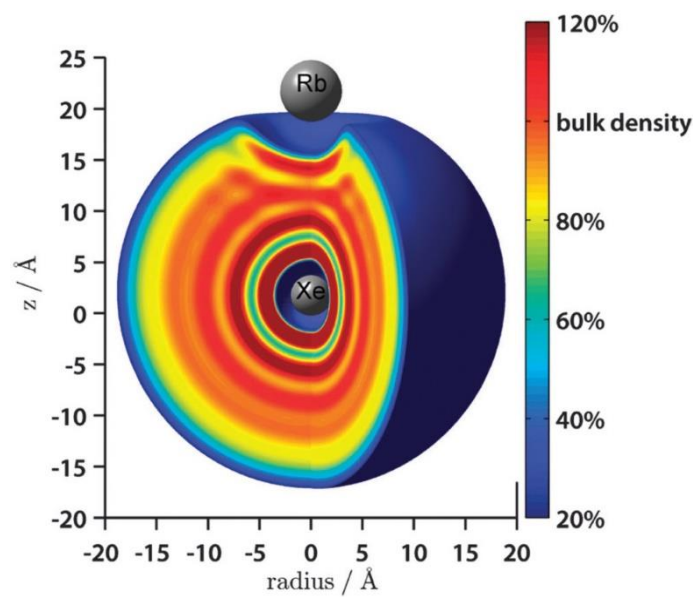

Figure 11 The helium density distribution is shown for a $He_{500}$ droplet doped with heliophilic Xe and heliophobic Rb. The distribution was calculated using the Orsay-Trento density functional [112]. Reproduced from Ref. [113] with permission from the PCCP Owner Societies.

The co-doping of a small HND with both Xe and Rb has been studied with density functional theory [113]. The behavior of heliophilic Xe and heliophobic Rb are shown together with the predicted helium density in Figure 11. For these species the Rb-Xe van der Waals attraction with a potential well depth of around 12.4 meV is comparable to the repulsion caused by the different helium solvation energies. In total, the Rb-Xe binding energy in the HND is reduced to ∼ 4.5 meV with a barrier of 2 meV impeding dimer formation.

## 3.6 Growth of large clusters

Large dopant clusters are of particular interest in applications such as catalysis, adsorption of high-energetic molecules for energy storage, sensors, etc. Naturally, large HNDs are needed for the production of large clusters, since every atom/molecule attached to the clusters releases its binding energy to the cluster and leads to the evaporation of an amount of helium atoms that is proportional to this energy. Large clusters of propyne, $C_3H_4$, with $10-10^4$ molecules were prepared in HNDs of $10^4-10^7$ He by Mozhayskiy et al. [114]. In the same study, gold and silver clusters, and mixtures thereof, with around 500 atoms were prepared in HNDs with $2.5\times10^6$ helium atoms, and a core-shell structure was proposed. A flux of $3\times10^{11}$ atoms/(cm²s) could be achieved for depositing metal atoms formed in HNDs on a surface, corresponding to an atomic monolayer in 100 minutes. This makes

HNDs an alternative to other deposition techniques, with the advantage of controlled cluster growth, composition, and size, along with soft landing conditions.

### 3.7 Manipulation of dopants

Laser induced alignment of molecules in HNDs has been pioneered in the group of Henrik Stapelfeldt [115]. Molecules were impulsively aligned with a laser pulse that was shorter than the timescale of the rotation of the molecule. Alignment of $CH_3I$ could be achieved 17−37 ps after a 0.45 ps laser pulse with $\lambda$ = 800 nm. Relaxation to random orientations then took around 70 ps. The orientation of the molecules has been deduced from velocity map imaging after an intense 30 fs probe pulse that leads to multiple ionization followed by Coulomb explosion. The induced alignment is much slower than for isolated molecules in vacuum and in contrast to isolated molecules, revivals and permanent alignment were not observed at first in HNDs. Recently, Shepperson *et al.* reported revivals of the alignment in HND using a fs laser at low fluence, where the molecule rotates together with its solvation shell [116]. For high fluence, rapid, gas-phase like dynamics is observed at first, but revivals later on disappear. The results were interpreted with the help of the angulon quasiparticle theory [117, 118].

## 4 Nanoparticles and nanowires

The addition of multiple atoms or molecules to HNDs leads to agglomeration to form clusters. With large droplets clusters with nanoscale dimensions can be grown. This, combined with unique properties of HNDs such as the low temperature, the very fast cooling and the ability to add more than one type of dopant sequentially, makes it possible to conceive of growing many novel nanostructures inside HNDs. This in itself is an exciting prospect. Furthermore, successful removal of these nano-objects from the droplets paves the way for generating new and useful nanotechnology. This section reviews this rapidly burgeoning aspect of HND research.

### 4.1 Large helium nanodroplets

Suppose that the target of an experiment is to grow a cluster consisting of 100 metal atoms. Since such structures will have dimensions on the nm scale, we will here onwards refer to these as nanoparticles. Furthermore, let us assume that the atoms bind together via metal-metal bonds with dissociation energies exceeding 1 eV. Since 1 eV is dissipated by evaporation of 1600 helium atoms (assuming 0.6 meV is removed per helium atom), then the minimum acceptable HND size will approach $10^6$ helium atoms. For larger nanoparticles even larger HNDs will be required. This is a different size regime from most of the experiments described elsewhere in this review.

To generate very large HNDs, it is necessary to go beyond the gas condensation mechanism used for making small droplets (< $10^5$ helium atoms). This means colder expansion conditions, leading to operation in either the supercritical expansion regime or, in the extreme case, liquefaction of the helium and production of droplets by breakup of a liquid microjet. In the supercritical case, which begins at nozzle temperatures near 10 K (the exact temperature is pressure-dependent), the droplets derive from fragmentation of the liquid phase, formed by expansion cooling of supercritical helium, as it exits the nozzle [36]. The supercritical expansion generates mean droplet sizes spanning from roughly $5\times10^4$ through to $5\times10^8$ He atoms [48], which show an exponentially decaying size distribution [44]. At nozzle temperatures below the critical temperature of $^4$He, 5.2 K, the liquid phase can fully form behind the pinhole and the helium is driven out as a narrow jet, a microjet [39]. The stability of the microjet is short-lived on account of Rayleigh oscillations and results in droplets a short distance downstream of the pinhole. The breakup mechanism delivers droplet dimensions comparable to the aperture diameter and so droplets of diameters of several microns with in excess of $10^{12}$ helium atoms can form. The expectation in this regime is of relatively monodisperse droplets. These exceptionally large HNDs were first accessed by using a helium bath cryostat to cool the nozzle, which was capable of achieving temperatures as low as 1.5 K [39]. However, it is now possible to achieve temperatures below the critical point of $^4$He with closed cycle cryostats. Indeed, with these cryostats an almost continuous span of droplet size regimes, from the very small through to micron sizes, is accessible.

## 4.2 Soft Landing

Although there is much interesting science to be gained by studying nanoscale objects inside HNDs, any practical application of this technology necessitates removal of the particle from the helium. This can be achieved by collision of a doped HND with a substrate. Growth of unique types of particles in liquid helium would offer little benefit if the deposited particles were then to undergo a significant change in structure as a result of the deposition process. Ideally one would want to be able to *soft-land* the particle, such that it is delivered onto the surface with a kinetic energy well below the threshold for re-structuring.

A simple calculation suggests that this is viable with particles embedded in liquid helium. Consider a HND with $10^8$ helium atoms moving at a speed of 200 m s$^{-1}$, which is typical for a beam composed of such large droplets. The total kinetic energy of the droplet is an impressive 84 keV, which is energy that must be dissipated on collision with a solid surface. However, the energy per atom is low. For example, with a gold cluster embedded within a HND moving at this speed the collision energy per Au atom is ∼ 0.04 eV, which is sufficiently small to be classified as a soft-landing [119].

Although this crude calculation suggests that soft-landing is plausible, a more detailed microscopic understanding of the collision process would be highly beneficial. For example, if the droplet begins to disintegrate as it collides with the surface then part of the helium may be propelled backwards, providing an additional 'cushioning' effect for the approaching dopant nanoparticle. At the same time this splashback could reduce the sticking probability of the nanoparticle. A computational model based upon molecular dynamics might help to clarify such issues. However, a purely classical MD simulation is unlikely to be satisfactory because a HND is a strongly quantum object. The first attempts to provide a quantum treatment of HND-surface collisions have recently appeared [120, 121] and have made use of time-dependent density functional theory to describe the quantum behavior of the helium. This initial work focused on collisions of relatively small droplets (300 He atoms) with TiO$_2$ [120] and graphene [121] substrates. In the case of TiO$_2$ the quantum simulation was compared with a fully classical MD simulation and distinct differences were found. Classical MD showed extensive fragmentation of the droplet and substantial splashback. Very different behavior is predicted by the quantum model. Here the droplet spreads across the surface on a timescale of ∼ 20 ps and no significant splashing occurs. The next significant step in such calculations will be to incorporate a dopant cluster inside the HND and determine exactly how this lands onto the surface and thereby how much energy is deposited into the particle.

## 4.3 Deposition experiments: metallic nanoparticles

Metallic clusters have been the subject of many HND studies. The focus here is on clusters consisting of at least tens of metal atoms, which were first grown and identified by mass spectrometry as far back as 1998 [122]. More recently larger metallic clusters, consisting of more than 1000 metal atoms, have been made. Clusters of Cd, Zn and Mg, were explored via both electron ionization and femtosecond photoionization mass spectrometry to determine features such as electronic and geometrical shell effects [123, 124]. Reasonably large alkali metal clusters have also been studied, see Section 3.5 [110, 111]. All of the studies mentioned in this paragraph focused on the growth and behavior of the particles in HNDs and no attempt was made to remove these clusters from the helium.

The experimental deposition of nanoparticles grown in HNDs was pioneered by Vilesov and co-workers. Crucial groundwork began in 2007 with a study of the optimal pick-up conditions for growing large clusters [114]. The idea was initially tested with a gaseous dopant, propyne, to establish the dopant pressure that generated the maximum flux of embedded molecules for a given HND size. Attention was then turned to gold and silver dopants, which could be grown as nanoparticles that should survive deposition onto a surface. The metal atoms were added via Au- or Ag-coated tungsten coils, which were resistively heated to generate the metal vapor. Downstream of this pick-up zone a quartz crystal microbalance measured the deposition rate. Combined with information gleaned from the propyne measurements, the deposition of $Au_n$ and $Ag_n$ particles composed of several hundred atoms was inferred and the rate of deposition was determined. However, in this early study no attempt was made to view the deposited nanoparticles with microscopy and so there was no direct evidence of their sizes and shapes.

The first explicit confirmation of nanoparticle formation in HNDs was made in 2011 and was derived from the same research group [125]. The basic experimental arrangement is illustrated in Figure 12. In this particular experiment, which was concerned with silver nanoparticles, an alumina oven took the place of the coated tungsten coils and the quartz microbalance was replaced with a substrate holder consisting of several transmission electron microscopy (TEM) grids. With this apparatus it was now possible to image any deposited silver nanoparticles by removing the TEM grids from the apparatus and taking them to an external TEM system. Adopting the maximum flux conditions discovered in their earlier experiments, $Ag_n$ particles with mean sizes of *ca.* 300 and 6000 Ag atoms were expected for mean HND sizes of $2.4 \times 10^6$ and $4 \times 10^7$ atoms. The larger particles were confirmed through the TEM observations, which showed a mean particle size of 6400 Ag atoms and an RMS deviation of 5000. This width is dictated primarily by the exponential distribution of droplet sizes in the supercritical expansion regime.

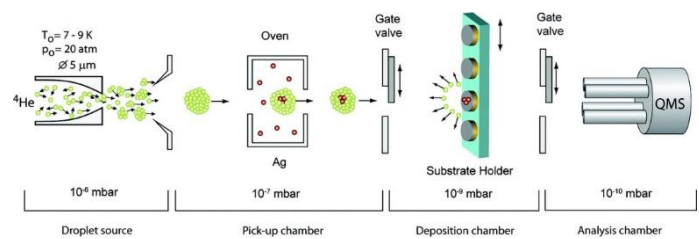

*Figure 12 Illustration of the apparatus used to grow and deposit silver nanoparticles by the Vilesov group. The deposition target can be moved up and down so that different TEM grids can be brought into the line-of-site of the HND beam. Reprinted with permission from Ref. [125]. Copyright 2011 American Chemical Society.*

The fabrication and deposition of nanoparticles using HNDs has seen growing activity in the past four years, with several groups entering this area of research. These studies have started to address the fundamental growth mechanism and have explored various basic properties of the deposited particles. A significant finding, which is perhaps rather surprising given the very low temperature and rapid cooling in superfluid helium, is the formation of crystalline rather than amorphous nanoparticles. The first crystalline structure was reported by Boatwright *et al.* as part of a study of core-shell nanoparticles and was obtained from high resolution TEM images of particles grown from sequential addition of Ag and then Au [126]. Given the very similar lattice parameters of Ag and Au the core-shell structure could not be established but clear evidence for a face-centered cubic (fcc) structure was obtained from diffraction fringes in the TEM images.

A detailed TEM study of deposited particle structures was reported by Volk *et al.* for pure Ag nanoparticles [127]. These particles had mean diameters in the 2-4 nm range and consisted of several hundred Ag atoms. The most abundant particles showed crystalline (fcc) structures. However, particles with non-crystalline, but highly ordered, icosahedral and dodecahedral structures, were also observed and representative TEM images are shown in Figure 13. By recording images for many particles the abundance ratio was found to be fcc > dodecahedral > icosahedral, which is opposite to that expected on energetic grounds. Molecular dynamics simulations have provided insight into this kinetic effect [128]. It seems that it is icosahedral and dodecahedral particles that initially grow inside the HNDs but the smaller particles tend to rearrange into fcc clusters on contact with the substrate, whereas somewhat larger ones can survive intact. This explanation fits nicely with the smaller sizes seen for fcc nanoparticles in TEM images [127].

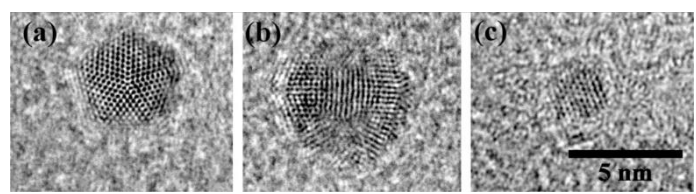

*Figure 13 TEM images of (a) decahedral, (b) icosahedral and (c) face-centered cubic silver nanoparticles. Reprinted from Ref. [127], with the permission of AIP Publishing.*

One of the appealing features of HNDs for the growth of nanoparticles is that materials can be independently added to the droplets through the use of multiple pick-up cells. Provided the droplets are not too large (see Section 3.3), the timescale for agglomeration of dopant A will be complete before the droplets arrive at a second pick-up cell to acquire dopant B. When combined with the low droplet temperature, layer-by-layer structures will be frozen into place, delivering core-shell nanoparticles. In principle one can form a huge variety of core-shell nanoparticles by simply switching pick-up dopants. However, will the core-shell structure be retained after collision of the droplet with a surface? There is evidence for the retention of core-shell structures in deposited nanoparticles. For example, Boatwright *et al.* have presented X-ray photoelectron data for Ni/Au core-shell nanoparticles that suggest survival of the core-shell structure rather than alloying between the two metals [126]. More direct evidence has come from very recent work by Haberfehlner *et al.*, which has presented exceptionally high quality TEM images of both single core and double core Ag/Au core-shell nanoparticles [99]. Structure and composition of individual HND grown core-shell Ag-Au nanoclusters were resolved with atomic resolution [99]. Images of such clusters are shown in Figure 14.

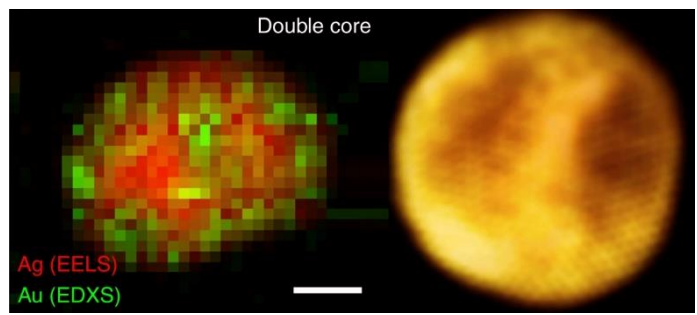

Figure 14 Two images resolving the composition of a double core Ag-Au cluster, recorded for Ag by electron energy-loss spectroscopy (EELS) and for Au by energy dispersive X-ray spectroscopy (EDXS). Adapted from Ref. [99] licensed under CC BY.

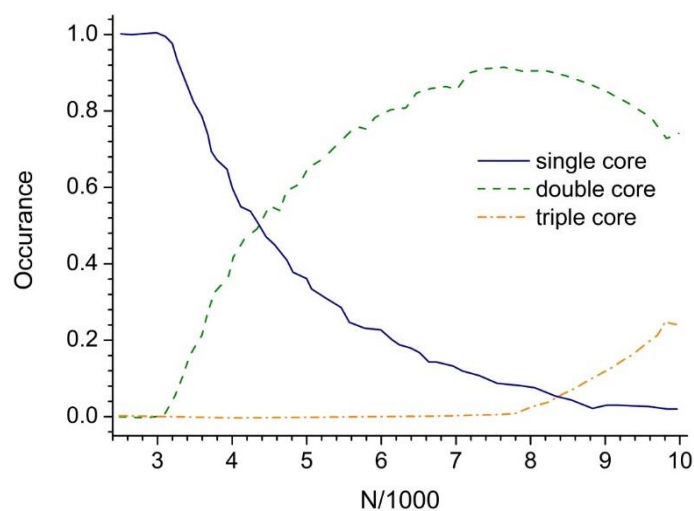

Figure 15 Occurrence of single, double and triple core clusters with increasing cluster size N (number of atoms). Data taken from Ref. [99].

In the work by Haberfehlner *et al.*, the clusters were grown by sequential pickup of first silver and then gold, which leads to a core of silver surrounded by gold [99]. As shown in Figure 15, the authors observed the emergence of double and triple core clusters with increasing HND size. These were attributed to multiple separate nucleation centers in the droplet. The cross over between single and double core growth was given for HND sizes with approximately $4–5×10^7$ He atoms. The clusters were deposited on a carbon grid for transmission electron microscopy. The survival of the double core structure upon impact indicates that melting and alloying is suppressed by efficient helium cooling, which is granted if helium remains in its liquid phase around the cluster. The interaction with the surface deforms the cluster to a lenticular shape. The smoothness of the cluster surface was attributed to surface diffusion after landing. These experimental and theoretical studies of deposited nanowires have been extended to include Cu and Ni [129]. Their melting behavior on heated supports is nicely captured also by theory [130]. Recently, the stability of Ni-Au core-shell nanoparticles on heatable transmission electron microscopy grids was investigated [131]. It was found, that a heating cycle from room temperature to 400 degrees C and subsequent cooling leads to inversion of the core-shell structure from Ni-Au to Au-Ni. Temperature-dependent alloying processes were also studied for Ag-Au and Au-Ag core-shell structures on heatable carbon substrates leading to fully mixed alloys around 573K [132].

The work described above has been concerned with the characterization of individual nanoparticles on sparsely covered deposition targets. A different philosophy has been adopted by Lindsay and co-workers, who have used a HND source to deliberately create nanocluster films. The principal target here has been magnesium films, where the motivation lies in creating new highly energetic materials with applications such as explosives and flares. Magnesium deposition rates well in excess of 1 μg cm$^{-2}$ hour$^{-1}$ were demonstrated and the possibility of achieving rates in excess of 100 μg cm$^{-2}$ hour$^{-1}$ was predicted [133, 134]. In highly energetic materials, a metal such as magnesium will act as a fuel and must be combined with an appropriate oxidizer. A nanostructured mixture of these components brings the reactants into close proximity, enabling faster reaction when initiated. Lindsay and co-workers conjectured that formation of such materials in HNDs, utilizing both the low temperature and the soft-landing, might assist in delivering high quality pre-reactive mixtures that cannot be made by any other means. Preliminary experiments seem to bear this out. One example explored was a nanofilm derived from a mixture of magnesium and a perfluoropolyether (Fomblin) [135, 136]. Using temperature-programmed thermal desorption the films grown from HNDs were shown to consist of unreacted magnesium cores and perfluoropolyether shells, whereas films grown from gaseous deposition of magnesium and Fomblin led to reaction. In another study the same group attempted to form films composed of copper-coated magnesium nanoparticles but found that the core-shell structure becomes inverted, with Cu moving into the core even though it was added to the HNDs after Mg pick-up [137]. This inversion appears to happen before deposition, *i.e.* takes place inside the HNDs.

The nanoparticle deposition experiments described above have delivered nanoparticles with diameters ≤ 5 nm. One might imagine that creating larger nanoparticles is simply a case of scaling up the droplet size so that more dopant atoms/molecules can be added. However, two factors complicate this scaling process. The first is multi-center growth, which becomes significant for droplet sizes > 10$^6$ helium atoms, as discussed in Section 3.3. As well as multi-center growth, a second feature to consider in large HNDs is the possible impact of quantum vortices. This is discussed in the next section.

### 4.4 Deposition experiments: nanowires and vortices

Superfluids are said to be irrotational because they do not undergo a collective rotation of the entire fluid when subjected to a rotational torque, in contrast to normal fluids. When a superfluid, such as $^4$He, does carry angular momentum, it does so in the form of quantum vortices. The axis running through the vortex core is a line of zero helium density about which fluid circulates. Crucially, the angular momentum of the rotating fluid about this vortex line is quantized. This behavior is well known in bulk superfluid helium, as well as in weakly interacting Bose-Einstein condensates.

Until recently the existence of quantum vortices in nanoscale helium was conjectural. This changed in 2012 with the observation by Gomez *et al.* of narrow track-like deposits of Ag nanoparticles formed from relatively large HNDs [96]. These tracks began to appear when droplets with diameters larger than 300 nm were used and can be explained by anisotropic growth of Ag clusters along a quantum vortex. A quantum vortex provides a short-range attractive force which, although very weak [138], is sufficient to pin particles to the core at the low temperature of superfluid helium. Thus, as Ag atoms or small Ag particles, perhaps formed by multi-center growth, migrate through the droplet they eventually meet and become attached to a vortex. The threshold droplet size for seeing this vortex effect does not seem to be a consequence of any fundamental physics: theoretical studies suggest that quantum vortices could exist in relatively small HNDs, although no experimental evidence for vortices in small droplets was found so far [139-141]. However, Gomez *et al.* suggested that the 300 nm size limit may simply be a consequence of the way that droplets of this size, and larger, are made. A schematic of the process is shown in Figure 16 together with a transmission electron microscopy image of the deposited elongated clusters. The theoretical treatment of vortices in HNDs is complicated and beyond the scope of this review; an introduction into this topic may be found in Ref [142].

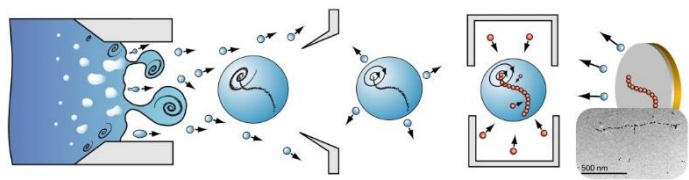

*Figure 16 Formation of HNDs with vortices by skimming, evaporation, doping with silver and deposition of the linear structure created along the vortex. In the bottom of the figure a TEM micrograph from an elongated Ag structure created by this procedure is shown [96]. Reprinted figure with permission from Ref. [96]. Copyright 2012 by the American Physical Society.*

In particular, the fragmentation of the helium as it breaks into droplets, perhaps assisted by collisions with the walls of the nozzle, might be the source of the necessary angular momentum and the vortices survived fragmentation and the transition to superfluidity upon evaporative cooling in vacuum. The presence of branched structures in the TEM images of Gomez *et al.* indicate a variety of vortex geometries from curved lines to rings and parallel vortices. Subsequent work by Spence *et al.* has shown that the constituent silver particles in the deposited tracks may be spherical (diameter ~ 10 nm) or rod-like, depending on the droplet size used, with rod-like particles appearing in the largest droplets [143]. However, the particle separation, measured from the mid-point of each particle, whether rod or sphere, is roughly the same. By depositing the droplet onto a heatable substrate, Volk *et al.* have recently provided compelling evidence that the formation of distinct particles is a consequence of the fracturing of a continuous silver nanowire via Rayleigh breakup after deposition [144].

Latimer *et al.* have demonstrated that continuous nanowires can be grown in HNDs for a variety of metallic elements [145]. The case of nickel is illustrated in Figure 17. Here the progressive elongation of Ni nanoparticles can be seen as the droplet size is increased, with the departure from a spheroidal shape happening for droplets as small as 55 nm. This work showed that the length of the wires could be controlled by varying the dopant pick-up pressure for a fixed HND size but the width of the wires remained roughly constant at around 4-5 nm.

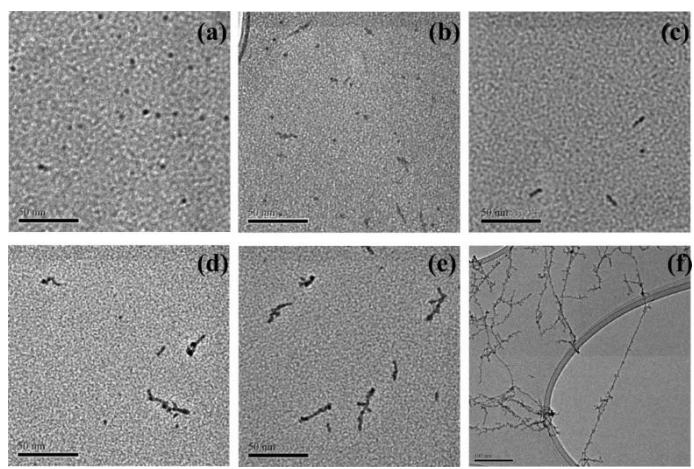

*Figure 17 TEM images of deposited silver nanostructures from HNDs with the following estimated diameters: (a) 55 nm, (b) 63 nm, (c) 135 nm, (d) 245 nm, (f) 790 nm and (g) 1.7 μm. For all of these experiments the Ni oven temperature was fixed. Reprinted with permission from [145]. Copyright 2014 American Chemical Society.*

Studies on single HNDs doped with Xe atoms via ultrafast coherent x-ray diffraction imaging point towards a significant contribution of the dopant to the total angular momentum of the droplets [146]. Recently, the kinematics of these combined vortex-cluster systems in dependence on the total angular momentum was theoretically investigated by Bernando and Vilesov [147].

Two studies from the Ernst group have provided new insight into the growth mechanism for nanowires. The first of these showed that it was possible to form core-shell nanowires in HNDs [148]. This work focused on a combination of silver and gold, picked up sequentially. High resolution STEM equipped with element specific detection (energy-dispersive X-ray spectroscopy and electron energy loss spectroscopy) showed that distinct core and shell layers could be made, with the choice of core and shell being dictated by the order in which the metals were added. Given the miscibility of gold and silver and their very similar lattice parameters, the authors pointed out that it would be difficult to achieve this separation into a distinct core and shell by any other means of fabrication. Figure 18 shows STEM images of the Ag and Au content of a single wire, as well as a color-coded view of the two element distributions. The authors interpret the bulges in this image as the product of a multicenter growth process in which distinct particles migrate to a vortex and eventually partly fuse. A more recent study has explored the nanowire growth process in more detail, providing both further experimental information and a computational model [149]. The experimental work looked at both pure Ag and pure Au nanowires and focused on the effect of doping rate on the internal structures of the wires. The data suggest that the wires form by aggregation of particles that are initially formed by multicenter growth. At low doping rates particle nucleation favors icosahedral and dodecahedral particle symmetries while higher doping rates favor fcc particles. The individual particles then aggregate along the vortices but do not fully melt as they fuse, preserving much of the original particle morphology.

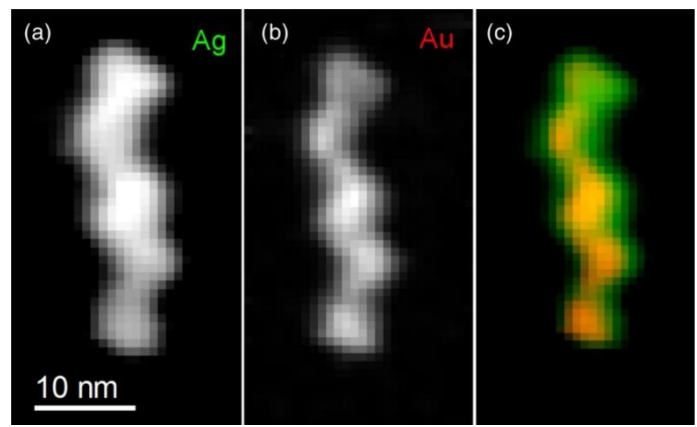

Figure 18 STEM images of (a) the Ag content of a nanowire, (b) the Au content of the same wire, and (c) a color-coded overlay of the images in (a) and (b). Details of how the element-specific images were obtained can be found in the original publication. Reprinted figure from Ref. [148]. Copyright 2014 by the American Physical Society.

Finally, we note that while the existence of quantum vortices in HNDs has been inferred from the observation of deposited nanochains and nanowires, these studies do not in themselves provide concrete proof of their existence. However, Gomez *et al.* have recently managed to extract more direct evidence for these vortices using ultrafast X-ray diffraction of individual HNDs with $N_{He} = 10^8 - 10^{11}$ [50], as was discussed briefly in Section 2. The non-spherical shapes possessed by many of the pure HNDs are consistent with the injection of considerable quantities of angular momentum into the droplets. Even more striking was the observation of distinct and regular diffraction spots arising from some xenon-doped HNDs. An analysis of these diffraction patterns suggests that Xe clusters are pinned to an ordered array of quantum vortices.

# 5 Ionization of pure helium nanodroplets

The high ionization energy of helium makes single ionization with a standard light source impossible. Consequently, the first ionization experiments on HNDs were performed utilizing electron ionization [36, 61-63]. If a HND is thought of as a dense helium gas, then Beer's Law can be used to estimate the penetration depth of the incident electron:

$$I(x) = I_0 e^{-\sigma \rho x}. \qquad (3)$$

Here $I(x)$ is the electron current at the penetration depth $x$, $I_0$ is the electron current before entering the HND, $\sigma$ is the ionization cross section for helium ($\sigma_i$ = 3.0×10$^{-21}$ m$^2$ for 70 eV electrons [150, 151]) and $\rho$ is the density of liquid helium ($\rho$ = 0.020 Å$^{-3}$ [152]). For $I/I_0 = 1/e$, a penetration depth of about 17 nm is obtained. This means that for HNDs containing fewer than 10$^5$ helium atoms (which corresponds to a radius of 10.6 nm), a uniform probability for the ionization of a random He atom can be expected. In contrast, for a HND larger than 10$^5$ helium atoms, the ionization rate in the part of the droplet that faces the incoming electron beam should be higher compared to the other regions in the HND.

For a HND containing less than 10$^5$ He atoms, the majority of the He atoms are located close to the surface (less than three atomic layers from the surface) and thus the probability for forming an initial He$^+$ near the surface of the HND is considerably larger than an ionization event close to the center. In the center of the HND, however, the electrostatic potential for a charged species is lowest and thus there should be a migration of ions towards this favorable position.

## 5.1 Charge localization and transport in a helium nanodroplet

In bulk liquid He the mobility of charged species has been investigated extensively [153]. Surprisingly low values for the mobility of cationic species were determined, which did not match the expected hydrodynamic radii of the ions [154]. This is because the mobility of ions in liquid He is dictated by the formation of a charged He cluster, commonly referred to as a "snowball". The formation of this snowball is driven by the balance of electrostrictive attraction [154-156] and short-range repulsive exchange forces [157]. Atkins calculated the size of a snowball by depicting helium as a classical dielectric continuum with a surface energy equivalent to the surface tension of liquid He [155]. The high pressure arising from the electrostriction suggests a solid structure in the immediate vicinity of the positively charged center of the cluster. Density functional theory studies identified a linear triatomic He$_3^+$ ion as the core of an isolated snowball [158]. However, the initial step for the development of a snowball is the formation of He$_2^+$, which is a relatively slow process due to the large difference in the internuclear separation in the ground states of the neutral and charged dimer (see Figure 19). Thus there will be competition between charge tunneling to other atoms via resonant charge transfer and the approach of the nearest neighbor neutral atoms toward the charge under the attractive ion-induced dipole interaction. The resonant charge transfer mechanism was introduced by Atkins [159] and further developed by Buchenau *et al.* in the case of HNDs [77]. A calculation of the mean time between charge transfer steps resulted in a value of the order of 20 fs [160]. On this time scale the motion of the atomic nuclei is small, especially since the atoms initially find themselves on a flat region of the potential energy curve (see Figure 19). Thus the charge may be assumed to hop several times before dimerization takes place. Based on measurements with doped HNDs a more elaborate model of the hole-hopping mechanism has been developed [161]; it will be explained in more detail in section 7 on doped HNDs. From a classical model a mean free path of 34 Å for the charge hopping in helium was calculated, which corresponds to approximately 11 hops before charge self-trapping occurs. This is similar to the hopping numbers deduced by other research groups using different models [162].

## 5.2 Formation of a strongly bound ionic core

Ionization of a rare gas dimer results in the formation of a covalently bonded molecular cation, since an antibonding electron is removed. This gain in binding energy leads to a substantial reduction of the adiabatic ionization energy, as depicted in Figure 19. However, electron ionization, as well as photoionization, is a vertical process for which the changes in the electronic system occur on a time scale during which the nuclei can be expected to be frozen.

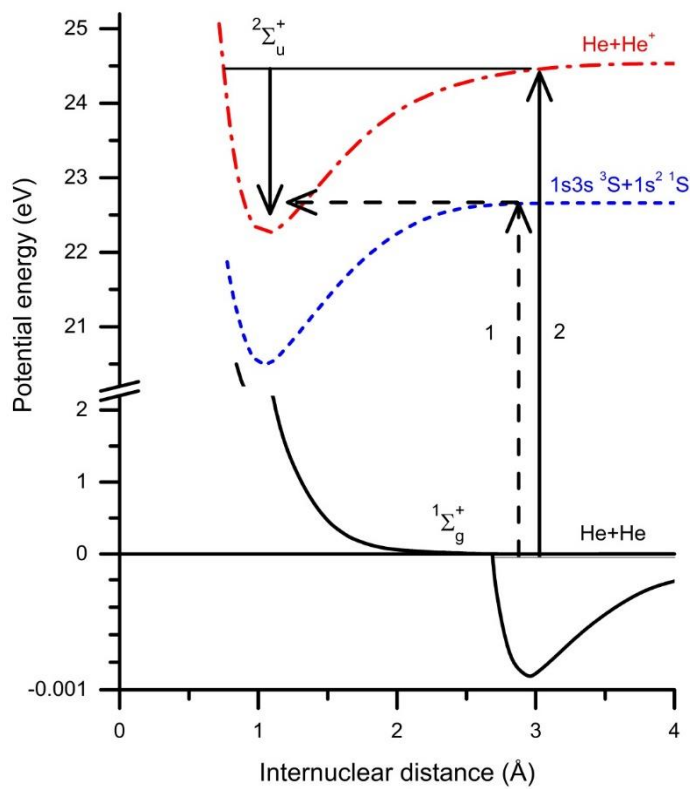

Figure 19 Potential-energy diagram for a helium dimer and its corresponding cation. The arrows indicate indirect (dashed line, 1) and direct (solid line 2) ionization mechanisms. Data taken from Ref. [77, 163].

The potential energy curves of neutral helium dimer and its corresponding cation are shown in Figure 19 and illustrate the energetic situation in the case of a vertical ionization event [164]. The adiabatic ionization energy corresponds to the energy difference between the vibrational ground state of the neutral dimer and the vibrational ground state of the ion [165]. Ionization of helium clusters is characterized by a dramatic change of the geometry. For the neutral helium dimer the equilibrium distance between the two He atoms of 3.58 Å is shortened to 1.08 Å for the cationic dimer [71]. Another dramatic change is the different binding energies of the neutral and the cation. The neutral $He_2$ dimer has a binding energy of only 0.11 µeV [166], whereas for $He_2^+$ the binding energy is 2.35 eV [167]. Thus one can expect that the adiabatic ionization energy for the dimer will be lower than that of the monomer by an amount equivalent to the binding energy of the cationic dimer. This gives an expected adiabatic ionization energy of $He_2$ of 22.2eV.

The vertical ionization energy (appearance energy) is the energy separation between the vibrational ground state of the neutral dimer and the potential energy curve of the cluster ion at the internuclear distance of the former. Because of the Franck-Condon principle electron ionization and also photoionization probe the vertical process, *i.e.*, the adiabatic value cannot be reached directly. Nevertheless, the adiabatic ionization energy can be approached in the case of small clusters via initial excitation of autoionizing Rydberg states, followed by relaxation of the atomic distance as well as structural relaxation of the atoms in the vicinity of the excited atom and loss of an electron leading to an ionized cluster (associative ionization within the cluster) [168]. This mechanism is only possible if the cluster radius does not exceed the orbital radius of the Rydberg electron [169]. This process was proposed to be responsible for the electron and photo-ionization threshold behavior of small cluster ions of various other rare gases, namely Ar [170, 171], Ne [172], Kr [171], and Xe [173, 174].

### 5.3 Threshold energies of small $He_n^+$ ions

Denifl *et al.* investigated the threshold ionization behavior of small helium cluster ions (cluster size n = 2–10) formed via electron ionization of neutral HNDs and derived appearance energies for mass-selected cluster ions using a nonlinear least-square-fitting procedure [163]. The apparatus used for the measurements was a hemispherical electron monochromator combined with a quadrupole mass spectrometer. This experiment demonstrated that helium clusters are not only formed via direct ionization above the atomic ionization potential, but also indirectly via autoionizing Rydberg states of $He_2^*$. In the case of photoionization of HNDs, Fröchtenicht *et al.* [71] observed $He_2^+$ and $He_3^+$ signals a few eV below the atomic threshold. They assigned weak peaks centered at 23 and 23.8 eV to the contribution of the $3^1P$ and $4^1P/5^1P$ states of the helium atom. The threshold energy of the 23 eV peak of $He_2^+$ is located at 22.8 eV and is in perfect agreement with the threshold value reported by Denifl *et al.* [163] for $He_2^+$ formed via electron ionization. For the ion efficiency curves of $He^+$, Denifl *et al.* [163] reported an onset at 21.31 eV, followed by a nearly constant ion yield up to 24.6 eV, where the yield increases sharply at the atomic threshold (see Figure 20 open circles). A similar behavior was reported for the ion efficiency curve of $He_6^+$, *i.e.*, a first threshold at 21.13 eV, followed by a nearly constant ion signal up to 23.49 eV and then an additional contribution at 24.37 eV.

Buchenau *et al.* [77] reported a threshold of 21 eV for $He^+$ formed via electron ionization from large neutral HNDs (> $10^4$ helium atoms), which is 3.5 eV below the onset of $He^+$ formed via electron ionization of He atoms (dashed line in Figure 20). The authors in Ref. [77] explained the occurrence of $He^+$ at an energy far below the atomic threshold by the formation of a $He^*$ exciton in the cluster after the initial electron impact, which captures another He atom and leaves the cluster. The $He_2^*$ is finally ionized by a second electron (however, a check of this hypothesis by varying the electron current was not successful) or by a collision with a metal surface, which leads to ejection of $He^+$. This mechanism was questioned by Callicoatt *et al.* [167], who observed a strong decrease of the $He_2^+$ signal for $He_nAr$ clusters. This is surprising since the Ar atom(s) should be located near the center of the cluster whereas production of $He_2^*$ is assumed to be near or at the cluster surface. Buchenau *et al.* [77] also observed a threshold energy at 21 eV for $He_2^+$, which they explained via formation and ejection of $He_2^*$ and subsequent ionization via a collision with a metallic surface (see Figure 21 left).

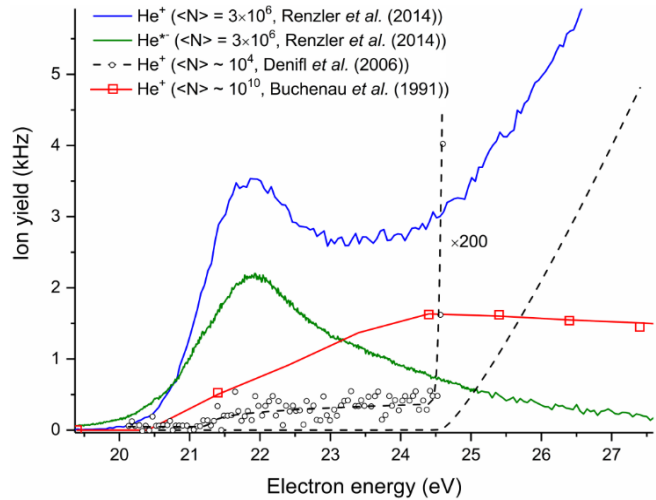

*Figure 20 Ion efficiency curves for He$^+$ and He*$^−$ formed via electron ionization of undoped HNDs. All curves exhibit a more or less pronounced contribution below the ionization energy of He (24.59 eV). Renzler et al. [175] propose a model where the HND is hit by two electrons that form He* and two low-energy electrons. One of these electrons recombines with the neighboring He* and forms He*$^−$, which then polarizes and attracts a second He*. The collision of He*$^−$ with He* forms He$^+$ (see Figure 21 right).*

With the discovery of the efficient formation of He*$^−$ (for more details see Section 6) via a strong resonance at 22 eV and two weaker resonances at 23 eV and ∼ 25 eV [176-178], Renzler *et al.* [175] proposed a mechanism for formation of He$^+$ at electron energies below the ionization energy of the helium atom that involves He*$^−$. An incident electron with a kinetic energy of 22 eV can form He* and the inelastically scattered electron will then form a vacuum bubble via Pauli-repulsion, a so-called "electron-bubble" [179, 180]. By electron-induced-dipole interaction the pair can attract each other and then react to form He*$^−$. If another He* is present the highly mobile He*$^−$ will move towards it and produce He$^+$, which is then ejected as a result of a highly exothermic reaction in the surface region of the HND. The quadratic dependence of the ion signal of He$^+$ formed at 22 eV on the electron current supports the proposed model. It is interesting to note that the ion efficiency curve of He$^+$ measured via electron ionization of gas phase He exhibits narrow Fano resonances that can be assigned to doubly-excited states [181] that form with the projectile electron short lived transient negative anions. The lifetimes of these autoionizing states may be strongly affected by the surrounding He atoms.

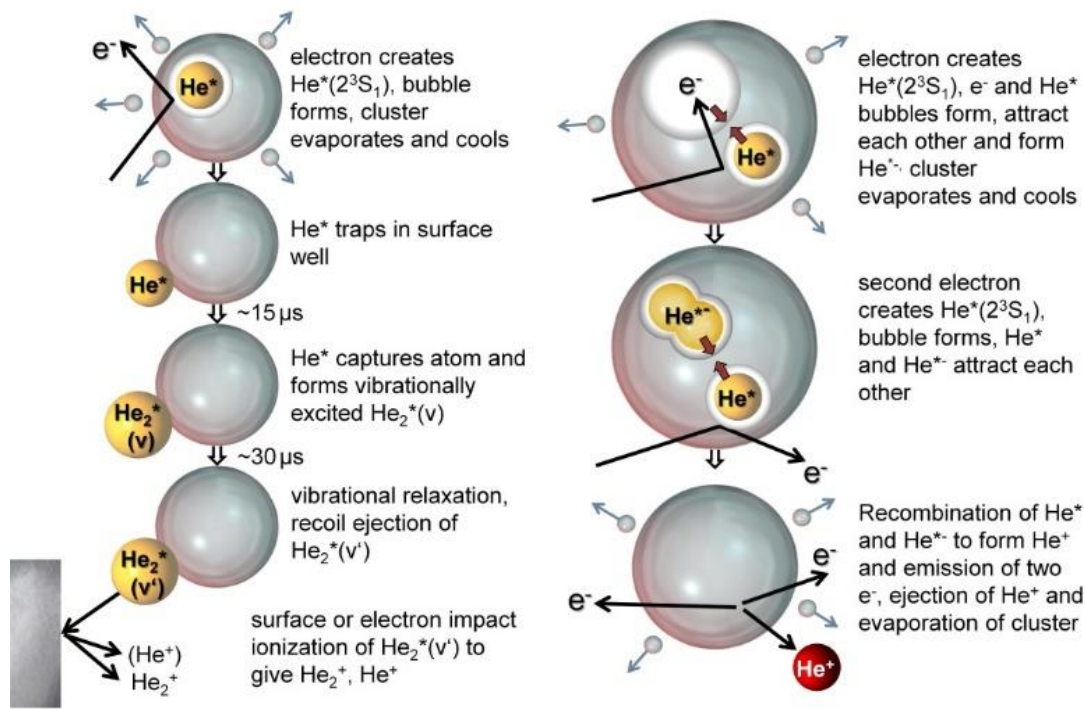

*Figure 21 Left part: reaction scheme proposed by Buchenau et al. [77] for the formation of He$_2^+$ and He$^+$ below the ionization energy of atomic helium. Right part: reaction sequence proposed by Renzler et al. [175] for the formation of He$^+$ via a two-electron process and the formation of an intermediate metastable helium anion, He*$^-$.*

## 5.4 The possibility of multiple ionization of helium nanodroplets via free electrons

Generally, the possibility that a single HND might be struck by multiple electrons should not be ignored, particularly for large droplets. The droplet geometric cross section scales with $N^{2/3}$ and typical electric fields in ion sources and mass spectrometer systems have only a small effect on the velocity of heavy HND with $N>10^6$. Thus the HND will cross the ionizing electron region with roughly a constant speed that originates from the nozzle expansion and is in the range of about 200 m/s [11, 48]. When passing a well-focused electron beam with a beam diameter of about 1 mm and a current of 100 µA, a HND containing $10^6$ He atoms will be hit by an average of 5.1 electrons, while for droplets ten times larger this rises to 23.8 electrons. At high enough electron energies a single electron may even ionize several He atoms on its way through a HND. The effect of multiple ionization events on a large HND has been studied indirectly by measuring the abundance of low-mass fragments, *i.e.*, He$_2^+$, which makes up about 15% of the low-mass ions (m < 400) [49]. The Vilesov group observed that multiple ionization of large HNDs leads to the emission of a large number of He$_2^+$ ions ejected from the droplet (see Section 2). Given the electron flux and the average droplet size used in their experiment, they deduce that the number of measured He$_2^+$ ions is about a factor 40 lower than the number of electron impact events. Thus, either not all electron collisions lead to an ionization event or the majority of charges remain in heavy cluster ions that were beyond the operating range of the mass spectrometer [49]. In principle, a time of flight mass spectrometer system has an unlimited mass range, if the frequency of the ion extraction pulses is matched to the heaviest ion of interest. The low velocity of heavy ions often leads to detection problems. The probability for the emission of a secondary electron upon the collision of an ion with the surface of a secondary electron multiplier depends on its velocity (kinetic emission [182]) and its potential energy (potential emission) [183]. As kinetic emission of slow heavy ions is very low, often special post-acceleration devices, such as Daly detectors, are utilized [184]. However, the potential energy of He*, He*$^-$ and He$^+$ is far larger than the work function of metallic or semiconducting surfaces utilized in secondary electron multipliers, and provides potential emission. Thus, also very massive, singly-charged HNDs can be detected without excessive post-acceleration. As described in the section on the formation and size distribution of HNDs (Section 2), charged massive cluster ions have been observed and analyzed by various deflection techniques [43, 44].

## 5.5 Photoionization of undoped helium nanodroplets

Fröchtenicht *et al.* were the first to perform single photoionization experiments on HNDs. The extreme ultraviolet photons were provided by the BESSY synchrotron in Berlin [71]. As mentioned above, threshold energies were measured for the ion efficiency curves and an indirect ionization process for $He_n^+$ clusters was proposed via electronically excited He* states. The first photoelectron spectroscopy experiments on pure He droplets were carried out using synchrotron radiation by Neumark and co-workers [185]. Various peculiarities of directly excited or ionized He droplets were identified, such as the emission of electrons having almost zero-electron kinetic energy from pure droplets [185, 186], the indirect ionization of dopants by charge transfer or by excitation transfer out of excited and relaxed states of He, and the development of a conduction band structure in large He droplets [187].

Strong-field ionization of heavy rare gas clusters has become a very active field of research [188, 189], triggered by the ground breaking work of Corkum [190]. Pure He droplets are less attractive targets for strong-field ionization studies due to the high threshold intensity needed for photo ionization of He ($1.5 \times 10^{15}$ W cm$^{-2}$ [191] for photon energies below the ionization energy of He) and the low number of electrons which each He atom can contribute to a nanoplasma. However, He droplets loaded with dopants of low-ionization energy have recently revealed quite diverse strong-field ionization dynamics related to the large differences in ionization energies of dopants and the surrounding He matrix [28, 192]. Furthermore, the controlled localization of dopants inside or at the surface of the droplets [193] provides another novel aspect to the study of strong-field ionization.

Free electron lasers (FELs) provide intense light pulses in the extreme ultraviolet (EUV) and X-ray regime. This enables the investigation of the dynamics of nanoplasmas at increasingly high photon energies up to the keV range. At such high photon energies, however, the physics of cluster ionization is very different from the plasmon-enhanced charging of clusters illuminated by near-infrared pulses that were frequently utilized for the ionization of heavier rare gas clusters [194, 195]. Multistep photoionization becomes the dominant absorption process and plasma heating has no significant effect [196]. In recent experiments performed in a collaboration at the FEL FERMI@Elettra in Trieste [197], using intense tunable EUV radiation, the autoionization dynamics of multiply excited HNDs was studied [29, 198]. Extremely high ionization rates were measured upon excitation below the He ionization threshold, as well as photoelectron spectra which are characteristic of thermal electron emission. These observations indicate that a novel, many-body, autoionization process is active which is related to inter-atomic Coulombic decay [198, 199]. Due to the low ponderomotive energy of electrons in the EUV field, which is of the order of meV, this regime would be classified as the weak-field regime. Nevertheless, owing to the high intensities available at FELs, the majority of He atoms are electronically excited within tens of fs. The subsequent ultrafast collective autoionization of the whole cluster generates a highly ionized nanoplasma similar to that formed with strong-field near-infrared ionization [194]. Buchta *et al.* [200] recently studied the ionization dynamics of pure HNDs irradiated by EUV radiation using velocity-map imaging photoelectron-photoion coincidence spectroscopy. Below the autoionization threshold of He droplets, they find evidence for multiple excitation upon absorption of many photons and subsequent ionization of the droplets by a Penning-like process. At high photon energies inelastic collisions of photoelectrons with the surrounding He atoms in the droplets are observed. Ongoing EUV experiments on doped clusters, including fs time-resolved measurements, promise exciting insights into the details of such extreme states of matter. A recent article by Mudrich and Stienkemeier provides a comprehensive overview on recent advances in photoionization of pure and doped HNDs [201].

Very recently Fano resonances were observed in photoelectron spectra measured upon XUV irradiation of pure HNDs, similar to doubly-excited states in atomic He [202]. However, the resonances are broadened and blueshifted compared to an atomic target [202]. Electron correlation is an important aspect of doubly-excited states and thus these highly-excited species are particularly sensitive to a perturbing environment. In a double excitation resonance, the ionization cross section exhibits an interference between the direct ionization and autoionization pathways [203]. Fano resonances were also reported by Fiegele *et al.* upon electron impact of gas phase He [181] and thus doubly-excited states are expected to also contribute to the ionization processes upon electron bombardment of pure and doped HNDs.

## 5.6 The special behavior of $He_4^+$

In electron ionization of large ($<N>> 10^5$) HNDs, $He_4^+$ is usually the most abundant small cluster ion [36, 77, 78]. This observation is puzzling at first sight, since it is well established [72-76] that the most stable ionic helium clusters consist of a tightly bound ionic core of three (or sometimes two) atoms, with other He atoms attached only by polarization forces. Thus one would expect $He_3^+$ to be at least as abundant as $He_4^+$ in experiments where the ionization process imparts considerable internal energy to clusters and resulting evaporation leads to an enhancement of the yield of the most stable cluster ions. It has been suggested that the observed special stability of $He_4^+$ may arise from the presence of a metastable electronically excited state [78]. Photodissociation of $He_4^+$ formed from large HNDs can be achieved with much lower photon energies than of $He_4^+$ formed from smaller HNDs [78]. In addition, the fragment kinetic energy release is unusual, indicating that $He_2^+$ is virtually the only ionic product [78]. Knowles and Murrell [75] investigated theoretically the structure of $He_4^+$ consisting of a dimer ion and an excimer which was proposed by von Issendorff *et al.* [78]. As the reaction pathway for the formation of $He_4^+$ via $He_2^*$ and $He_2^*$ becomes very favorable for large HNDs Gomez *et al.* [48] proposed a method for the determination of the average size of a beam of neutral HNDs based on the ratio of the ion yields of $He_4^+$ and $He_3^+$ or $He_2^+$. Recently, further experiments with pulsed electron bombardment of HNDs revealed an enhanced $He_4^+$ production in large droplets for a characteristic pulse duration of ~ 10μs [79]. In the same study a qualitative model is presented in which two surface-bound, metastable excited $He_2^+$ share a Rydberg electron to form $He_4^+(^4A_2)$.

# 6 Helium anions

## 6.1 Negatively charged helium nanodroplets

Negatively charged HNDs were first reported by Gspann in 1991 [204]. Negative helium cluster ions were observed for clusters with more than 2 million helium atoms per cluster. Before these experiments, bound states of an excess electron to the surface of helium clusters were calculated using a simple one-body potential for the electron interacting with a helium cluster [205]. The minimum cluster size $N$ was determined to be 5×10$^5$ with a binding energy of 4.96×10$^{-6}$ eV (0.04 cm$^{-1}$) of the excess electron. In the same year as the first experimental findings on negatively charged HNDs, the surface barrier for penetration of electrons into a helium cluster was reported [206]. Via electron scattering the surface barrier was determined to be in the range of 0.6–1.1 eV depending on the cluster source temperature at a fixed helium pressure of 20 bar. From these measurements, the core density of the helium clusters was determined to vary between $1.2\times10^{22}$ cm$^{-3}$ and $2.2\times10^{22}$ cm$^{-3}$ for $T_0$ = 14 K and $T_0$ = 11 K, respectively.

After these studies the interaction of low-energy electrons with HNDs was investigated further. Electron capture was employed to measure the distribution of cluster ions formed from a supercritical expansion of helium gas [43]. It was found that the ion yield for negatively charged helium clusters is about two orders of magnitude lower than for positively charged ones. More detailed investigations on negatively charged HNDs have revealed a threshold size below which no ions have been detected [207, 208]. Results from infrared photodetachment studies of negatively charged HNDs have revealed a single broad peak at 1.5 μm (0.8 eV), which can be explained by a calculated 18.7 Å bubble in a 0.7 eV potential energy well [209]. Later, the threshold size of $N \approx 5\times10^5$ was reported for negative helium cluster formation derived from deflection experiments [210]. Calculations for the threshold size of surface bound electrons revealed about the same cluster size. However, the ions were much more stable in electric fields than expected [211], which indicated that the charge was not due to surface bound electrons. It was therefore suggested that the electron is bound in a bubble state.

Electron impact experiments were then carried out to observe both positively and negatively charged HNDs [212]. The negative cluster series could be fitted with a log-normal distribution and indicated singly charged HNDs from threshold up to clusters containing $10^8$ helium atoms. Experiments involving high electron energy resolution revealed an exponential size distribution for large charged helium clusters anions [44]. In these experiments the dependence of the ion yield on the initial electron energy was investigated and two resonance structures were observed in the electron energy range from 0 eV to 30 eV. The first feature has its threshold at $\sim$ 1 eV, which reflects the energy needed for a free electron to penetrate the HND and localize to form a bubble. The second feature, above 20 eV, is a superposition of three narrow peaks which were attributed to single or successive electronic excitations of the droplet that results in a zero-kinetic-energy electron, which then also localizes in an internal bubble. A detailed summary of the aforementioned findings was reported in [213]. The first anion formation in doped HNDs upon electron impact was reported in 2006 [214] and is summarized in Section 9.

## 6.2  Atomic and molecular helium anions in helium nanodroplets

In 2014 it was shown that atomic and molecular helium anions He*$^-$ and He$_2$*$^-$ can be formed in HNDs upon electron impact, provided the HNDs exceed a certain size [176]. Since both atomic and molecular ground state helium cannot bind a free electron, they first have to be electronically excited (He* and He$_2$*), hence the notation He*$^-$ and He$_2$*$^-$ for the anions. Using the experimental setup described in Section 2.1, the most dominant anion in the mass spectrum of pure HNDs with an average size of about $2.4\times10^6$ helium atoms and electron impact energy of 22 eV was He*$^-$. He$_2$*$^-$ was also observed, but with a detected abundance two orders of magnitude below that of He*$^-$ (see Figure 22). Inspection of the anion efficiency curve for forming He*$^-$ revealed a resonant feature around 22 eV which could be reproduced by a triple-Gaussian function, yielding maxima at $22.0 \pm 0.2$ eV, $23.0 \pm 0.2$ eV, and $25.1 \pm 0.5$ eV. A later study utilizing a hemispherical electron monochromator with higher electron energy resolution confirmed the previous values ($21.98 \pm 0.05$ eV, $22.96 \pm 0.07$ eV, and $25.78 \pm 0.1$ eV [178]. Similar values have been reported earlier for the resonant formation of anionic HNDs [44] and also for anion formation of or from various dopants inside HNDs, *e.g.* (D$_2$O)$_2^-$ [215], anionic fragments of chloroform [216] or the dehydrogenated thymine parent anion [214].

The formation of atomic helium anions inside HNDs proceeds via a two-step reaction. First, the electron penetrates the surface of the HND, which is estimated to require roughly 1–2 eV [207], as discussed earlier. This is followed by the excitation of a ground state helium atom to He* (which is equivalent to He(1s2s $^3$S) unless stated otherwise) and a subsequent electron capture leading to the formation of He*$^-$. The dependence of the helium anion yield as a function of the droplet size has revealed the minimum (average) droplet size for the detection of helium anions in HNDs to be $1.8\times10^5$ helium atoms [176]. This size is necessary to accommodate the electron bubble, as discussed in the previous section. The anion efficiency curve for He$_2$*$^-$, on the other hand, showed a resonance behavior which could be fitted by a double-Gaussian function with maxima at $22.9\pm0.2$ and $24.8\pm0.5$ eV [176]. The formation process of He$_2$*$^-$ cannot proceed via the interaction of a He*$^-$ with a ground state helium due to a barrier in the potential energy curve and so an alternative mechanism will be presented in Section 6.3.1.

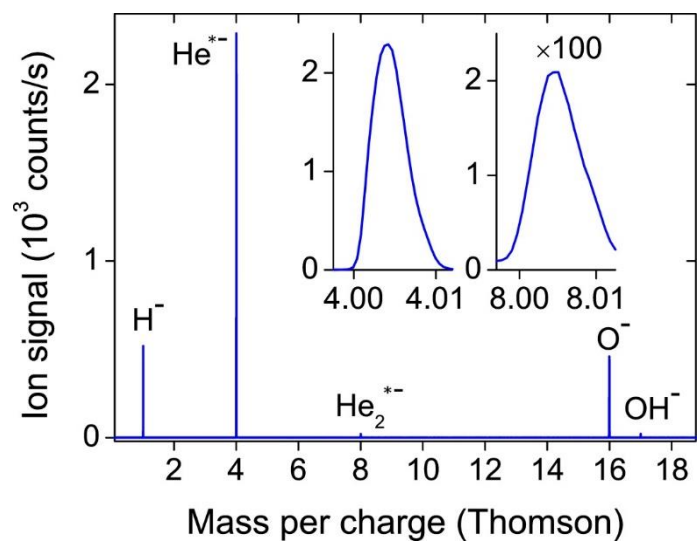

*Figure 22 He*$^-$ and He$_2$*$^-$ in the mass spectra of HNDs formed by electron impact at 22 eV.*

The mobility of the anionic species He*$^-$ and He$_2$*$^-$ was investigated via charge transfer reactions with SF$_6$ doped in HNDs [176]. SF$_6$ is known to be heliophilic and preferentially locates near the center of a HND [15]. Comparison of the ion yield of He*$^-$, He$_2$*$^-$ and SF$_5^-$ as a function of the SF$_6$ pressure in the pick-up chamber revealed that the increase of the SF$_5^-$ ion yield correlates with an exponential decrease of the atomic helium anion yield. Since the formation of He*$^-$ is a resonant process, it is expected that its formation occurs near the surface of the droplet at a depth of about $D = (2\pm0.5)$ nm, as was estimated from absolute elastic electron scattering cross sections [217]. In comparison to the average HND size of $N = 5\times10^6$, the helium anion will have to cover 15-25 times the penetration depth to reach the dopant in the center [176]. The decreasing He*$^-$ yield with respect to the increasing SF$_5^-$ yield can be explained either by a direct interaction of the electron with the foreign molecules in the HND or a charge transfer reaction of a highly mobile helium anion. The latter seems more likely, since the polarizability of SF$_6$ (6.55 Å$^3$)[218] is much lower than the polarizability of He* (46.77 Å$^3$)[219]. It is therefore reasonable to assume that the electron will first combine with He* to form He*$^-$ and this will then travel the distance to the center to transfer one electron to the dopant. In contrast, the ion yield of He$_2$*$^-$ is hardly influenced by the number of SF$_6$ atoms present in the HND. This indicates that the molecular helium anion is much less mobile than its atomic analog. Following the discussion of the mobility of these anions, He*$^-$ was assigned as the exotic fast negative carrier (EFNC), which first was reported in liquid helium experiments in 1969 [220]. It is noteworthy, until the recent detection of helium anions in HNDs, the EFNC found in the ion mobility experiments could not be assigned. In later studies, it was suggested that the EFNC could be He*$^-$[221], as well as bubbles containing two electrons [222, 223] or He$_2$*$^-$ [224].

To explain the expulsion of helium anions from the HNDs, the ion yield of He*$^-$ produced at 22 eV electron impact energy was investigated as a function of HND size for two different electron currents, 160 μA and 10 μA [176]. Both curves show a similar size dependence, but the onset for He*$^-$ is shifted to smaller cluster sizes in the case of the higher electron current. The shift is related to the probability of more than one electron hitting a HND. This is an indication that He*$^-$ is driven out of the HND based on Coulombic repulsion. Inspection of the ion yield of He*$^-$ as a function of the electron current at fixed electron impact energies and HND size has revealed a quadratic dependence, which also supports the expulsion due to Coulombic repulsion from another negatively charged species, such as an electron bubble, He*$^-$ or He$_2$*$^-$. It was suggested, that this will also apply to other anions embedded in HNDs. On the other hand, the ion yield of He$_2$*$^-$ as a function of electron current at fixed electron impact energy and HND size has shown a linear dependence, which indicates that the molecular helium anion resides on the surface of the HND, rather than being solvated inside. This is in agreement with the observation of its low mobility and points towards simple desorption from the surface due to the weak interaction energy of He$_2$*$^-$ with the neutral HND.

## 6.3 Quantum chemical considerations of anionic helium species

### 6.3.1 Formation of anionic species

Both, He*$^-$ and He$_2$*$^-$ have received substantial amount of interest from experiment [225-227] and theory [228-234]. For both helium anions, fast, non-relativistic Coulomb autodetachment on the fs-time scale is forbidden [226], which explains their metastability and thus long life times of (359 ± 0.7) μs (for total angular momentum quantum number j = 5/2) for the atomic helium anion [226] and (135 ± 15) μs for the molecular helium anion [235].

Recently, *ab initio* calculations have been carried out to try and explain some of the properties of He*$^-$ and He$_2$*$^-$ at the CCSD(T) level [236]. The basis sets used in the calculations were derived from Dunning's correlation consistent basis set series x-aug-cc-pVXZ with x = s, d and X = D, T, Q, 5 [237, 238] which yields an improved convergence behavior in the self-consistent field procedure [239]. A thorough test via calculating various properties such as the polarizability of He* [219], the excitation energy E(He → He*) [240], the electron affinity of He* with respect to an anionic state of (1s2s2p $^4$P) configuration [241], and the ionization energy of He [242], revealed the q-aug-cc-pVTZ basis set as minimum size of basis set for all properties to be regarded as converged. The most sensitive quantity with respect to basis set convergence was the electron affinity of He*. However, with the q-aug-cc-pVTZ the accuracy was within 95% of the high-level theoretical estimate of 77.1 meV [241]. With these tools at hand, the interaction of He$^+$, He, He* and He*$^-$ with ground state He was investigated by calculating the potential energy surface [239]. The potential energy surfaces of He* – He and He*$^-$ – He show a local minimum and a substantial barrier before they reach the covalent bound systems He$_2$* and He$_2$*$^-$. The barrier, which separates the chemical bound molecular configuration from the weakly interacting polarization-bound complex has been assigned to the deformation of the molecular orbitals as the inter-atomic distance decreases. Only after the barrier has been overcome do the molecular orbitals of the helium anion begin to span a large volume of space around the two helium nuclei. In the cold environment of the HND these barriers are too high to be overcome, therefore the formation of the molecular species He$_2$* and He$_2$*$^-$ cannot proceed via the interaction of a ground state helium with He* or He*$^-$, respectively [239]. Indeed, a study on the formation of excited helium dimers in HNDs has shed light on this very issue [243]. Using the equation-of-motion coupled cluster with single and double substitutions (EOM-CCSD) method [244] and the q-aug-cc-pVQZ basis set [239], excited states of the diatomic interaction potential have been calculated. These calculations have shown that the formation of a helium molecule in an excited state is hindered by a barrier, if the electron is excited into the 2s or 3s orbital (and concurrent spin-flip). This is due to the fact that upon approach of a ground state helium towards a helium atom in an excited state the orbitals overlap, which gives rise to a large Coulomb integral and therefore the energy increases. The corresponding barrier heights range from 16 meV to 294 meV. The situation is different for excitation into He(1s2p $^3$P). If the two helium atoms approach each other, only the occupation of a p-orbital perpendicular to the molecular axis results in a vanishing overlap. Although a vanishing overlap does not necessarily result in a vanishing Coulomb integral, it can be assumed to be at least considerably smaller than in the other cases. Similar arguments can be applied for excitation into He(1s3p $^3$P). Indeed, the calculations reveal two excited states with a 0 eV barrier, namely He(1s2p $^3$P) + He(1s$^2$ $^1$S), which results in He$_2$(1$\sigma_g^2$1$\sigma_u$2$\sigma_u$ $^3\Pi_g$), and He(1s3p $^3$P) + He(1s$^2$ $^1$S), which results in He$_2$(1$\sigma_g^2$1$\sigma_u$2$\pi_u$ $^3\Pi_g$), respectively. The corresponding excitation energies are 20.9 and 23.0 eV. The energy difference between these two states is in reasonable agreement with the experimental findings [176] and indicates the formation pathway of the molecular helium anion. The other exited states yielded barriers in the range from 12 meV to 980 meV. Although some barriers are very small, the probability for overcoming these barriers at 0.37 K is virtually zero [243].

6.3.2   Anionic species in helium nanodroplets, the ratio Ξ

The molecular species He$_2$, He$_2$*, He$_2^+$, and He$_2$*$^-$ have also been investigated by scanning their potential energy landscape with a ground state helium atom [239]. The He$_2$ – He system revealed a peanut-shaped potential energy surface, with binding energies in the range of 0.81 meV (linear arrangement) to about 1.6 meV (triangular arrangement). For both linear and rectangular arrangements, all bond lengths were found to be equidistant. The potential energy surface of He$_2$*$^-$ – He has shown a rather spherical shape with binding energies in the range of 0.12 meV (linear arrangement) to 0.11 meV (rectangular arrangement). A large repulsive void with a diameter of about 6.5 Å surrounds He$_2$*. In this void no ground state helium can reside due to the positive interaction energy. In contrast, yet not surprising, the repulsive void of He$_2^+$ was found to be rather small, where repulsive forces only act below a distance of 1.7 Å between He and the center of mass of He$_2^+$. The interaction energy for a linear arrangement was found to be 124.71 meV and is substantially higher than for the rectangular arrangement, which is about 39.83 meV. Most noteworthy is the very large repulsive void which was found for the He$_2$*$^-$ – He system. The potential energy surface was found to be very asymmetrical and resembling the $\pi$-type HOMO of the molecular helium anion. The lowest binding energy was found for linear arrangement at a separation distance of about 8.6 Å with respect to the center of mass of He$_2$*$^-$ and a binding energy of 0.2 meV. For the rectangular configuration, the repulsive part was found to reach as far as 13.5 Å. From these results it is clear that the interaction between He$_2$* and He$_2$*$^-$ and any other He atom in its ground state is very small and dominated by dispersion and polarization. Both He$_2$* and He$_2$*$^-$ are characterized by a large repulsive void.

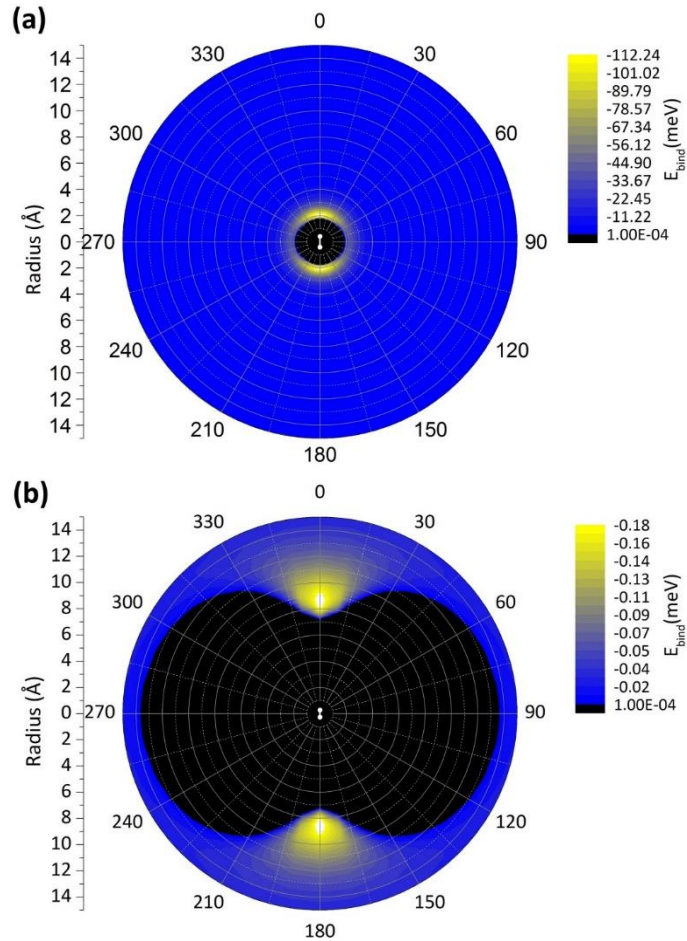

Figure 23 Polar plots of the interaction energy for (a) $He_2^+ - He$ and (b) $He_2^{*-} - He$. The black areas represent the repulsive part of the interaction. For each plot, the lowest energy value of the scales represents 90% of the lowest energy obtained in the potential energy scans. The positions of the dimers are indicated by the white dots.

To examine the different mobilites of $He^{*-}$ and $He_2^{*-}$ reported in the experiments [176], the ratio $\Xi$ was introduced [239]. This ratio is defined as the energy gain to energy cost ratio. The energy gain (energy lowering) is due to the immersion and associated attractive dispersion forces between the dissolved helium species and the surrounding (ground state) helium atoms. The energy cost arises from the formation of the surface around the repulsive void of the dissolved helium species. These values have been estimated in a way similar to the simple arguments given by Ancilotto *et al.* [104]. The surface was defined as the region where the interaction potential with a ground state helium atom becomes zero. Due to the symmetry of the various species, the respective surfaces were calculated by applying Guldin's rule for surfaces of rotation. The energy cost was then defined as the product of the surface area times the surface tension of liquid helium, $\sigma = 0.179$ cm$^{-1}$ Å$^{-2}$ [245]. The energy gain was calculated in a similar manner where the first shell of helium atoms surrounding the repulsive void was taken into account. The number density of liquid helium, $\rho = 0.022$ Å$^3$, was used. For $\Xi < 1$ the respective helium species is not expected to be solvated in HNDs, since the energy cost for the formation of the repulsive void is larger than the energy gained by surrounding it with helium. This was found to be the case for $He^*$ ($\Xi = 0.91$), $He_2^*$ ($\Xi = 0.74$), and $He_2^{*-}$ ($\Xi = 0.31$) [239]. Indeed, $He^*$ as well as $He_2^*$ are known not to reside inside HNDs but are weakly bound on its surface [67, 246]. As for $He_2^{*-}$, it will reside on the surface of the droplet in a 'head-on' position, where the molecular axis is perpendicular to the surface. On the other hand, the $\Xi$ for $He^{*-}$ was reported to be 2.75, in comparison to $\Xi = 2.30$ for He. This was taken as indication that $He^{*-}$ indeed is highly mobile and dissolved in the HND [239].

Path integral Monte Carlo calculations have been employed to investigate the geometry and energetics of atomic and molecular helium anions with up to 32 solvating helium atoms [247-249]. The results support the idea that He*$^-$ is solvated inside the HND, whereas He$_2$*$^-$ is located on the surface. Further investigations were dedicated to quantum features of these anionic species in small helium clusters with focus on the He$_3$*$^-$ system [250]. A detailed review on these phenomena was published recently [248].

## 6.4 The importance of helium anions in helium nanodroplets

### 6.4.1 Single charge transfer from He*$^-$ to dopants

The findings and explanation of the high mobility of He*$^-$ inside HNDs [176, 239] has ample implications on the interpretation of anion formation in HNDs upon electron impact for electron energies of approximately 22 eV and multiples thereof. As mentioned above, single charge transfer from He*$^-$ to SF$_6$ has been observed [176], but some findings have not been explained. For instance, in the absence of a foreign molecule in the HND, the ion efficiency curve of He*$^-$ peaks at 22 eV, with a tail to higher electron energies [177]. Upon doping the HND with SF$_6$ prior to ionization, a significant change in the He*$^-$ signal was observed. Due to electron transfer from the atomic helium anion to the dopant, the overall intensity of the He*$^-$ signal decreases substantially. However, the two lower energy resonances at ∼ 22 eV and ∼ 23 eV decreased more strongly than the higher energy resonance at ∼ 25 eV. It was therefore concluded that the helium anion signal at 25 eV is due to Rydberg-like states [177]. The dependence of the He*$^-$ signal on the droplet size showed a decrease of the higher energy resonances for smaller droplets, which would support an assignment of the 25 eV peak to a diffuse Rydberg state of He*$^-$. In light of recent experimental and theoretical investigations it seems at least plausible that the so-called satellite peaks can be explained via the formation of He*$^-$.

Ion efficiency curves of anions formed by electron interactions with doped HNDs exhibit repetitions of the resonances that are displaced by ∼ 21 eV (see below) [251] For example, this was observed for (D$_2$O)$_2^-$ as well as (D$_2$O)$_2$He$^-$, as shown in Figure 24. Both ion efficiency curves show a resonance around 1.5 eV and a satellite peak at 22.5 eV [215]. This feature was attributed to the electronic excitation of a helium atom by an energetic electron, which loses thereby at least 19.8 eV of kinetic energy. Subsequently, the thermalized electron can react with the dopant trapped inside the HND. However, given the high polarizability of He* (see section 6.2), it seems more likely that the electron will interact with He* first to form He*$^-$, which then transfers its charge to the dopant. The effect of He*$^-$ on dopants in HNDs was observed for electron attachment to nitromethane embedded in HNDs [252]. For (M-H)$^-$, the ion efficiency curve showed no resonance around 1 eV, yet a previously attributed satellite peak could be observed at 22 eV. However, since there is no peak 21 eV lower in electron energy, the resonance around 22 eV cannot be a satellite peak. Closer inspection of the peak shape indicates a similarity to the ion efficiency curve of He*$^-$, which clearly points towards a charge transfer reaction from He*$^-$ to the embedded dopant.

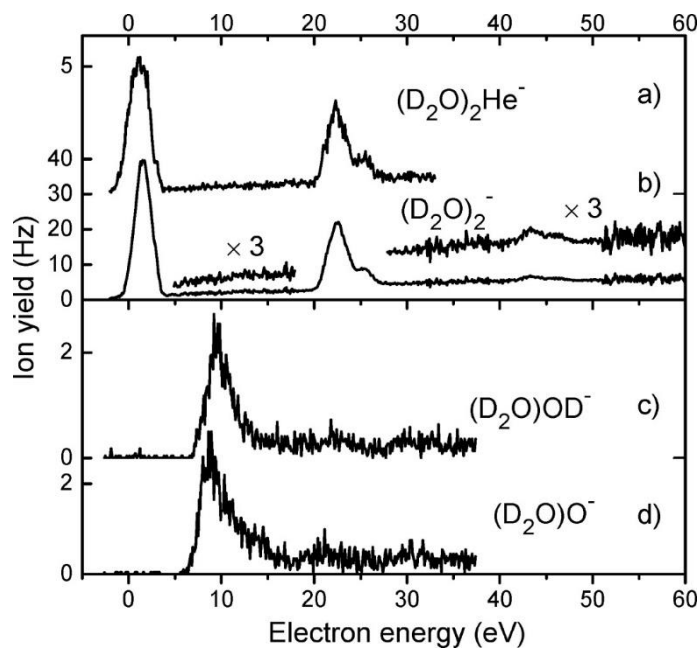

Figure 24 Ion efficiency curves of $(D_2O)_2He^-$ and $(D_2O)_2^-$ (upper panel). Reprinted with permission from Ref. [215]. Copyright 2008 American Chemical Society.

### 6.4.2 Double charge transfer from He*$^-$ to dopants

In addition to single electron transfer from atomic helium anions to dopants, double electron transfer has recently been observed [239]. This is of special interest given that multiply charged anions in the gas phase often decay by electron autodetachment or Coulomb explosion[253]. In the mass spectra obtained at an electron energy of 22 eV for HNDs doped with $C_{60}$, the dominant cluster series is due to $(C_{60})_n^-$, which could be followed up to n = 19 [177, 254]. In addition to this cluster series, also the cluster series $(C_{60})_n^{2-}$, although much weaker in intensity, was observed. The smallest observable cluster dianion in these experiments was the fullerene pentamer [239]. The same was observed for $C_{70}$ cluster dianions. The formation of these dianions in HNDs cannot proceed via sequential electron attachment due to the strong Coulombic repulsion after addition of the first electron. Inspection of the ion efficiency curves for dianionic fullerene clusters has revealed a striking similarity to the ion efficiency curve of He*$^-$ with a relatively narrow resonance peaking at 22 eV. This has indicated that the formation of dianionic fullerene clusters occurs via double charge transfer from the helium anion, which consists of a He$^+$ core surrounded by two loosely bound electrons [255]. In contrast, anion efficiency curves for singly charged fullerene clusters revealed a much broader resonance structure ranging over more than 20 eV, although even here there is a resonance peak at 22 eV indicating a role for He*$^-$ [239]. The energetic accessibility was examined by calculating the double vertical electron affinities with the density functional B3LYP [256, 257] and the Pople's 6-31+G(d) basis set [258]. These calculations revealed that the electron affinity curves for singlet and triplet states of dianionic fullerene clusters from monomer to pentamer are very similar. Taking into account the double ionization energy of He*$^-$, which was estimated to be 4.69 eV based on the first ionization energy of helium (24.6 eV), the excitation energy of helium (19.8 eV) and the corresponding electron affinity (77 meV). In combination with the calculated electron affinities it was argued that two-electron transfer from the He*$^-$ ion to $(C_{60})_n$ clusters with n < 4 is energetically forbidden. However, given the uncertainties in the calculations a more detailed investigation needs to be carried out. After the transfer of the two electrons to the fullerene cluster, it is necessary to explain how the resulting cation and dianion separate. Given the surrounding helium environment,

He$^+$ is expected to form He$_2^+$, which will release 2.5 eV [239, 259] of energy because of the strong bond in the dimer cation. In the case that all excess energy is converted into kinetic energy, the two ions could overcome the Coulomb barrier at a distance as short as 12 Å [177]. This formation pathway of dianions from doped HNDs is expected to be transferable to other types of dianions [239].

### 6.4.3  He$^+$ subthreshold formation via He*$^−$

He*$^−$ may also be involved in the formation of He$^+$ at subthreshold ionization energies [175] (see also Section 5.2). Earlier possible explanations for He$^+$ formation at subthreshold energies, *i.e.* below 24.6 eV, include multiple excitations, *e.g.* Penning ionization of He* by another He* [77, 163]. However, the ejection of He$^+$ from the droplet is difficult to explain [200]. A comparison of the ion yield of He$^+$ and He*$^−$ at subthreshold energies reveals a close similarity [175] (see Figure 20). A possible scheme for forming sub-threshold He$^+$ is proposed in the right hand reaction sequence in Figure 21.

### 6.4.4  Multiple ionization events via He*$^−$

The formation of multiply charged cations by He*$^−$ has also been proposed. Multiply charged fullerene cluster cations have been observed from HNDs at electron impact energies of 85 eV and an electron current of 82 μA [177]. As well as doubly charged ions, triply and quadruply charged ions were also detected. Threshold sizes have been identified for the different charge states and they were similar to values reported from the gas phase [260, 261] (see also chapter 7.3). As in the case of dianionic fullerene clusters, the smallest doubly charged fullerene cluster cation reported to be stable is the pentamer. In a theoretical study it was shown, that the doubly charged fullerene pentamer is not thermodynamically stable, but kinetically [262]. Multiply charged cations in HNDs cannot be formed via initial formation of He$^+$ as the electron impacts the HND. Although the charge can find a (neutral) dopant inside the HND, Coulomb repulsion will prevent any further ionization events via additional He$^+$ present in the HND. As an alternative possibility, Penning ionization via He* was suggested [162], *i.e.*

$$\text{He*} + (C_{60})_n \rightarrow \text{He} + (C_{60})_n^+ + e^-. \quad (4)$$

Because of attractive forces due to polarization between the cation and He*, further reactions could occur

$$\text{He*} + (C_{60})_n^{z+} \rightarrow \text{He} + (C_{60})_n^{(z+1)+} + e^-. \quad (5)$$

Nevertheless, it is known that He* is heliophobic and resides on the surface of the HND while most dopants will be located near the center. An alternative route to multiply charged cations is reactions involving He*$^−$. Since He*$^−$ is heliophilic and of high mobility, it can interact with dopants located near the center of the HND [239]. Furthermore, the strong Coulombic interaction between a cation and He*$^−$ will drive the reaction very efficiently, *i.e.*

$$\text{He*}^- + (C_{60})_n^{z+} \rightarrow \text{He} + (C_{60})_n^{(z+1)+} + 2e^-. \quad (6)$$

The reaction is viable in droplets which are sufficiently large to (i) accommodate the necessary electron bubble to form He*$^−$ and (ii) have a large enough cross section for multiple hits from electrons.

# 7 Ionization of doped helium nanodroplets

## 7.1 Charge transfer from He⁺

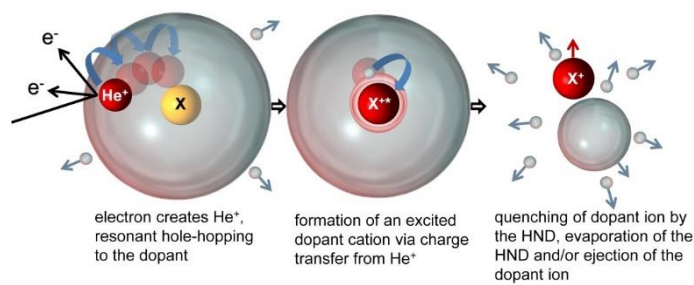

*Figure 25 Reaction scheme for the ionization of a dopant via charge transfer from an initially formed He⁺ ion. A random He atom close to the surface is ionized via electron (or photon) ionization. The positive charge approaches the dopant X via resonant hole-hopping. Ionization by charge transfer can lead to internal excitation of the dopant cation, $X^{+*}$. This excited cation can be quenched via energy transfer to the surrounding He matrix. Complete evaporation of small HNDs or ejection of the dopant ion $X^+$ from larger HNDs leads to bare cations, which are the dominant dopant ions in the mass spectra from doped HNDs.*

The size of a dopant (cluster) in a typical HND is orders of magnitude smaller than the diameter of the doped HND. Thus, both electron and photoionization at energies above the ionization potential of He leads almost exclusively to the formation of He⁺. Long-range electrostatic interaction between the dopant and the He⁺ attracts the two reaction partners and the charge approaches the dopant via resonant hole-hopping [160-162]. A simple model predicts a limit of about 11 hops before He⁺ reacts with an adjacent He atom to form vibrationally excited $He_2^{+*}$. Vibrational relaxation releases 2.35 eV of energy, which is transferred to the HND and leads to the ejection of $He_2^+$. As this ion departs it may capture one or more He atoms at it exits from the HND, leading to the detection of $He_n^+$ cluster ions with n ≥ 2 [161]. The charge transfer reaction

$$He^+ + X \rightarrow X^+ + He \qquad (7)$$

is highly exothermic for all neutral dopants. The difference of the ionization energies of He and X has to be accommodated initially by the dopant. For polyatomic molecules having an ionization energy of typically around 10 eV, almost 15 eV of excess energy is transferred to the dopant. This amount of excess energy can break almost every molecular bond unless the surrounding superfluid He matrix can rapidly cool the excited dopant. For small average droplet sizes ($<N>$ about $10^4$) an excess energy of 10 eV is sufficient to vaporize the entire helium solvent, which results in the detection of bare dopant ions. However, the energy required to vaporize a droplet larger than $10^6$ He atoms exceeds 600 eV (0.6 meV binding energy per He atom). Ion induced dipole interaction between cations and He results in the formation of so called snowballs [153-157] (see Section 5.1) which prefer to stay inside the HND. This should result in the absence of bare dopant ions in mass spectra obtained via ionization of large doped HNDs. Figure 26 shows the ion yield of selected product ions formed upon electron ionization of HNDs doped with water as a function of the average droplet size. The maximum yield for a protonated water cluster depends strongly on the average droplet size. As expected, a large HND picks up more water molecules due to the enhanced capture cross section. Protonated water cluster ions can be observed even for the largest average droplet size. For $H_3O^+$ which originates from the doped HNDs, the data indicate an average droplet size larger than 1.2 million. The abundance of $He_n^+$ cluster ions (dashed lines) strongly increases with the size of the neutral HND, similar to the observation reported by Gomez *et al.* for undoped HNDs [48].

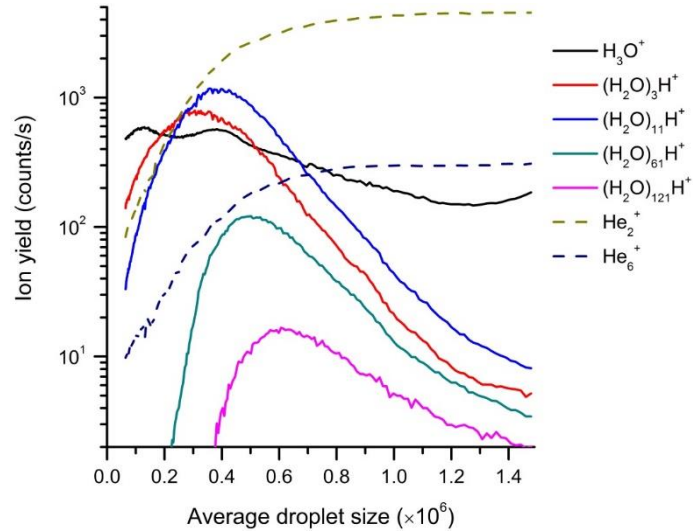

*Figure 26 Ion yields of selected product ions formed upon electron ionization of HNDs doped with water as a function of the average droplet size <N>. The conditions for this measurement were: $p_{He}$ = 2.1 MPa, 8.5 K < $T_{He}$ < 10 K, $p_{water}$ = 2.2 mPa, electron current 45 µA, electron energy 70 eV.*

## 7.2 Penning ionization

The Dutch physicist Frans Michel Penning first reported, back in 1927, the occurrence of the ejection of an electron caused by the transfer of internal excitation energy from metastable Ne* or Ar* to an atom or molecule with an ionization energy lower than the excitation energy of the respective rare gas atom [263]. Now known as Penning ionization, this process results in the production of a cation plus an electron and can trigger the ionization of activated gases in high-temperature chemistry and plasma physics [264, 265]. The high excitation energy and the exceptionally long lifetime (see Table 1), particularly of some helium triplet states, triggered many studies of the reaction

$$\text{He*} + X \rightarrow X^+ + \text{He} + e^- \tag{8}$$

*Table 1 Excited states of atomic He at n = 2 with corresponding energy and lifetime*

| State | Energy (eV) | Mean lifetime (s) |
|---|---|---|
| $2\ ^3S$ | 19.82 | 7870 [266] |
| $2\ ^1S$ | 20.61 | $2 \times 10^{-2}$ [267] |
| $2\ ^3P$ | 20.96 | $1.1 \times 10^{-7}$ [268] |
| $2\ ^1P$ | 21.22 | $5.6 \times 10^{-10}$ [269] |

Jesse and Sadauskis [270] studied the increase in ionization due to α particles in gaseous He caused by the addition of spurious amounts of other gases. The collision of He$^{2+}$ with He leads to He*, which is formed in 90% of collisions in the singlet $2\ ^1S$ state. The resulting cross sections for Penning ionization are dominated by this metastable singlet state. Sholette and Muschlitz [271] measured ionizing collisions of He* in both spin states ($2\ ^3S$ and $2\ ^1S$) with the rare gas atoms Ar, Kr and Xe, as well as the diatomic molecules $H_2$, $N_2$, $O_2$ and CO. For the rare gas atoms, the cross section for Penning ionization was found to be independent of the metastable state and to scale directly with the geometric size of the atom and not with the exothermicity of the charge transfer reaction. Metastable helium atoms He* and molecules $He_2$* were investigated in bulk superfluid by Hill *et al.*

[246] via infrared absorption spectroscopy of electron-bombarded superfluid helium. They confirmed that both species are located in stable voids or so-called bubbles.

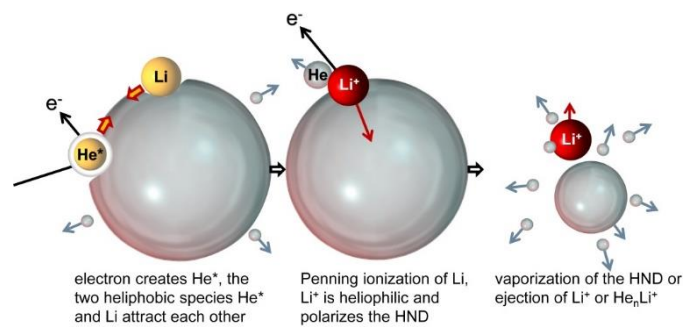

Figure 27 Reaction scheme for Penning ionization of Li by metastable He*. He* is formed by inelastic scattering of an electron at a He atom and the resulting He* then moves to a dimple site at the surface of the HND. Li is also heliophobic and located at the surface. The two highly polarizable (compared to ground state He) heliophobic species attract each other and Li is ionized by the collision with He*, which relaxes to the ground state. Excess energy may lead to the vaporization of a small HND or the ejection of Li+ or He$_n$Li+ ions, which can be detected via mass spectrometry.

Recent calculations by Huber and Mauracher confirmed these bubble states for metastable helium [239]. The high energy cost to form these relatively large bubbles squeezes these species out of the HND and leads to a location above a dimple formed at the surface of the HND. In Figure 27 the reaction scheme for Penning ionization of the heliophobic species lithium attached to the surface of a HND is shown. Since the He* is preferentially located near to the droplet surface, Penning ionization of a heliophilic dopant (most likely located at the center of the HND) can only occur for small droplets. The first investigation of Penning ionization by metastable He in HNDs was performed by Fröchtenicht et al. [71]. They measured SF$_5^+$ from a small HND ($<N>$ = 8000) doped with one SF$_6$ molecule upon irradiation with monochromatic synchrotron light. Although the threshold energy for photoionization of SF$_6$ is 16 eV [272], the ion efficiency curve of SF$_5^+$ exhibits a first resonant feature centered at 21.6 eV which can be assigned to Penning ionization via an electronically excited He* in the 2 $^1$P state. Thus, even for these relatively small HNDs the probability for photoionization of the dopant is much lower than the probability for photoexcitation of one of the surrounding He atoms, even though the cross section for the latter may be three orders of magnitude lower.

Penning ionization upon electron irradiation of doped HNDs was first investigated by Scheidemann et al. [67]. Li vapor was picked up by HNDs with an average size $<N>$ estimated to be between 2000 and 5000. All product ions that contain Li exhibit a threshold energy of roughly 19 eV, which agrees reasonably well with the excitation energy for He* in the 2 $^3$S state. This indicates that the Penning ionization process is involved. By picking up Na by HNDs with an average size of about $10^5$ He atoms, the group of Kresin was able to observe Na$_n^+$ cluster ions containing up to 13 sodium atoms [273]. The threshold and shape of the ion efficiency curves for Na$_n^+$ indicate that the neutral precursor clusters are located at the surface of the HND and that Penning ionization is operative. The authors also report a pronounced odd-even oscillation in the intensities of the Na$_n^+$ ion peaks with a particularly high abundance for Na$_9^+$. Such oscillations are well known in mass spectra of metal clusters. They reflect the higher stabilities of even-electron systems, as explained, for example, by the jellium model [274, 275]. A similar odd–even pattern has been seen in silver clusters in HNDs [276, 277] and has been interpreted as evidence for considerable fragmentation accompanying ionization of the doped HND by 70 eV electrons. As mentioned later in Section 7.1, the main source for excess energy in the dopant cluster is the highly exothermic charge transfer from an ionized He atom. The pickup process is of a purely statistical nature and should result in a Poisson (for single sized neutral HNDs) or log-normal distribution (for a log-normal size distribution of the neutral HNDs). Thus Vongehr et al. [273] concluded that Penning ionization also causes significant ion fragmentation, even though the ejected electron could carry away excess energy released by this often very exothermic process.

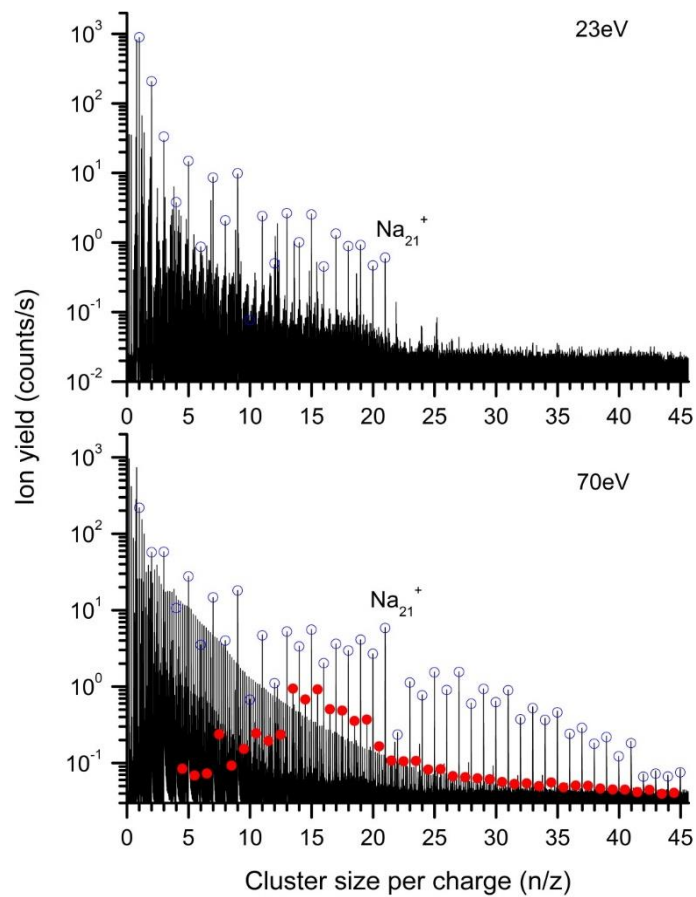

*Figure 28 Mass spectrum of HNDs doped with sodium. Conditions: stagnation pressure 2.5 MPa and cluster source temperature 5.7 K (which gives a mean droplet size of $1.4 \times 10^9$ [48]), temperature of the sodium oven: 533 K, electron current 9.2 μA. The peaks for singly charged $Na_n^+$ are designated by the open circles and odd-numbered doubly charged clusters $Na_{2n-1}^{2+}$ by solid circles. At 23 eV electron energy the $He_n^+$ ion series is missing and the $Na_n^+$ cluster ions are formed via Penning ionization or $He^{*-}$ when submerged into the droplet for n > 20. At an electron energy of 70 eV doubly-charged sodium clusters are also formed. The smallest doubly-charged sodium cluster is $Na_9^{2+}$.*

Combining a HND source with a high resolution time-of-flight mass spectrometer system (see Figure 3, Section 2.1) An der Lan *et al.* were able to observe cluster ions of Na [110] and K [111] containing up to 100 atoms. A section of two mass spectra for sodium doped HNDFigurs, measured at 23 eV and 70 eV is shown in Figure 28. As in the case of the previous studies [273] pronounced magic numbers due to spin pairing and electronic shell closures were found for both electron energies. The main emphasis of the studies by An der Lan *et al.* was the experimental determination of the location of large neutral alkali clusters within or upon a HND. According to a simple theoretical model by Stark and Kresin [278] the interaction of an alkali cluster with a HND has three contributions, *i.e.*, (i) van der Waals attraction, (ii) short range Pauli repulsion and (iii) the surface tension energy of the cavity wall. For clusters exceeding a critical size the van der Waals attraction will exceed the contributions of the repulsive interactions and thus the energetically lowest energy position for such an alkali cluster should be inside the HND. The critical sizes for submersion calculated by Stark and Kresin are 21 and 78 for $Na_n^+$ and $K_n^+$, respectively [278]. An der Lan *et al.* measured ion efficiency curves for $Na_n^+$ [110] and $K_n^+$ [111] clusters and observed a sudden transition of the threshold energy for $Na_n^+$ clusters from about 20 eV to 24 eV at a critical size of n = 21 [110], in perfect agreement with the value calculated by Stark and Kresin [278]. Also the mass spectrum measured at an electron energy of 23 eV (upper diagram of Figure 28) reveals a lack of $Na_n^+$ cluster ions beyond n = 21, as these clusters after submersion are not ionized via Penning ionization. The few peaks above the noise level can be assigned to product ions formed upon electron ionization of pump oil and other potential trace contaminants in the residual gas (note the logarithmic vertical scale). In the case of potassium, Penning ionization can be observed for all cluster sizes but at around n = 75 the relative contribution for Penning ionization strongly decreases by more than a factor of four [111]. This cluster size is also in reasonable agreement with the prediction made by Stark and Kresin [278], suggesting $K_n$ clusters of this size and larger submerge into the HNDs.

In a recent study we showed that alkalis can be induced to submerge into HND when a highly polarizable co-solute is additionally picked up [279]. When adding $C_{60}$ to an alkali doped HND all sodium clusters, and probably single Na atoms, enter the HND. Even clusters of cesium, one of the most extreme heliophobic species, seem to dissolve in liquid helium when $C_{60}$ is added. Only atomic Cs cannot be pulled into the HND by $C_{60}$ and remains at the surface. Hauser and de Lara-Castells interpret these experimental results in light of a quenched electron-transfer reaction between the fullerene and the alkali dopant, which is additionally hindered by a reaction barrier resulting from the necessary extrusion of helium upon approach of the two reactants [280]. In this theoretical study a combination of standard methods of computational chemistry with orbital-free helium density functional theory was used. The authors assumed that the electron transfer process from the Cs (monomer and dimer) to the $C_{60}$ is considerably quenched, but lets the dimer react and prevents the atom from approaching the fullerene.

### 7.3 Magic numbers in positively charged dopant cluster distributions

For dopants other than alkali metals the cluster size distribution obtained via ionization of heavily doped HNDs reveal magic numbers, too. In all cases these magic numbers perfectly match the intensity anomalies reported for cluster ion distributions obtained via ionization of isolated clusters of the same material. Cluster growth via gas aggregation or supersonic expansion initially forms boiling hot clusters, which may lead to intensity anomalies of the resulting neutral cluster sizes. However, every heliophilic atom that is picked up by a HND will remain inside the droplet and the rapid cooling should mean that the resulting neutral cluster size distribution will be free of any intensity anomalies or magic numbers [273]. Diederich *et al.* [123], however, have speculated that magic numbers of neutral clusters may also contribute to the intensity anomalies observed in mass spectra from Mg doped HNDs. Irrespective of the ionization method, *i.e.*, electron ionization as well as nanosecond and femtosecond multiphoton excitation, the mass spectra show similar characteristic features.

Table 2 summarizes all magic numbers reported for clusters obtained via ionization of doped HNDs and compares these with magic numbers reported for mass spectra obtained via ionization of isolated clusters in the gas phase. In some cases, sodium in particular, magic numbers are shifted by one unit. This is easily explained. In contrast to ionization of $Na_n$ embedded in HNDs, bare sodium clusters can be ionized softly, without fragmentation. The magic numbers of sodium clusters are related to the number of valence electrons [274, 275], thus magic cations require an extra atom compared to neutral $Na_n$. Electronic shell closures of metal cluster ions are often explained via the Jellium model [275, 281] whereas geometric packing often explains shell closures of van der Waals cluster ions [69]. In the case of adamantane clusters, $(C_{10}H_{16})_n$, from doped HND we observed pronounced magic numbers for singly, doubly and triply charged clusters that reveal a transition from icosahedral packing to fcc packing between n = 19 and n = 38. Another remarkable fact is that the magic numbers are essentially the same for all three charge states [282]. Both, magic numbers and multiply-charged dopant cluster ions suggest Coulomb explosion of initially highly–charged dopant clusters as the underlying mechanism in cluster formation. This also explains why cations which are strong heliophilic species are pushed out of large HNDs that cannot be vaporized by the energy provided by the ionizing projectile.

*Table 2 Intensity anomalies in the cluster ion size distribution for various materials from ionization of doped HNDs and isolated clusters of the respective substance. Values refer to local maxima (numbers in bold), minima (italic), and abrupt drops (underlined)*

| material | HNDs | isolated cluster |
|---|---|---|
| $Ar_n^+$ | *20*, **55** [283] | *20* [284, 285], **55** [286] |
| $Kr_n^+$ | **13**, **16**, **19**, **22**, **25**, **29** [286] | **13**, **16**, **19**, **22**, **25**, **29** [287] |
| $(H_2)_nH_3^+$ | **3**, **6** [288] | **3**, **6** [289, 290], |
| $(H_2O)_n^-$ | **2**, **6**, **11**, **16** [215] | **2**, **6**, **11**, **13** [291] |
| $(Serine)_n^+$ | **8** [292] | **8** [293] |
| $(CH_4)_nCH_5^+$ | **2**, **13**, **20**, **22**, **25**, **28**, <u>53</u> [294] | **2**, **7** [295] |
| $(NaF)_nNa^+$ | **4**, **13**, **22**, **38** [296] | **4** [297, 298] |
| $(C_{10}H_{16})_n^+$ | **13**, **19**, **38**, **52**, **61**, **70**, **75**, **79**, **82**, **86**, **90**, **94**, **98**, **104**, **108**, **112**, **116**, **120**, **124** [282] | - |
| $Mg_n^+$ | **10**, **20**, **35**, **47**, **56**, **69** [123] | **4**, **9**, <u>19</u>, **34**, **46**, **55**, **69** [299] |
| $Cd_n^+$ | **10**, **20**, **35**, **46**, **54**, **70** [123] | **10**, **20**, **35**, **46**, **54**, **69** [300, 301], |
| $Zn_n^+$ | **10**, **20**, **35**, **46**, **54**, **69** [123] | **10**, **20**, **35**, **46**, **54**, **69** [301] |
| $Ag_n^+$ | **3**, **5**, <u>9</u> [276] | **3**, **5**, <u>9</u> [302] |
| $Au_n^+$ | **5**, **7**, <u>9</u> [303] | **3**, <u>9</u>, <u>21</u>, **35**, <u>59</u> [304] |
| $Na_n^+$ | <u>3</u>, <u>9</u>, <u>21</u>, <u>41</u> [110] | <u>8</u>, <u>20</u>, <u>40</u>, <u>58</u> [305] |
| $K_n^+$ | <u>3</u>, <u>9</u>, <u>21</u>, <u>41</u> [123] | <u>3</u>, <u>9</u>, <u>21</u>, <u>41</u> [306] |
| $In_n^+$ | <u>7</u> [123] | 7 [307] |
| $Cr_n^+$ | <u>7</u>, 13 [303] | 7, 14 [308] |

## 7.4 Photoionization of doped helium nanodroplets

Photoionization of doped HND was covered in a recent review by Mudrich and Stienkemeier [201]. Multiphoton ionization is much more efficient for dopants than He and can be achieved with commercial laser systems [309-311]. In direct photo- or multiphoton-ionization of dopants the effect of the HND is reduced to a heat bath with a small matrix shift. However, as soon as the photons are able to excite or ionize helium, all processes mentioned above become relevant. The group of Neumark investigated photoionization of HNDs doped with $SF_6$ [312] and the rare gases Kr and Xe [187]. These experiments were performed using VUV synchrotron radiation from the Advanced Light Source providing ~ $10^{13}$ photons/s in the range of 20–26 eV with 18 meV bandwidth. Both, the kinetic energy of the resulting ions and photoelectrons were analyzed. Penning ionization of the dopants by metastable helium in various states was observed. The photoelectron spectra revealed four distinct mechanisms for photoelectron ejection. Sharp pairs of features at high photoelectron energy indicate that these electrons do not exchange energy with the HND and that some He* convert from the 2p $^1$P to the 2s $^1$S state via He – He* collisions within the HND. In addition, Neumark and coworkers reported on a broad, intense feature in the photoelectron spectra, representing electrons that undergo significant energy loss, and a small amount of ultraslow electrons that may result from electron trapping at the droplet surface. For very large average droplet sizes ($<N>$ up to $2.5×10^5$) this broad feature shows an onset around 1 eV, which coincides with the conduction band edge, $V_0$ in liquid He. Electron scattering experiments [112] as well as electron attachment studies of HNDs [44, 214-216, 313] have also shown evidence for a barrier in the range of about 1 - 2 eV. Utilizing monochromatized VUV light from the free electron laser FERMI@Elettra in Trieste the autoionization dynamics of multiply excited HNDs was studied [29, 198]. Extremely high ionization rates were observed for photon energies below the ionization energy of He via Penning ionization that the authors assign to a novel many-body autoionization process which is related to inter-atomic Coulombic decay [199]. Very recently doped HNDs were also investigated at this free electron laser facility [202], and first results on magnesium doped HNDs confirm the presence of a theoretically predicted [314, 315] and recently experimentally observed process [316, 317], electron transfer mediated decay. Doubly-charged magnesium cluster ions are formed at photon energies higher than the ionization energy of He and the resulting photoelectron spectra measured in coincidence with the (decaying) Mg cluster are strong indications for an electron transfer mediated decay process [202].

## 7.5 Multiple-ionization of dopants

For several species, X, the ionization energy of He is higher than the energy required for the formation of $X^{2+}$. The reaction

$$He^+ + X \rightarrow He + X^{2+} + e^- \qquad (9)$$

has been investigated in the gas phase utilizing ion flow tubes for X = $C_{60}$ and $C_{70}$ [318], and polycyclic aromatic hydrocarbons [319] such as naphthalene [320]. Selected ion flow tube (SIFT) experiments indicate a reaction of He$^+$ with $C_{60}$ at about room temperature, which leads to single (<90 %) and double ionization (>10 %) and no fragmentation [318]. In the case of naphthalene, only 3% of the reaction products end up as the singly charged naphthalene ion $C_{10}H_8^+$ whereas 22% of the reaction products are doubly charged ions and the rest are singly charged fragments [320]. Double ionization of a homogeneous cluster requires much less energy if two individual monomer units are singly ionized instead of the double-ionization of one monomer. For example, the double ionization of one argon atom requires 43.39 eV [242] whereas the single ionization of two neighboring atoms in an argon cluster requires only 35.5 eV (2×15.76 eV [321] plus the Coulomb energy of roughly 4eV which is the potential energy of two elementary charges at a distance of 3.8 Å). A similar reduction of the appearance energy of multiply-charged clusters upon electron collision was reported for several isolated van der Waals clusters [322-327]. Rühl *et al.* [328] determined the double ionization potentials for differently sized Ar clusters below the critical size for doubly-charged clusters via single photoionization using tunable synchrotron radiation. The threshold energies of $Ar_n^{2+}$ are found to be between those of the atom and the solid exhibiting a linear relationship between threshold energies and the average cluster size.

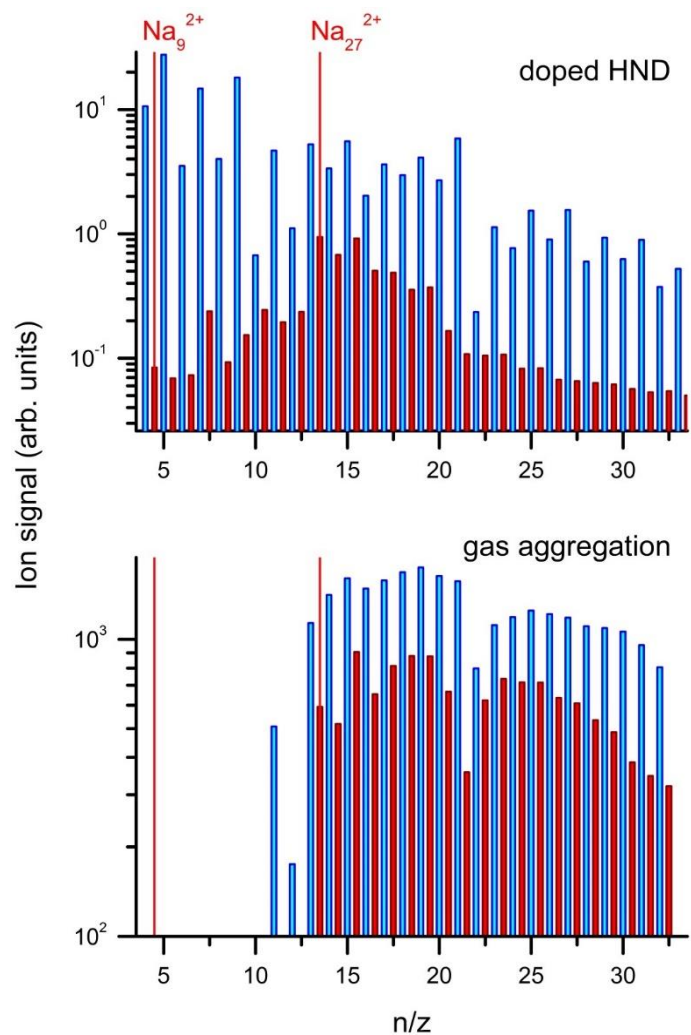

*Figure 29 Cluster size distribution of singly- (blue bars) and doubly-charged (red bars) sodium clusters formed via electron ionization of doped HNDs (upper diagram) and multistep photoionization of pristine Na clusters, followed by evaporative shrinking (lower diagram, values taken from Ref. [329]).*

If the electrostatic repulsion of the charged centers of a doubly- or multiply-charged cluster exceeds the binding energy, the cluster is prone to Coulomb explosion [330, 331]. With increasing cluster size, the distance and thus the repulsive Coulomb energy between the charged centers is reduced. Below a certain size, doubly charged cluster ions undergo Coulomb explosion to yield two singly-charged fragment ions. These charge-driven instabilities manifest themselves in the absence of multiply-charged clusters in mass spectra below an experimental appearance size $n_{exp}$, sometimes also referred to as the critical size, which was first documented by Sattler *et al.* [330]. Figure 29 compares the cluster size distributions of product ions formed upon multistep photoionization of gaseous Na clusters, as measured by Martin *et al.* [329] (lower diagram), with a mass spectrum obtained via electron ionization of HNDs doped with sodium [329] (upper diagram). In both cases a minimum stable cluster size is expected for doubly-charged clusters, but this is substantially lower in the case of the doped HNDs, *i.e.*, $n_{exp,HND}$ = 9 compared to $n_{exp,gas}$ = 27. Numerous reports have explored the energetics and dynamics of cluster ions in the vicinity of $n_{exp}$ (for reviews, see [332, 333]). Early work was directed at determining values of $n_{exp}$, fission barriers, and size distributions of fission fragments with a focus on van der Waals systems [324, 334-341]. Later work focused on metal clusters and the role of electronic shell structure [342-345].

Initially it was assumed that clusters of size $n_{exp}$ were near the Rayleigh limit at which the fission barrier vanishes. However, the observation of metastable fission, on the time scale of microseconds, revealed that fission near $n_{exp}$ is thermally activated and competes with evaporation of monomers [336, 338, 346-349]. In most experiments, cluster ions are "boiling hot;" their vibrational temperatures are set by the heat of evaporation and are difficult to vary [350]. Thus, in spite of the expected temperature dependence, experimental values of $n_{exp}$ have been highly reproducible, indicating a comparable vibrational temperature of the clusters ions under investigation. Huber and coworkers demonstrated that colder cluster ions can be produced in collisions with highly charged atomic ions [351]. For highly-charged sodium clusters with charge states up to ten they reported $n_{exp}$ very close to the Rayleigh limit but for low charge states the $n_{exp}$ values are substantially larger than the theoretically expected values [352]. The authors conjectured that the measured values for $n_{exp}$ were limited by the initial cluster temperature.

The enormous cooling power of superfluid He droplets provides a possibility to effectively form ions at temperatures below 1 K. In the case of multiply charged (metal) cluster ions $n_{exp}$ should reach the Rayleigh limit. The value of $n_{exp}$ for doubly-charged sodium cluster ions obtained from the mass spectrum shown in Figure 28 is indeed much lower than the values given in the literatures. For the alkali metals potassium and cesium the same observation was made and recently published [353]. An even larger discrepancy of $n_{exp,HND}$ with an experimentally determined critical size for dications of isolated clusters has been observed for coronene. Johansson *et al.* [354] obtained a value of $n_{exp,gas}$ =15 upon $He^{2+}$ collisions with coronene clusters formed via gas aggregation in an effusive jet. These dications are five times larger than the Rayleigh limit at which the calculated activation barrier for charge separation vanishes [355]. Mahmoodi Darian *et al.* report the mass spectrometric observation of doubly charged coronene trimers, produced by electron ionization of HNDs doped with coronene [356]. The observation implies that $Cor_3^{2+}$ features a non-zero fission barrier and that the helium environment efficiently quenches charge separation in doubly charged coronene clusters. It is interesting to note that for all van der Waals clusters studied so far, $n_{exp}$ obtained from doped HNDs is only marginally lower compared to values obtained from isolated clusters. In Table 3, $n_{exp}$ for all doubly- (or multiply-) charged cluster ions from doped HNDs are compared with values found in the literature for experiments with isolated clusters.

Table 3 Experimental appearance size ($n_{exp}$) for multiply-charged clusters obtained from doped HNDs and conventional techniques (isolated clusters)

| species | Charge state | Isolated cluster | Doped HNDs |
|---|---|---|---|
| $C_{60}$ | 2+ | 5 [261] | 5 [177] |
| $C_{60}$ | 2- | - | 5 [254] |
| $C_{24}H_{12}$ | 2+ | 15 [354] / 3 theory [355] | 3 [356] |
| $C_{10}H_{16}$ | 2+ | - | 19 [282] |
| $C_{10}H_{16}$ | 3+ | - | 52 [282] |
| $CO_2$ | 2+ | 43 [357] | 43 [358] |
| Na | 2+ | 27 [329] | 9 [357] |
| K | 2+ | 19 [343] | 11 [353] |
| Cs | 2+ | 19 [359] | 9 [353] |
| Cs | 3+ | 49 [359] | 19 [294] |
| $CH_4$ | 2+ | (78 theory [338]) | 70 [294] |

Coulomb explosion following the photoexcitation of small rare gas clusters provided the first experimental proof for interatomic Coulombic decay [360, 361], a mechanism that was theoretically predicted by Cederbaum and coworkers in the late 1990s [362]. An electronically excited ion transfers its excitation to a neighboring atom where it leads to the emission of an electron. This radiationless transition leads to the formation of two closely spaced ions which are separated via Coulomb explosion. In a later study the same group explored double ionization of atoms and molecules inside a HND [363]. According to their model, the double ionization cross section for Mg inside a HND is increased by three orders of magnitude compared with the double ionization cross section of the isolated Mg atom. This enhancement is achieved by an indirect process, where a He atom absorbs a photon and the resulting $He^+$ cation is neutralized quickly by a process known as electron transfer mediated decay, thereby producing a doubly ionized species. After initial formation of $He^+$ this mechanism is expected to be operative for any HND doped with a species that has a double-ionization energy lower than 24.59 eV. In a subsequent experimental study LaForge *et al.* confirmed electron transfer mediated decay for Mg clusters embedded in He droplets [364]. This decay channel was shown to be a dominant ionization mechanism for energies above the ionization threshold of He. For clusters larger than the tetramer, $Mg_4$, doubly ionized Mg clusters were observed following this ionization process that were stable within the time scale of the experiment (several tens of 10µs). For single Mg atoms embedded in HNDs, the electron transfer mediated decay channel is closed due to the formation of an equilibrated $He_2^+$ ion. Photoionization of doped HNDs via intense laser beams has been recently studied by the group of Stienkemeier and Mudrich [365]. For Xe doped HNDs high charge states of both $He^{2+}$ and $Xe^{21+}$ have been detected [366].

## 7.6 Sequential Penning ionization

When exploring the mass spectra of HNDs doped with methyl iodide, doubly charged ions of the form $He_nI^{2+}$ were discovered with relatively high abundance [367]. The energy required to form $I^{2+}$ from $CH_3I$ is 32.05 eV for the reaction

$$CH_3I + e^- \rightarrow I^{2+} + CH_3 + 3e^-, \quad (10)$$

or 31.97 eV if $CH_3^-$ is formed and two electrons are emitted [368]. Even when considering the solvation energy of the He atoms surrounding $I^{2+}$, a single $He^+$ ion cannot form $He_nI^{2+}$ ions, because more than 30 eV of energy is required. The ion efficiency curve for $I^{2+}$ exhibits two onsets, one at about 32 eV which can be assigned to the double ionization of gas phase $CH_3I$ and a second one at about 40 eV. All $He_nI^{2+}$ ions are formed exclusively via this second process, having a threshold energy that is twice the excitation energy of He. These observations point toward an ionization process where doubly-charged ions are produced by sequential Penning ionization. The stepwise ionization mechanism is illustrated in Figure 30 and the process itself can be described by the following reactions

$$e^- + HND \rightarrow e^-(-2 \times 19.81 eV) + 2He^* (2\ ^3S) @HND$$

$$CH_3I + He^* \rightarrow I^+ + CH_3 + He + 6.9 eV$$

$$I^+ + He^* \rightarrow I^{2+} + He + 0.7 eV \quad (11)$$

$$\rightarrow HeI^{2+} + 0.8 eV$$

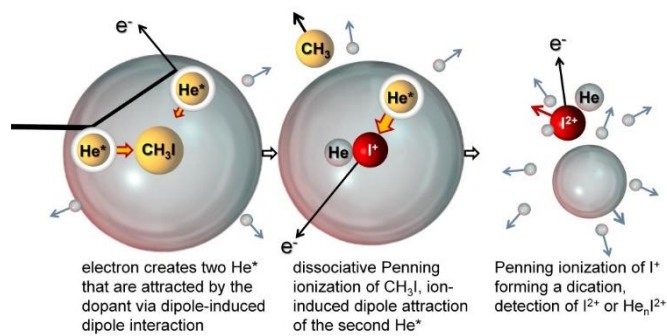

electron creates two He* that are attracted by the dopant via dipole-induced dipole interaction

dissociative Penning ionization of $CH_3I$, ion-induced dipole attraction of the second He*

Penning ionization of $I^+$ forming a dication, detection of $I^{2+}$ or $He_nI^{2+}$

*Figure 30 Sequential Penning-like ionization where two metastable He* (2 $^3S$) atoms are formed by one incident electron. Both He (2 $^3S$) atoms are weakly attracted via dipole-induced dipole interactions to the methyl iodide. One excited helium atom transfers its internal energy to the embedded methyl iodide, and a molecular or atomic cation is produced according to the reaction sequence above. Ion-induced dipole interaction attracts further electronically excited He* atoms efficiently. In a final step, the internal energy of the second metastable He* (2 $^3S$) is transferred to the atomic iodine cation, and this energy transfer can lead to a Penning-like ionization of $I^+$. For better visualization the sizes of the electrons, He* (2 $^3S$), and the neutral and charged dopants are strongly exaggerated compared to the size of the droplet.*

Optimum conditions for sequential Penning ionization of HNDs doped with $CH_3I$ were found for relatively small average droplet sizes of $<N> \approx 50000$ ($p_{He}$ = 20 bar and $T_{He}$ = 10 K). For larger HNDs the heliophobic He* will be less likely to interact with the heliophilic dopant $CH_3I$. In addition, the small average HND size prevents multiple electron collisions, whereas these cannot be neglected for large HNDs as the geometric cross section of the droplet scales with $N^{2/3}$ (see Section 5). The close proximity between He* and the dopant leads to a competition between Penning ionization and formation of $He^{*-}$, where the latter can also contribute to the formation of multiply charged dopants.

# 8 Solvation of ions in helium nanodroplets

In Section 5 we discussed the formation of snowballs upon injection of ions into liquid helium. There we focused on $He^+$, $He_2^+$ and $He_3^+$ while here we consider heterogeneous systems, *i.e.* atomic or molecular impurity ions in helium. Topics of interest are as follows. What is the evidence for formation of one (or perhaps even more than one) solvation shell? How large are the shells and how many He atoms do they contain? Is the solvation shell solid-like? Do quantum phenomena play a role? If solid-like, what is the structure in the shell? Is there a substructure within the solvation shell? What are the dynamics of solvation?

## 8.1 Size and structure of snowballs

The long-range interaction between a charge and helium is attractive. The resulting electrostrictive force leads to a local increase in the helium density. For large enough forces the density will exceed that of solid helium. In this situation the atoms in the solvation layer(s) may localize and form rigid and possibly ordered structures.

The strength of the electrostrictive force will depend on the size of the ion and its charge state (singly *versus* multiply-charged). It will also depend on the short-range interaction between the ion and the medium. The interaction between an electron and helium atoms is repulsive as a result of the Pauli exclusion principle; an excess electron injected into helium will reside in a bubble at an energy 0.22 eV above the vacuum value [179], (see also section 9). Atomic and molecular ions will lower their energy upon solvation in helium (with the exception of $H^-$ [369]) but the energy gain tends to be weak for singly charged alkaline earth elements because of the repulsion due to the remaining valence electron, and for anions. Thus, ions in HNDs may or may not be surrounded by a solid-like layer of He atoms.

When doped HNDs are ionized, one often observes ion series of the form $A^{\pm}He_n$, where A is an atomic or molecular ion (either the dopant species, a cluster of dopants, or a fragment). The degree of complexation with helium depends on the experimental conditions, and on the initial position of the dopant (dimple *versus* interior state). Early work showed that embedded clusters such as $Ar_n$ give rise to long-lived, detectable $Ar^+He_n$ ions but not $Ar_2^+He_n$; it was argued that the vibrational energy in $Ar_2^+$ would either boil off all attached He atoms or lead to the ejection of a bare $Ar_2^+$ from the droplet [66]. The difficulty of complexing molecular ions with helium has been overcome in experiments that involve much larger HNDs (typically, $10^6$ atoms rather than $10^3$ as in the early work [66]).

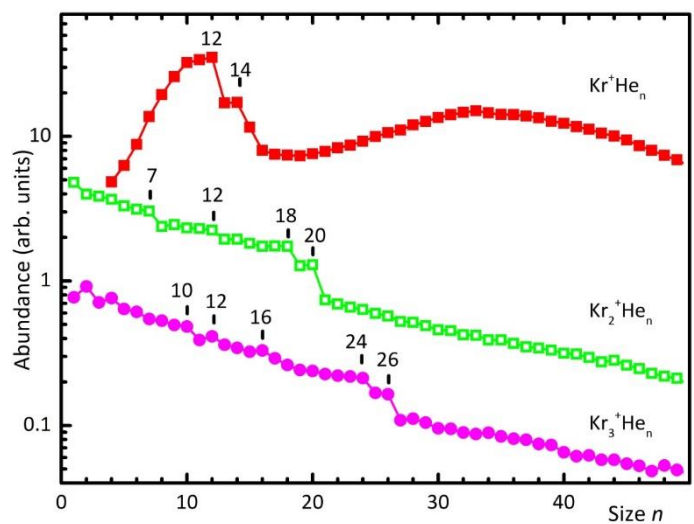

*Figure 31 Ion abundance of $Kr_m^+He_n$ for m = 1, 2, 3 [370]. Each series shows a distinct pair of steps, at 12 & 14 for $Kr^+$, 18 & 20 for $Kr_2^+$, and 24 & 26 for $Kr_3^+$. Several other minor abundance anomalies are also seen.*

The abundance distribution of $A^{\pm}He_n$ often shows local deviations from an otherwise smooth curve. An example is shown in Figure 31, which displays the abundance of $Kr_m^+He_n$ for m = 1, 2, 3 [370]. $Kr^+He_n$ exhibits abundance anomalies at n = 12 and 14, $Kr_2^+He_n$ at 18 and 20, $Kr_3^+He_n$ at 24 and 26. Several other weaker anomalies are also labeled in this figure. Note that the envelope of the $Kr^+He_n$ series differs from that of $Kr_2^+He_n$ and $Kr_3^+He_n$. These envelopes depend to some degree on the experimental conditions; experiments involving smaller HNDs produce series that decrease more rapidly. The anomalies may be less obvious on a rapidly decreasing distribution but they can be recovered by dividing the distribution by a smooth function (for example, a running average [371], or a low-order polynomial fit to the distribution [372]). Some authors prefer to plot the ratio of the abundance of adjacent cluster sizes, $I_{n+1}/I_n$ [32, 373, 374]; local minima in this quantity will reflect steps or local maxima in the abundance distribution. Whatever method one applies, one finds that the values at which steps or maxima (often referred to as magic numbers [69]) occur do not depend (or barely depend) on the experimental conditions (*cf.* the discussion of $He_n^+$ distributions in Section 2.1).

What do these anomalies tell us? As explained later in section 12 they mirror the size dependence of the dissociation energies $D_n$, *i.e.* the difference between the total energies of adjacent cluster sizes. In other words, $D_n$ equals the energy that is required to adiabatically remove the most weakly bound He from the complex. In one scenario [375, 376], the abundance directly tracks $D_n$. In Figure 32 we show two cases where the agreement between experimental abundances and calculated dissociation energies is particularly convincing, namely $Ar^+He_n$ and $Na_2^+He_n$ (panels a and b, respectively). The experimental data [370, 376] were scaled to match the computed adiabatic dissociation energies of $Ar^+He_n$ [377] and $Na_2^+He_n$ [378]. For $Ar^+He_n$ the agreement is excellent as far as the "magic numbers" are concerned: strong drops after n = 12, 32, 44 and an otherwise smooth curve. Likewise, the strong drops after n = 2 and 6 in the calculated dissociation energies of $Na_2^+He_n$ agree with measured ion abundances. Around n = 20 there is some disagreement but with increasing size n the accuracy of calculations and the statistical significance of experimental data will deteriorate (in contrast to $Ar_m^+He_n$, mass spectra of $Na_m^+He_n$ are affected by many impurity ions).

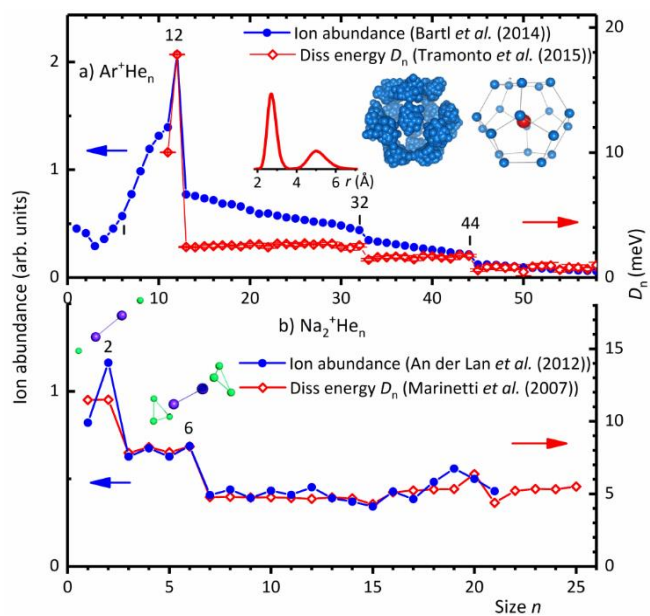

Figure 32 Ion abundances of $Ar^+He_n$ [370] and $Na_2^+He_n$ [376] (panels a) and b), respectively, left scale) compared with their calculated dissociation energies [377, 378] (right scale). Insets in panel a) show properties computed for the magic $Ar^+He_{32}$, namely the radial density distribution of $^4He$ surrounding $Ar^+$ (revealing two distinct solvation shells), the polymers associated with $^4He$ in the 2nd shell, and the structure if each polymer in the 2nd shell is replaced by its center of mass (the icosahedral 1st shell is not shown) [377]. The computed geometries of the magic $Na_2^+He_2$ and $Na_2^+He_6$ are indicated in panel b) [378].

The "contrast" in the computed dissociation energies is somewhat larger than in the abundance data. This may indicate limits in the assumption [375, 376] that the heat capacity of these species is negligible, inaccuracies in the computed dissociation energies, or the effect of finite temperature in the experiment. The computed dissociation energies are of the order of 10 meV (see Figure 32), and therefore the ions in the experiment are at about 4 K (see also the discussion in Section 12.1).

It is important to note that the anomalies do not necessarily indicate closure of solvation shells. One may use the numbers to suggest structural models, but most importantly they serve as benchmarks for computational studies. It was speculated that the sizes n = 12, 32, 44 for $Ar^+He_n$ indicate successive closure of three solvation shells of icosahedral symmetry around $Ar^+$, namely an icosahedron inside a dodecahedron inside an icosahedron [370]. It is interesting to note that $Cs^+$ and $H^-$ solvated in molecular hydrogen feature the same set of magic numbers [379, 380]. These geometries were indeed found in the path-integral study by Galli and coworkers at 0.5 K [377], probably the only case so far in which three distinct, solid-like solvation shells were obtained. The left-most inset in Figure 32a) shows the radial distribution function computed for $Ar^+He_{32}$. The two shells are distinct; the helium density in the intermediate region is essentially zero. The next inset shows the spatial distribution of the helium atoms in the 2nd shell (which is completed for 32 He atoms), computed for $Ar^+He_{128}$. In the right-most inset the polymers are represented by their centers of mass, revealing the dodecahedral structure of this 2nd shell [377].

Recently, adducts formed between small gold clusters and helium atoms, $Au_n^+He_m$, were reported [381]. For structures with n ≤ 7, planar gold configurations were determined, partly supported by quantum chemical calculations. For n = 8, the experimental findings point towards the formation of a planar structure rather than a 3D structure. However, this finding is in contrast to conclusions drawn from ion mobility spectrometry measurements [382, 383].

Things get more complicated when the embedded ion has structure. The anomaly at $Na_2^+He_2$ cannot possibly signal completion of a solvation shell. The interaction between each He atom and $Na_2^+$ is much stronger than the He-He potential leading to the formation of a quasilinear (HeNaNaHe)$^+$ ion [378]. The first solvation shell is complete when each Na atom is capped by three He atoms (see the insets in Figure 32b).

Table 4 compiles experimental and theoretical data for many ions but the list is by no means complete. Gianturco and coworkers have investigated several ions that are not listed, see [384] and references therein. Reviews of ion chemistry in helium have been published by Grandinetti [385, 386]. Also not listed are anomalies in the abundance of Ne$_m^+$He$_n$, Ar$_m^+$He$_n$, and Kr$_m^+$He$_n$ for m $\geq$ 3 [370, 387]. Most experimental data are extracted from mass spectra of doped HNDs but some data were obtained by injecting ions into a drift tube filled with helium gas at low density and low temperature (4.4 K) [388-390].

There are several systems listed in Table 4 where experiment and theory disagree. Disagreement is particularly striking for Na$^+$, for which theoretical studies predict enhanced stability for the icosahedral Na$^+$He$_{12}$ [391-394] while the experiment reports n = 9. In a recent study by Issaoui *et al.* vibrational delocalization was accounted for by using zero-point energy corrections at the harmonic or anharmonic levels [395]. The snowball effect, *i.e.* localization of the He atoms, was found to break down before shell completion, and n = 10 emerges as a possible magic number. Quite generally, vibrational delocalization decreases the size of the first solvation shell, i.e. the number of atoms that fit into the first solvation shell [396].

Some theoretical values listed in Table 4 refer to anomalies in computed dissociation energies, while others refer to the number of atoms in a solvation shell, usually extracted from computed radial distribution functions. The former quantity would be expected to appear in experimental data but the latter may be elusive. The minimum between the first and second solvation shell is often blurred [394]. For example, Gianturco and coworkers find two abrupt steps in the computed dissociation energy of Li$^+$He$_n$, at n = 6 and 8, although the first solvation shell is probably not filled until n = 10 [393].

C$_{60}^\pm$ and C$_{70}^\pm$ represent special cases of snowball formation because of their highly corrugated surfaces. As discussed in section 12.1, a strong (50 %) drop in the calculated dissociation energy occurs when all hollow sites are filled, at n = 32 and 37, respectively; a similarly strong drop is observed in the ion abundance [375]. Completion of the first solvation layer, which probably shows much less, if any, structural order, requires approximately 30 additional helium atoms.

The bosonic nature of $^4$He makes it challenging for theorists to establish whether or not a snowball is ordered and solid-like. The radial distribution functions do not provide a valid measure, nor do apparent localization in spatial or angular distribution functions [393, 397]. From the static density profile alone it is not possible to assess the degree of localization of the $^4$He atoms and to discriminate between solid-like and liquid-like behavior within individual solvation shells [398].

Table 4 Selected data related to helium snowballs formed around ions $A^{\pm}$. Experimental values mark the size (n) at which the abundance of $A^{\pm}He_n$ exhibits a local maximum or stepwise decrease. Theoretical values refer to sizes for which computed dissociation energies display maxima or abrupt drops, or to values obtained from the geometrical size of solvation shells inferred from radial density distributions. The latter depend slightly on the total number of He atoms in the complex. Anomalies in the ion abundances and computed dissociation energies may refer to subshell closures, or closures of higher shells. The first three columns refer to cations; columns 4 to 6 refer to related ions (e.g., dimers or anions).

| Cation | Experiment | Theory | Ion | Experiment | Theory |
|---|---|---|---|---|---|
| $H^+$ | 6, 13 [388], [399] | 6 [400] | $H_3^+$ | 10, 11 [388], 12 [399] | |
| $Li^+$ | | 8.2 [392], 10 [401], 10 [393] | $Li_2^+$ | | 6, 14, 16, 18 [378] |
| $Na^+$ | 2, 9 [376] | 16 [402], 9 [403], 12 [391], 12 [392], 11-12 [393], 12 [394], 10 [395] | $Na_2^+$ | 2, 6 [376] | 2, 6, 20 [378] |
| $K^+$ | 4 [373], 2, 12 [376] | 12 [403], 15 [391], 15 [393], 15 [394] | $K_2^+$ | | 6, 12, 17, 24 [378] |
| $Rb^+$ | 14 [373] | 19.2 [397] | | | |
| $Cs^+$ | 12, 16 [373], 17-18, 50 [374] | 17.5 [391], 21.4 [397], 18 [394] | $Cs_2^+$ | 8 [374] | |
| $Ag^+$ | 10, 12, 32, 44 [32] | | $Ag_2^+$ | 6, 10, 12 [32] | |
| $Au^+$ | 12, 14, 30 [381] | | | | |
| $Be^+$ | | 12 [404], 15 [391], 15 [394] | | | |
| $Mg^+$ | (19-20)[a] [32] | 19 [391], 20 [405], 18 [394] | $Mg^{2+}$ | 4, 8, (10-11)[a] [32] | 9 [405] |
| $Zn^+$ | 12 [32] | | | | |
| $Cd^+$ | 11 ± 1 [32] | | | | |
| $Al^+$ | 4 [406] | | | | |
| $Pb^+$ | 17 [28] | 17 [407] | $Pb^{2+}$ | 12 [28] | 12, 15 [407] |
| $F^+$ | 10.2 ± 0.6 [408] | | $F^-$ | 18.3 ± 0.9 [408] | 13.1 ± 0.1 [409] |
| $Cl^+$ | 11.6 ± 0.2 [408] | | $Cl^-$ | 19.5 ± 0.2 [408] | 17.3 ± 0.3 [409] |
| $Br^+$ | 13.5 ± 0.1 [408] | | $Br^-$ | 22.0 ± 0.2 [408] | 19.0 ± 0.2 [409] |
| $I^+$ | 16.3 ± 0.9 [408] | | $I^-$ | | 21.1 ± 0.3 [409] |
| | | | $I_2^+$ | 20.0 ± 0.1 [408] | |
| $Ne^+$ | 11, 13 [387, 389] | 7, 10, 12 [410] | $Ne_2^+$ | 12, 14 [387] | |
| $Ar^+$ | 12 [66, 283, 389], 2, 12, 32, 44 [370] | 2, 12, 14 [410], 12, 32, 44 [377] | $Ar_2^+$ | 7, 12, 14, 17, 20, 26 [370] | |
| $Kr^+$ | 10 [65], 12 [286, 389, 411], 12, 14 [370] | | $Kr_2^+$ | 12 [411], 7, 12, 18, 20 [370] | |
| | | | $Kr^{2+}$ | 12, 32 [412] | |
| $Xe^+$ | 4, 8, 15 [65], 12 [370] | | $Xe_2^+$ | 24 [370] | |
| $N_2^+, O_2^+, CO^+$ | 12 [390] | | | | |
| $H_3O^+$ | 3 [413, 414] | 3 [414] | | | |
| $CH_3I^+$ | 17.8 ± 0.2 [408] | | | | |
| $C_6H_6^+$ | | 34 [396] | | | |
| $C_{24}H_{12}^+$ | 38, 41, 44 [415] | 38, 41, 44 [416] | | | |
| $C_{60}^+$ | 32, 60 [375] | 32, 58 [375], 32, 60 [417] | $C_{60}^-$ | 32, 60 [418] | |
| $C_{70}^+$ | 37, 62 [375] | | $C_{70}^-$ | 37, 65 [418] | |

We end this section with a caveat. The size of solvation shells discussed above should not be confused with the size of snowballs as determined in ion mobility measurements [153]. Electrostriction, the attractive interaction between the ion and the polarizable helium, increases the density of the surrounding helium. Densities exceeding that of solid helium (which forms at pressures above 25 atm) are often taken as a criterion for snowball formation. Atkins performed ion mobility measurements; he estimated that most cations will drag a few dozen He atoms with them [156]; the larger the snowball the lower the ion mobility. Later mobility measurements showed large differences between nearly isobaric ions, *e.g.* $K^+$ and $Ca^+$ [419]. According to recent theoretical work $K^+$ will drag about 17 He atoms with it but $Ca^+$ only 4 [420] because the remaining valence electron in $Ca^+$ drastically weakens the ion-helium interaction; a similarly large difference, between $Cs^+$ and $Ba^+$, will be discussed below. As shown in Figure 33, the number of helium atoms in the first solvation shell of positively and negatively charged halogen ions may be used to estimate the ionic radii in helium [408]. The values increase with increasing atomic number, contrary to the computed number of He atoms in snowballs formed around halide ions in helium [420].

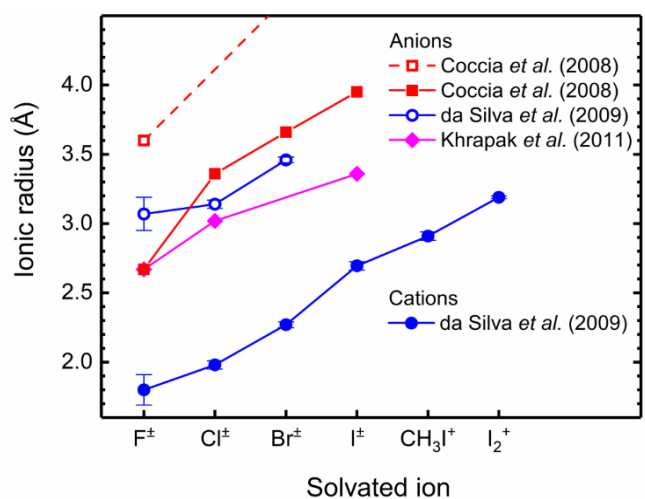

*Figure 33 Open and solid dots: Ionic radii of various anions and cations solvated in helium, deduced from anomalies in ion abundances [408]. Solid squares: ionic radii estimated from computed radial density distributions; open squares: computed bubble radii [409]. Diamonds bubble radii deduced from ion mobility measurements in superfluid helium [224]. Data taken from Ref. [408].*

## 8.2 Solvation dynamics

Experiments in the gas phase make it possible to probe the dynamics of snowball formation in HNDs. How rapidly does helium respond to the change in the force field that comes with the sudden formation of an ion? The changes are particularly dramatic for alkali atoms because in their electronic ground state, as well as most excited states, the neutrals reside in shallow dimple-like states on the surface of a HND. The states, bound by about 1 meV, result from a combination of the long-range van der Waals attraction and the strong short-range repulsion of the valence electron with the surrounding He [421]. When electronically excited from the ground state, atoms in dimple states usually desorb because of the very weak binding and exciplex formation [422]. However, Ernst and coworkers observed that Rb* and Cs* atoms excited into the $5^2P_{1/2}$ and $6^2P_{1/2}$ state, respectively, do not desorb [374]. When these systems are subsequently ionized in a pump-probe experiment, very large alkali-doped HND ions are observed.

In another study of Rb-doped HNDs, von Vangerow *et al.* observed that a second laser pulse could prevent Rb* from leaving the droplet [423]. The researchers first excited Rb with a femtosecond laser pulse to drive it into the perturbed 6p state. The interaction of Rb* with helium is strongly repulsive and the atom would normally rapidly desorb. However, upon ionization by a second femtosecond laser pulse the interaction becomes weakly attractive; the ion may turn around and become solvated. The point of no return was reached when the delay between the two lasers exceeded approximately 1 ps. In recent dynamic studies, photoinduced chemical reactions of cold, isolated $Cr_2$ molecules in HNDs have been investigated, taking advantage of the quantum state specific spatial separation of solvated states and surface states [424]. For atomic Cr atoms, ejection from the HND can be triggered by photoexcitation by populating metastable atomic states which cannot easily be reached from the ground state [425]. Measurements on electronic excitations of Cu in HNDs is in good agreement with computed spectra and also leads to subsequent ejection from the droplets [426].

In subsequent work the authors applied velocity map imaging to monitor the yield and velocity of the escaping Rb$^+$; further insight was gained from a simulation based on time-dependent density functional theory [91]. They observed that desorption of atoms excited into the 5p states was two orders of magnitude slower than for atoms in 6p states. Apparently the dynamics of atoms in excited 5p states proceeds in a transition regime between repulsive dissociation and statistical, evaporation-like desorption.

Zhang and Drabbels have studied solvation and desolvation of barium ions [427, 428]. As for alkalis, the neutral precursors reside in dimples on the surface of the HND consisting on average of 2700 He atoms. The beam of doped HNDs was intersected by two laser beams. The first laser ionized barium by single photon absorption in the UV. The second, tunable laser excited Ba$^+$ from the $6^2S_{1/2}$ ground state to the $6^2P_{1/2}$ and $6^2P_{3/2}$ states. A comparison of the measured absorption spectrum with spectra of Ba$^+$ in superfluid bulk He showed that the ions had become fully solvated within the time delay of 185 ns between the two lasers. Mass spectra revealed that the second laser depletes the number of ion-containing HNDs in favor of the appearance of bare Ba$^+$ or small Ba$^+$He$_n$ complexes. The change was attributed to non-thermal ejection of the excited Ba$^+$ because a thermal process could not possibly lead to a complete disintegration of the droplet.

The dynamics of Rb$^+$ and Cs$^+$ solvation in droplets containing 1000 atoms has been investigated by time-dependent density functional theory [397]. The calculations reveal that a snowball surrounding Rb$^+$ forms within 20 ps; rearrangement of the whole droplet is completed on the time scale of 100 ps. Analysis of the location and velocity of the ion with respect to the center of mass shows that Rb$^+$ is initially accelerated toward the center and reaches a maximum velocity of 112 m/s, which exceeds the Landau velocity [88]. The ion slows down and, after 3.3 ps, changes its direction. This oscillatory motion continues, with a period of about 3 ps, for 25 ps. Cs$^+$, too, is initially accelerated toward the center of mass. After turning around, however, it gradually moves away from the droplet. After about 90 ps it desorbs, carrying with it some 75 He atoms. Modeling the Cs$^+$ dynamics in larger droplets and on planar helium indicates that, like Rb$^+$, Cs$^+$ will become solvated in the droplet.

A similar theoretical investigation of the solvation dynamics of Ba$^+$ shows nucleation of a quantized vortex ring at the equator of a solid-like solvation structure that surrounds the ion, about 10 ps after ionization [429] (see Figure 34). The vortex ring slips around the ion and eventually detaches from it at 24 ps. The binding energy of Ba$^+$ in helium is much weaker than that of Cs$^+$ [397] and its oscillation period is much longer, about 100 ps. After the first oscillation its velocity remains below the critical Landau velocity. Consequently, the ion keeps oscillating until the end of the 350 ps simulation as translational energy can only be removed by the interaction with droplet surface oscillations and deformations.

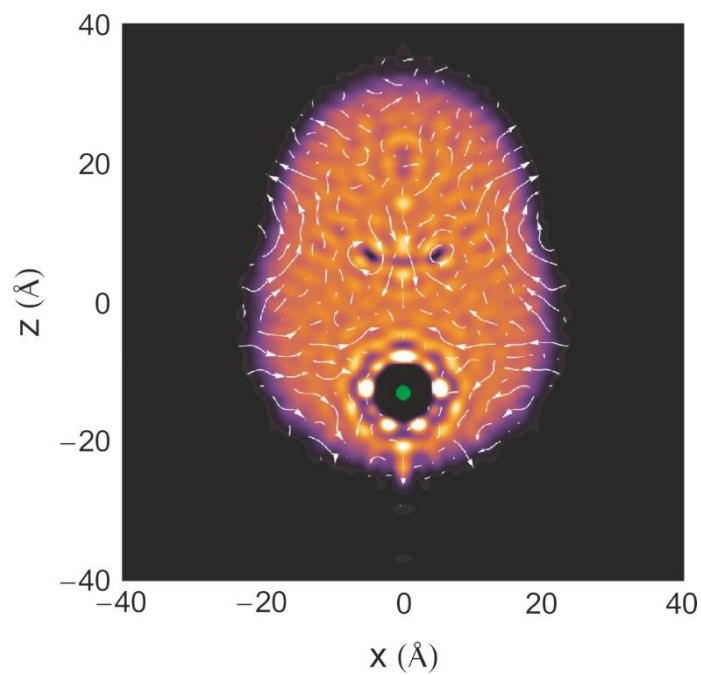

*Figure 34 Snapshot of the calculated helium density 45 ps after ionization of a Ba⁺ ion residing in a dimple state of He$_{1000}$. The helium density shows formation of a snowball. Superimposed are the circulation lines represented in white; they reveal a circular flow field around the quantized vortex ring. Reprinted from Ref. [429], with the permission of AIP Publishing.*

Another question is the time scale on which a metal cluster embedded in a HND will, upon ionization by an intense laser pulse of sub-picosecond duration, fragment into monomers and partly re-assemble as a result of caging by the helium matrix. Meiwes-Broer and coworkers have explored the dynamics of silver clusters in a non-resonant pump-probe experiment with a femtosecond laser [32]. They observed that the yield of doubly- and triply-charged clusters initially decreases with increasing pump-probe delay. The yield reaches a minimum value at about 30 ps but recovers for larger delays. The initial decrease is attributed to Coulomb explosion of the clusters into atoms or small fragments. However, the HND (containing > $10^6$ helium atoms) hinders the escape of fragments which start reassembling into a silver cluster after > 30 ps and thereby causing an increase in the yield of cluster ions.

## 9 Anions from doped HNDs

The capture of a low-energy electron by a HND, whether doped or undoped, has a particular characteristic due to the presence of an electronic surface barrier and the possibility of electron bubble formation. Hence, electron capture differs substantially from electron capture by a molecule in the gas phase. In the following discussion we will focus on electron attachment processes below the electron energy of 21 eV, *i.e.* reactions of He$^{*-}$ are excluded.

In general, the presence of the helium matrix affects the measured resonance energies of dopant anions, which therefore do not correspond to the electron energy in vacuum [251, 430]. The electron-helium interaction is dominated by short-range repulsive exchange forces because of the small atomic polarizability of helium [251]. A similar situation is encountered for other liquid-like droplets such as hydrogen and neon droplets. Considering the electronic band structure of the bulk (which is assumed to apply to droplets), the energetic difference between the bottom of the conduction band (representing the lowest delocalized state for an excess electron in the condensed medium) and the vacuum level (the energy of a free electron at rest far away from the medium), usually denoted as $V_0$, serves as an indicator of the balance between the repulsive and attractive interactions between an electron and the matrix. Interestingly, for bulk liquids $V_0$ decreases in the order He > Ne > Ar > Kr > Xe (1.06 eV > 0.67 eV > – 0.17 eV > – 0.37 eV > – 0.65 eV) while the electron mobility increases in the same order, *i.e.* these parameters turn out to be correlated [431]. Since $V_0$ > 0 for liquid helium, the surface of this medium presents an energetic barrier to an external electron, as illustrated in Figure 35.

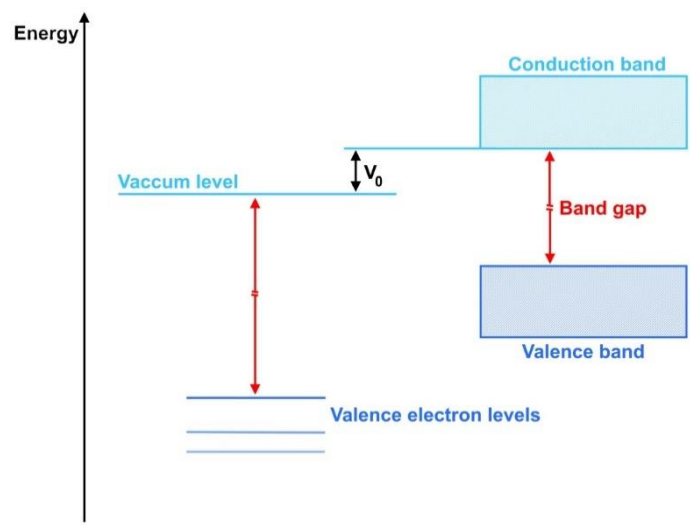

*Figure 35 Illustration of the difference in electronic energy for gaseous and bulk helium. See also Ref. [432].*

The generally accepted bulk value of $V_0$ was determined in the 1960s [433, 434], for example by studying the photoelectric effect with and without presence of the metal electrode in a liquid helium bath [434]. Later experiments were carried out studying electron impact on HNDs. Investigating the yield of electronically excited clusters as a function of the electron energy, the Toennies group derived a value of 1.03 ± 0.2 eV [206]. For smaller HNDs formed at a nozzle temperature of 14.9 K, they showed that the barrier lowers to $V_0$ = 0.68 ± 0.2 eV [206]. Later the same group obtained $V_0$ = 1.15 ± 0.025 eV, which was determined for droplet sizes from $2\times10^3$ up to $5\times10^4$ [44]. The higher barrier compared to the bulk was rather unexpected since the atom density becomes smaller when going from the bulk to smaller droplets and thus it is expected that $V_0$ should decrease with decreasing size [251]. All of these experiments show that in electron attachment studies with HNDs a blue-shift of resonances of at least $V_0$ for a molecule in the droplet should be expected compared to the vacuum situation.

When a low-energy electron enters liquid helium a fascinating dynamic starts: the electron thermalizes in a few picoseconds by creating a bubble with a radius of 17 Å (bulk value) [180] and the electron localizes in this cavity inside the liquid. The process of electron injection into liquid helium and the subsequent bubble formation was discussed in several theoretical works. Two of the most detailed articles focusing on the energetics and dynamics of excess electrons in HNDs were published by Rosenblit and Jortner in 2006 [179, 435]. They predicted that the electron bubble is metastable, since this electronic state will decay by electron tunneling back into vacuum after an oscillating motion of the electron bubble in the droplet. Hence, the survival probability of the electron bubble in the droplet was predicted to be lower for superfluid $^4$He compared to normal fluid $^3$He (the latter providing practically infinite lifetimes) [435]. Furthermore, Rosenblit and Jortner predicted a minimum droplet size for the formation of electron bubbles in HNDs ($N \approx 4500 \pm 700$). The latter size was obtained by considering the energetic stability (corresponding to the size dependent criterion: total electron bubble energy at equilibrium configuration < $V_0$) and dynamic stability (tunneling effects) [435].

From an experimental point of view, the metastability of electron bubbles implies challenges for the mass spectrometric observation of negatively-charged HNDs. Large droplets (with $N > 7.5 \times 10^4$) are necessary to observe pure, negatively-charged HNDs on mass spectrometric timescales (about µs) [44, 204, 207]. A first measurement of the anion yield as a function of the electron energy was carried out by Jiang *et al.* [207]. They observed a broad peak centered at 1.2 eV, but due to the broad energy distribution of the electron beam, the peak onset (which represents the minimum energy required to enter the HND, *i.e.* $V_0$) was found at almost zero eV energy. This result indicates that a high energy resolution is required to precisely determine $V_0$. Later, Henne and Toennies measured the electron energy dependence of $He_N^-$ ($N > 9.3 \times 10^4$) droplets [44]. For the smallest droplet size of $9.3 \times 10^4$, the anion efficiency curve rose at the threshold of 1.26 eV and reached a maximum at 1.8 eV. This maximum is gradually shifted with increasing droplet size up to 2.3 eV for the largest anion studied ($N = 1.5 \times 10^7$). In contrast, for the peak onset they reported only a small upward shift of 60 meV between the smallest and the largest droplet size measured [44]. It should to be noted that the onsets reported by Henne and Toennies are substantially above $V_0$. Accordingly, the authors claimed that additional activation energy is required for localization of the electron in the droplet in a bubble [44].

When we move from pure droplets to doped droplets, the energetics and dynamics of bubble formation will not significantly change; at least not for systems where the dopant is deeply buried at the center of the droplet and surrounded by many shells of helium. If this condition is not fulfilled, *i.e.*, dopants are present at the surface of the droplet, or embedded in a very small droplet, the incoming electron may also directly interact with dopants via long range polarization or dipole interactions. In this case the electron may quickly localize on the dopant prior to bubble formation. The dopant anion will be thermodynamically stable if the molecule or its fragments possess a positive electron affinity. This may lead to thresholds of resonances also observed below $V_0$, indicating that no conduction band is developed. Very recently the formation of the conduction band in HNDs was studied as a function of the droplet size by studying high resolution electron attachment to the water dimer [436]. The measured ion yield of the water dimer showed that the surface barrier does not continuously decrease with smaller droplet size. Instead, a lowest value of $V_0 = 0.76 \pm 0.10$ eV was obtained, corresponding to a droplet size of 1600 ± 900. Below that size, the well-known threshold peak at about 0 eV of neat $H_2O$ dimers appeared.

### 9.1 Electron attachment to the nucelobases thymine and adenine

The first mass spectrometric study of anions formed in doped HNDs was published in 2006 and was devoted to the biologically relevant nucleobases adenine and thymine [214]. These biomolecules are some of the building blocks of DNA and RNA. The interest in the stability of these species upon electron capture arose from the discovery that low-energy electrons cause DNA single and double strand breaks by resonant electron attachment [437, 438]. Low energy secondary electrons are formed abundantly upon the interaction of energetic radiation (α, β, γ, ions, *etc*.) with biological matter. Subsequently, numerous electron attachment studies with single building blocks of DNA and proteins have been carried out [439]. These mass spectrometric studies in the gas phase reported resonance energies, as well as fragmentations patterns, for biomolecular compounds upon electron capture. It turns out that isolated biomolecular anions formed by free electron attachment are particularly unstable towards dissociation and spontaneous electron emission. The drawback of these studies was the single collision conditions, which of course limits any prediction for a complex biomolecular target like, for example, hydrated DNA [440, 441].

Experiments with nucleobases embedded in HNDs allows the study of weak solvent effects on the electron attachment process to biomolecules [214, 442]. Both thymine (T) and adenine (A) have been looked at in these environments and compared with behavior in the gas phase. A remarkable change of the anion abundances was observed for nucleobases in HNDs. When electron attachment to a nucleobase takes place in HNDs, the molecular anion becomes observable on mass spectrometric timescales [214]. Such stabilization can be only observed because of the transfer of sufficient excess energy deposited in the temporary negative ion of the nucleobase (basically the kinetic energy of the incoming electron and the electron affinity of the anion formed) into the surrounding helium.

As mentioned above, for localization of the electron bubble at the dopant the electron affinity of the dopant must be positive. Nucleobases have a high dipole moment (*e.g.* the dipole moment of thymine is 4.34 D [443]), which allows the formation of a dipole bound anion with a positive electron affinity [444, 445]. Furthermore, it was also recently suggested that nucleobases can capture an electron in a weakly bound valence state which, in contrast to the neutral molecule and dipole bound anion state, has a non-planar geometry [446]. Previously, photoelectron spectroscopy of nucleobase anions when bare or solvated by another molecule (for example by one water molecule), showed a remarkable transformation from a dipole bound anion in isolation to a valence bound anion in solvation [447, 448]. In HNDs the anion efficiency curves (*i.e.* ion yield as function of the electron energy) at energies < 10 eV for the parent (cluster) anions showed a main resonance at low electron energies. For example, the dimer anion of adenine formed in helium is shown in Figure 37. While at the vacuum level a zero eV resonance can be expected, the peak for the nucleobase anions stabilized in HNDs was shifted by about 1.6 eV for the reasons mentioned above. However, one should note that the measurement was carried out with a standard Nier type ion source which lacks precision at electron energies close to about zero eV, because the high electron current provided by these standard ion sources (typically in the μA range) will lead to space charge effects in the electron beam.

Isolated nucleobases (NB) form transient intermediates [NB⁻]* that may undergo autodetachment or dissociate into dehydrogenated NB [NB-H] according to equations (12) and (13)

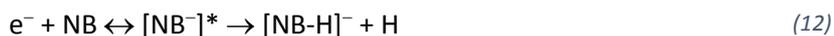

$$e^- + NB \leftrightarrow [NB^-]^* \rightarrow [NB\text{-}H]^- + H \qquad (12)$$

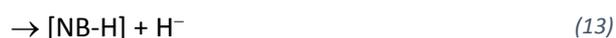

$$\rightarrow [NB\text{-}H] + H^- \qquad (13)$$

Channel (12) has been observed to dominate for all DNA bases at electron energies around 1 eV while resonances for the formation of H⁻ and other product anions attributed to the further dissociation of [NB-H]⁻* are observed around 5 and 7 eV with much lower cross sections (see Figure 36). When embedded in superfluid HNDs thymine and adenine respond quite differently to low-energy electrons. Figure 36 shows the results observed, along with those from gas phase measurements. Only [NB-H]⁻ anion formation is observed in the droplet experiments and shows two resonances, a low-energy one at 1.2 eV (also observed for the molecules in the gas phase) and a strong and broad resonance that extends to electron energies as high as 15 eV. The strong broad resonance far exceeds the corresponding resonance observed for isolated molecules and was attributed to the *quenching* of the dissociation of [NB-H]⁻* that occurs in the gas phase.

A more detailed look into the dissociation mechanism follows. Above the electron energy of about 4 eV, where electronic excitation of the target accompanies the capture process, ring cleavage reactions are highly efficient. The total anion yields for gas phase adenine and thymine plotted in Refs. [449] and [450] indicate that these core excited resonances are only about one order of magnitude weaker than the ion yield below about 4 eV. In the case of thymine, for example, the most abundant fragment anion is NCO⁻ [451]. The latter anion possesses a closed electronic shell and therefore NCO has a very high electron affinity (~ 3.6 eV), which is in the range of halogen atoms. As with any other fragment anion involving ring cleavage processes, the NCO⁻ formation becomes completely quenched in HNDs. Instead, the resonances leading to ring cleavage in the isolated molecule show up in the dehydrogenated anion in HNDs, as shown in the upper panel of Figure 36 and Figure 37. The channeling into (NB-H)⁻ can be explained by the different dissociation mechanism for H-loss and ring cleavage: the former occurs directly upon dissociation along a purely repulsive potential energy surface of the electronically excited state of the transient negative ion. After that an isolated vibrationally hot (NB-H)⁻, now in the electronic ground state, further decays with cleavage of multiple bonds and considerable rearrangement of the molecule. The exact reaction pathways of this ring cleavage reaction were recently determined in [451]. NCO⁻ formation is a time-consuming process, which was also observed by the metastable decay of the (NB-H)⁻ on the microsecond time scales [451]. Such slow decay can be effectively quenched by the helium environment due to fast transfer of excess energy into the cold environment.

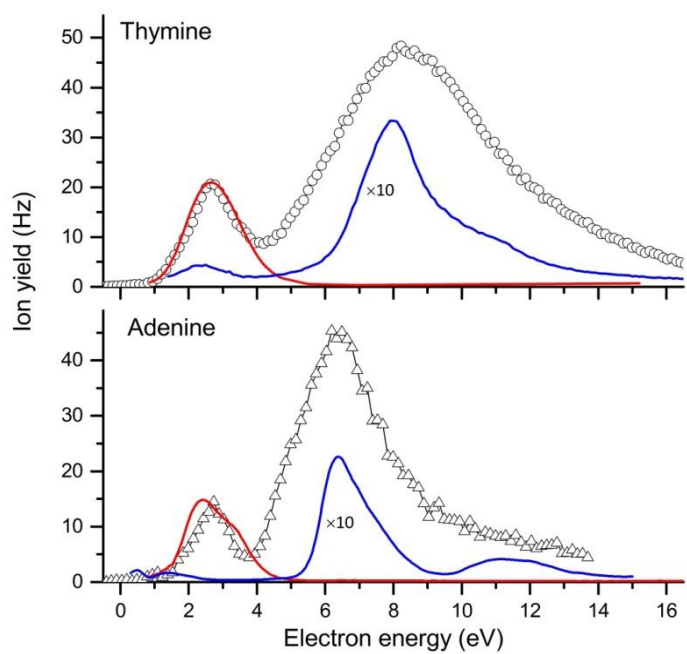

*Figure 36 [NB-H]⁻ anion yields measured as a function of electron energy upon dissociative electron attachment to the nucleobases (NBs) embedded in He droplets at 0.37 K (circles and triangles). The corresponding results obtained for the isolated NBs at 400 K are displayed for production of [NB-H]⁻ anions (solid red lines) and the sum of all other product anions (solid blue lines). The measured electron energies for the droplet experiments have been lowered by 1.5 eV to account for the energy required for the electrons to penetrate the droplets. Data taken from Ref. [442].*

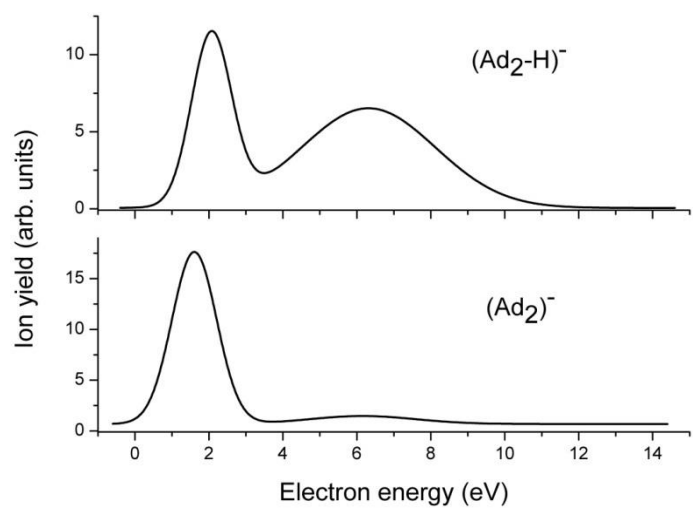

*Figure 37 Ion yield as function of the electron energy for the dimer anion (lower panel) and the dehydrogenated dimer anion of adenine (Ad) (upper panel) formed in HNDs. See also Ref. [214].*

This first study with nucleobase anions formed in HNDs revealed one further interesting aspect: dopant anions leave the droplet and can be observed as bare anions. The anions formed, for example by associative electron attachment, have low initial kinetic energies and hence this observation may *a priori* be surprising for closed-shell anions. However, the recent experiments on He*⁻ [176] showed clearly that HNDs are prone to capturing more electrons and hence anions formed from a dopant may face the Coulombic repulsion of another negatively charged species present in the same droplet and are therefore ejected from the droplet.

## 9.2 Electron attachment to TNT and nitromethane

Electron attachment to the explosive trinitrotoluene exhibits striking differences in anion formation between the gas phase and He droplets [452]. While the total anion yield for electron energies between 0 and 12 eV shows similar resonance features for both, the anion population that is observed mass-spectrometrically is completely different (Figure 38). Dissociative attachment dominates in the gas phase but is completely absent in the droplet. The complete suppression of dissociation for TNT@He is even unique among other molecules investigated in this way. Clearly the excess energy that accompanies the initial attachment is dissipated in the droplet by collisions with He (and neighboring TNT molecules) before energy redisposition within the transient anion can cause bond dissociation. On the other hand, in the gas phase where collisional quenching is absent, extensive bond breaking and bond making occurs and a large variety of dissociative product anions are produced. These reactions can be observed up to several microseconds after the interaction with the electron [453, 454]. Especially dramatic in the droplet environment is the absence even of C-N cleavage that leads to $NO_2^-$ in the gas phase. The cold droplet has completely frozen any dissociative processes in $TNT^-$.

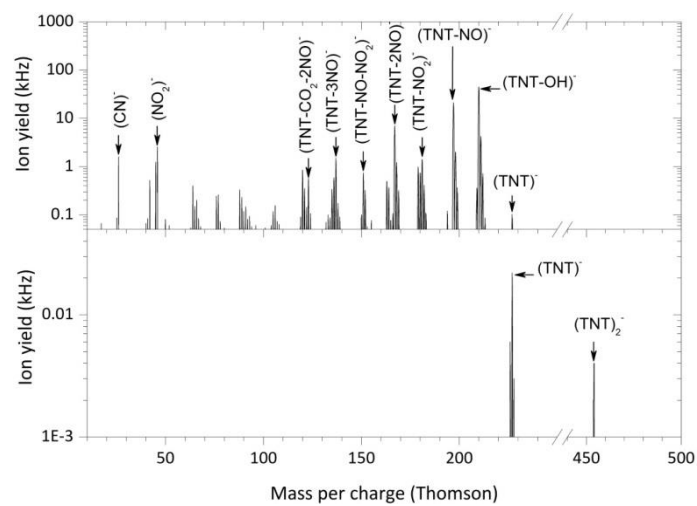

*Figure 38 Negative ion mass spectra from gas phase TNT recorded at an electron energy close to 0 eV (upper panel) and TNT embedded in He droplets recorded at 2 eV electron energy (lower panel). Reproduced from Ref. [452] with permission from the PCCP Owner Societies.*

In comparison, dissociation of the C-N bond in $CH_3NO_2^-$ to form $NO_2^-$ (along with small amounts of the clusters $NO_2^-$ $(CH_3NO_2)_n$) was observed to occur in low energy (2 eV) electron attachment experiments with He droplets doped with nitromethane ($CH_3NO_2$). The parent anion and anion multimers $(CH_3NO_2)_n^-$ were also observed [252, 455]. No parent anion was detected in the gas-phase electron attachment experiments; $NO_2^-$ was the dominant fragment anion that was observed, along with 15 additional fragment anions, in the electron energy range 0 - 16 eV. So, while freezing occurs, it is incomplete for the dissociation of $CH_3NO_2^-$ in He droplets, perhaps due to the smaller size and number of degrees of freedom available for energy redistribution.

## 9.3 Anions formed upon electron attachment to chloroform in He droplets

Dissociative electron attachment appears to be even more prevalent with chloroform, an even smaller molecule, embedded in He droplets. Chloroform ($CHCl_3$) belongs to the class of halogenated hydrocarbons that are of relevance in atmospheric chemistry and also find technical applications in plasma etching and gas phase dielectrics [216]. For these fields, knowledge on anion formation by electron attachment is important because halogenated hydrocarbons have high capture cross sections for low energy electrons [456, 457]. This can be ascribed to the high electron affinity of the halogen atoms (EA(X) = 3.40, 3.61, 3.36, and 3.06 eV for X = F, Cl, Br, and I, respectively) [458]. In contrast, studies with simple hydrocarbons are rather scarce, because their attachment cross sections are quite low [459, 460]. This can be attributed to the fact that for hydrocarbons no resonance close to zero eV electron energy is available, which could drive efficient production of a stable negative anion or efficient DEA via formation of a temporary negative ion. Instead fragment anions of hydrocarbons turn out to be formed upon core-excitation with resonances close to 6-10 eV [459, 460]. Hydrocarbon anions are more efficiently formed via ion pair formation, where the production of a negatively charged fragment and its positively charged counterpart can be observed. However, the threshold for ion pair formation is close to the ionization threshold of the fragment, *i.e.* it can be found only above 10 eV [460].

The situation turns significantly upon halogenation of hydrocarbons. Such halogenated compounds always dissociate upon electron attachment (most efficiently below 1 eV), with the excess charge localized at the halogen atom. Previously, a systematic study with halogenated methane was carried out by Chu and Burrow [461]. They observed that for $CH_{4-x}Cl_x$ (x = 1–4) the relative cross sections at low electron energy vary by six orders of magnitude. The highest electron attachment cross section was observed for CCl$_4$, where Cl$^-$ is formed in an exothermic dissociation reaction initiated by s-wave electron capture (*i.e.* without a centrifugal barrier) with a $\frac{1}{\sqrt{E}}$ dependence [462]. Furthermore, Cl$^-$ is formed in a p-wave resonance close to 0.94 eV [456]. Electron attachment to the chloroform molecule, CHCl$_3$, leads also to Cl$^-$ as by far the most abundant anion [463, 464] (in addition, nine other fragment anions have been observed [464]). In contrast to the CCl$_4$ case, some activation energy is required to drive the dissociation of the molecule in the vibrational ground state by electron capture. However, for molecules in a vibrationally excited state the reaction becomes exothermic, with s-wave capture close to zero eV electron energy [463].

Due to these interesting properties, chloroform was chosen as a model system for electron attachment studies in HNDs [216]. The results indeed demonstrated the strong influence of the cold He matrix on the electron attachment process and subsequent dissociation. As for aromatic molecules discussed above, a stable parent anion (as well as stable parent clusters anions) can be observed, indicating that excess energy can be efficiently dissipated in the helium matrix. No parent anions were seen in electron attachment experiments with isolated molecules of CHCl$_3$ in the gas phase, where quenching is absent [464]. Ten fragment anions were observed instead. The smaller number of degrees of freedom offered by CHCl$_3$ compared to CH$_3$NO$_2$ and TNT lead to the least amount of collisional quenching and the strongest dissociative electron attachment process in He droplets. In this case the fragment anion series (CHCl$_3$)$_n$Cl$^-$ was observed in high abundance, which is likely a direct effect of the high electron affinity of the chlorine atom. In this case the excess electron occupies the anti-bonding $\sigma^*$ (C-Cl) orbital, which leads to dissociation in a fast electronic decay. This fragment anion, as well as the parent cluster anions, showed a resonance at the same energy of about 1.8 eV, *i.e.* the shift relative to the vacuum level due to the electronic surface potential of HNDs. Moreover, the same resonance energy for fragment and parent anion indicates that the same electronic state may be involved in the formation of these anions. This state is purely repulsive close to the equilibrium distance of the neutral, but it possesses a weak minimum at larger equilibrium distances [465]. In this case the thickness of the surrounding helium layer decides if fragmentation can be quenched or not. Therefore, the parent cluster anions may be formed from larger droplets of the HND size distribution and fragment clusters anions from the smaller ones. In the ion yields of (CHCl$_3$)$_n$Cl$^-$ (the case for n = 0, *i.e.* Cl$^-$, is shown in Figure 39), which are also formed in the region of core excited resonances close to 10 eV, one feature is noteworthy: the low-lying resonance and the core-excited resonances have yields of the same order of magnitude.

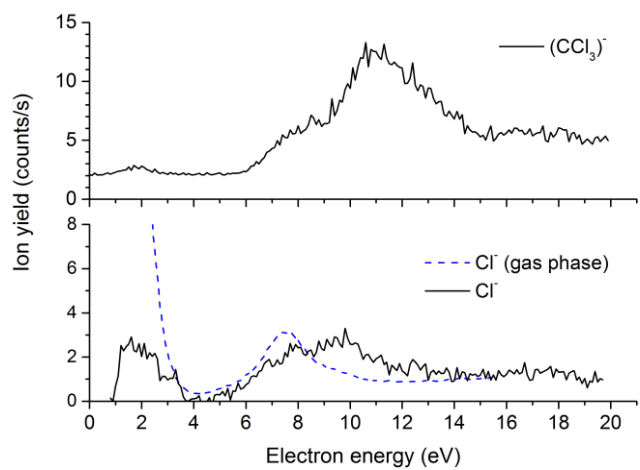

*Figure 39 Ion yield for $CCl_3^-$ (upper panel) and $Cl^-$ (lower panel, solid line) formed upon electron attachment to $CHCl_3$ in HNDs. The dashed line in the lower panel shows the $Cl^-$ ion yield for electron attachment to a single chloroform molecule in the gas phase. Data taken from Ref. [216].*

The same observation is made for the dehydrogentated parent anion (see upper panel in Figure 39), which is not observed in the gas phase [464]. In HNDs the anion yield of core-excited resonances exceeds the low-energy features. Therefore, the high ion yields ascribed to s-wave reaction under vacuum conditions are obviously quenched by embedding the molecule in a HND. This effect may be ascribed to the non-zero ground-state energy of the electron in the bubble (see discussion by Rosenblit and Jortner in Ref. [179]), and hence electron attachment by electrons with kinetic energies approaching 0 eV is not possible in the droplet. Such an effect would account for all systems exhibiting an s-wave electron attachment reaction if the localization of the dopant (cluster) occurs inside the droplet. Indeed, such a result was observed for the fullerene $C_{60}$ monomer and dimer anion formed in HNDs [466]. These studies with cold $C_{60}$ (and co-doped with water) have revealed other aspects of anion formation by electron attachment, namely the influence of internal energy of the neutral precursor and the competition of autodetachment with dissociation [466]. The anion yield for the bare $C_{60}$ molecule is remarkable, because it is characterized by a sharp s-wave peak close to zero eV [467] (it should be mentioned that there was a disagreement in the literature for several years over whether this resonance has s-wave or p-wave character), and weaker ion yield with remarkably broad maxima extending up to about 11 eV.

For fullerenes in HNDs, anion-yield measurements at electron energies from 0 up to almost 40 eV show that $C_{60}^-$ anions persist to much higher electron energies than for isolated $C_{60}$ molecules in the gas phase, *i.e.* to more than 40 eV for the former compared to less than 20 eV for the latter. This difference has been attributed to a combination of the much lower temperature of 0.37 K for $C_{60}$ in the He droplets and the suppression of electron autodetachment (which is thermally activated [468]) by collisional quenching. $(C_{60})_2^-$ and $C_{60}D_2O^-$ production was also monitored. The onset of a bump in their electron efficiency curves around 22 eV can be attributed to the onset of He*$^-$ formation, and the ensuing depletion, more rapid than that observed for $C_{60}^-$, is thought to arise from the lower activation energy for the dissociation of $(C_{60})_2^-$ and $C_{60}D_2O^-$ than that for autodetachment of the electron.

## 9.4 Dissociative electron attachment to acetic and amino acids

Interest in DNA strand breaking by low-energy electrons has also motivated electron attachment experiments with amino acids embedded in cold He droplets. Glycine, alanine and serine have been investigated [469], as well as acetic acid [470]. In each case low-energy electrons have been seen to attach and to induce bond dissociation to a greater or lesser extent.

The electron-attachment chemistry at around 2 eV of glycine, alanine and acetic acid is similar in that parent cluster anions $AA_n^-$ (where AA = amino or acetic acid monomer) dominate as products with some occurrence of bond dissociation to form $[AA_n-H]^-$ and, with the amino acids, also $[AA_n-OH]^-$ anions. The yield of parent monomer anions was lower by more than an order of magnitude and appeared to be mainly the dehydrogenated anion $[AA-H]^-$, at least in the case of acetic acid.

Anion efficiency curves at electron energies up to 40 eV also were reported. These exhibit a number of resonances for non-dissociative and dissociative attachment at similar positions but different relative intensities.

The anion chemistry of serine (S) was observed to be different. OH loss is seen to be the major product channel; the signature OH group at the β carbon of serine is a facile leaving group not available to the other amino acids. $[S_n-OH]^-$ is seen to be the major product anion for the trimer and larger clusters, although H loss and no loss anions are still present in minor amounts. OH loss is a minor channel for the dimer and absent for the monomer. The predominance of OH loss for the trimer and larger clusters has been attributed to increasing zwitterion behavior that can disrupt intra-cluster hydrogen bonding with the OH of the $CH_2OH$ group in serine.

## 9.5 Anions formed upon electron attachment to water clusters in He droplets

The temperature of clusters as well as the presence of a surrounding matrix (caging or cooling) affects the formation of anions upon electron attachment. In the gas phase the cluster temperature differs from experiment to experiment, depending on the experimental conditions (seeded vs. non-seeded expansion). Water clusters have always been a species of high interest related to electron attachment [215, 471]. For example, in radiation biology the hydrated electron is the final stage of secondary electrons formed by the action of ionizing radiation on cell tissue [471]. From a radiobiological point of view, this species is thought to be chemically inactive, because it is bound to the cellular water environment [439]. However, before entering this stage, the electron is in a prehydrated state, when DNA damage was suggested to occur [472]. Water cluster ions with an excess electron have been suggested as model systems for studying the properties and dynamics of bulk water [215]. Photoelectron spectroscopy has provided valuable information on the location of an excess electron in water clusters with finite size (n < 200) [473]. Different localization places for excess electrons were suggested: external surface states as well as internal states [473]. Coe *et al.* suggested that those internalizing states act as embryos of bulk hydrated electrons [474]. With respect to the surface state it has been known for more than 20 years that the water dimer anion is formed in a dipole-bound state [475]. Only later, based on more recent photoelectron spectroscopic studies, was it suggested that dipole bound states may extend to least n = 16 [474]. However, at these cluster sizes intermediate states with higher vertical binding energy also appear, which represent a step towards internalization. In contrast, electron attachment to the single water molecule leads to dissociation or autodetachment, because the electron affinity of water is negative and therefore the transient negative ion is short- lived [476].

The first studies of pure water clusters in the gas phase were performed by Haberland *et al.*, who injected electrons into a beam of water molecules [477]. With this technique, water molecules agglomerate around the electron, which acts as a condensation nucleus. They managed to observe $(H_2O)_n^-$ cluster anions for n ≥ 11. Later, Recknagel and co-workers [478] used the well-known supersonic expansion technique to form neutral water clusters and attached free electrons in a crossed electron-cluster beam experiment. In agreement with the results by Haberland *et al.*, they also observed water cluster anions at a minimum size of 11. This threshold size was ascribed to the fact that the adiabatic electron affinity of small clusters does not exceed the dissociation energy of the cluster, *i.e.* autodetachment is favored over evaporation of single water molecules [215].

In gas phase experiments, water cluster anions smaller than n = 11 can only be observed if the neutral water clusters formed in the expansion are sufficiently cold [479]. Intact water cluster anions were observed by electron attachment of electrons close to zero eV [479]. Furthermore, fragment anions formed by loss of H and 2H can be observed in the electron energy range between 5 and 10 eV [478]. In this energy range the water molecule has three core-excited resonance states (at 7.0, 9.2. and 11.8 eV, respectively, which are of Feshbach type) [480]. The formation of $H^-$ and $O^-$ fragment anions via these states was predicted by theory [476]. However, in crossed beam experiments of electrons and neutral water molecules also $OH^-$ was observed, which was then ascribed to the secondary ion-molecule reaction $H^- + H_2O \rightarrow OH^- + H_2$ [481] or more likely to electron attachment to a small fraction of molecular dimers in the neutral beam [482]. In contrast, the most recent electron attachment study with an effusive beam of $H_2O$ in the gas phase concluded that $OH^-$ can indeed also form from single water molecules [476]. For (heavy) water embedded in HNDs, the abundance of fragment anions from water is rather surprising: the fragment anion $(D_2O)_{n-1}O^-$ was strongly enhanced relative to the fragment anion $(D_2O)_{n-1}OD^-$, which is a trend opposite to previous predictions. It is well known that frictional and caging effects can occur in a cluster matrix like helium which leads to quenching of fragmentation [215]; however, how these effects could lead to an increase of $(H_2O)_{n-1}O^-$ is still not known.

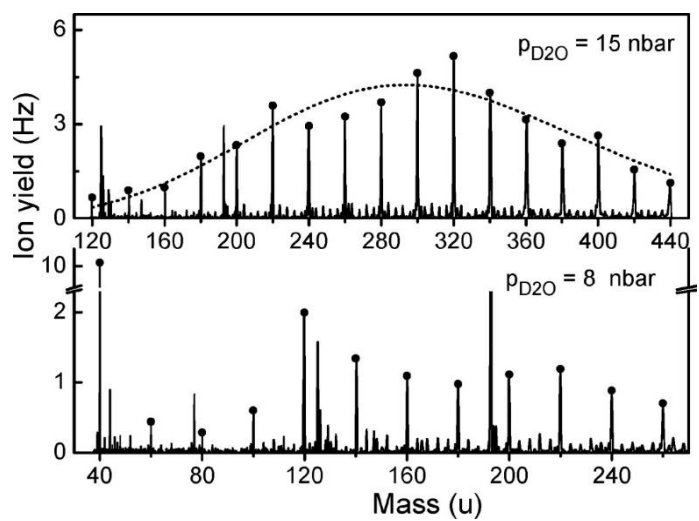

*Figure 40 Mass spectra of water cluster anions formed by attachment of a low-energy electron (energy about 2 eV) to $(D_2O)_n$ clusters embedded in HNDs; grown with high (upper panel) and low (lower panel) water-vapor pressures, respectively. $(D_2O)_n^-$ are marked by dots. The dashed line in the upper panel shows a fitted Poisson distribution with a size dependent coagulation coefficient. Reprinted with permission from Ref. [215]. Copyright 2008 American Chemical Society.*

When water cluster anions are formed in HNDs, a remarkable observation is the rather smooth abundance of intact water cluster anions for n ≥ 6 [215], which is illustrated in Figure 40. In previous studies with pure water clusters a pronounced size dependence was observed for small water cluster anions, where (i) the dimer was abundant (also true in HNDs; see Figure 40), (ii) sizes with n = 3, 4, 5, 8, 9 and 10 were just above the detection limit, while (iii) n = 6, 7, 11 and n ≥ 15 showed similar abundances and, (iv) n = 12, 13, 14 showed significantly lower intensities compared to n=11 and 15 [475, 479, 483, 484]. Such distinct differences between bare water clusters and water clusters formed within helium was tentatively explained by the formation of linear metastable water cluster structures [215]. The formation process of the dopant cluster in HNDs can be understood as sequential monomer addition to the growing clusters. The surrounding cold helium thereby allows the freezing of metastable structures (see section 3.2 and 11.2.1). It was suggested by Zappa *et al.*, that the rather smooth abundance distribution of water cluster anions formed in HNDs for n ≥ 6 may be ascribed to growing chains of water molecules [215]. A minor fraction of clusters grown in more traditional structures is responsible for slight variations within the smooth abundance. However, it should be noted that this explanation is still tentative and awaits verification [215].

## 9.6 High resolution electron attachment to carbon dioxide clusters in HNDs

In mass spectrometric studies of dopant anion formation in HNDs it turns out that the resulting anion yields are low compared to gas phase studies. This tendency makes utilization of monochromatic electron beams extremely difficult, since high resolution is, in principle, in conflict with the high sensitivity and long-time stability which is required for the measurement of anion signals in the Hz or sub-Hz regime. However, the first successful study utilizing an electron monochromator was reported in 2014 [313]. The experimental setup, shown in Figure 41, was a crossed electron/droplet beam experiment consisting of a HND source with pick-up stage, a home built hemispherical electron monochromator and a quadrupole mass spectrometer for detection of the anions.

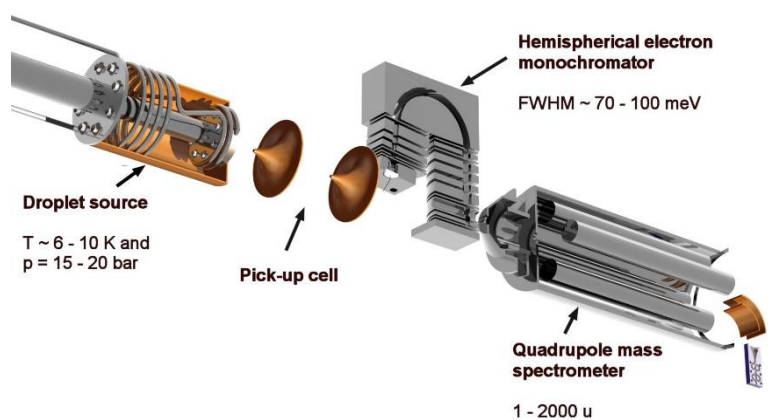

*Figure 41 Illustration of the experimental set-up utilized for high resolution electron attachment studies with doped HNDs. From left to right: HND source, the pick-up stage, the hemispherical electron monochromator and the quadrupole mass spectrometer.*

Utilizing this setup, electron attachment to carbon dioxide embedded in HNDs, was studied by Postler *et al.* [313]. The electron beam resolution was between 90–120 meV. The interest in anion formation of $CO_2$ has a long history, because $CO_2$ is an important molecule in atmospheric chemistry and astrochemistry [485]. Its role in electron attachment is also significant because the linear $CO_2$ molecule becomes bent when an excess electron is added: the bond angle decreases from 180° to 138° [485]. Compton *et al.* observed a negative adiabatic electron affinity for the $CO_2$ molecule (-0.6 eV) [486]. However, photodetachment studies reported positive vertical binding energies [487], in contradiction to [486] and results from quantum chemical calculations [488]. The current knowledge concerning this issue is that the adiabatic electron affinity is indeed negative, but a metastable parent anion is possible due to an energy barrier (of about 0.3 eV) between the electronic ground state of the neutral and the anion having a considerably larger equilibrium CO bond length [485]. This metastable state of the parent anion cannot be stabilized in the isolated system. The situation changes for clusters, where $(CO_2)_n^-$ cluster anions with n ≥ 1 were observed below electron energies of 4 eV [489]. Studies with high resolution revealed remarkably sharp structures in the ion yield of $(CO_2)_n^-$, which were assigned to a virtual state at zero eV (related to the infrared inactive symmetric stretching mode at threshold) and vibrational Feshbach resonances up to a few hundred meV [489]. These features involved symmetric stretch and bending vibrations of the molecule. Furthermore, the anion yield of the $CO_2$ tetramer anion revealed a broad unstructured peak from about 1 up to 4 eV [489]. The analysis by multiple Gaussian peak fits in Ref. [313] allowed the identification of three resonances with maxima at 1.4, 2.2, and 3.1 eV (see upper panel of Figure 42). The latter resonance was ascribed to the well-known $^2\Pi_u$ resonance [490]. Molecular dissociation of the anion at the upper tail of this resonance is energetically open and leads to formation of $O^-$ and $(CO_2)_nO^-$ homologues [489]. The peak at 2.2 eV was ascribed to excitation of the symmetric stretching vibration (165.8 meV) and two quanta of the bending vibration (each 82.7 meV). These nearly degenerate vibrational states mix and form two perturbed vibrational states (= a Fermi dyad) at 159 and 172 meV. The excitation of the higher member corresponds to the broad peak at 2.2 eV (indicative of a σ* resonance) [489].

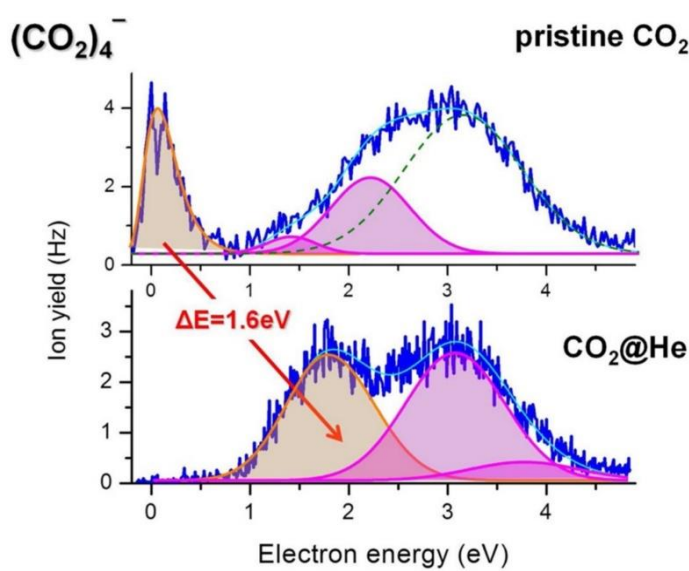

Figure 42 Comparison of the $(CO_2)_4^-$ ion yield for electron attachment to pristine $CO_2$ clusters (upper panel) and $CO_2$ clusters embedded in HNDs (lower panel). The ion yields were determined at the electron energy resolution of about 100 meV. Reproduced from Ref. [313] licensed under CC BY 4.0.

The ion yield below 4 eV changes significantly upon attachment of the electron to $CO_2$ embedded in HNDs, as shown in Figure 42. As expected from the discussion above, all features below 1 eV disappear and are shifted by about 1.6 eV. The sharp vibrational features observed for the clusters formed in the gas phase are not prevalent in HNDs because the peak is unstructured within the measurement statistics. The second resonance observed at ~ 1.4 eV for bare clusters is found at ~ 3.1 eV for HNDs. Much weaker signal was also observed at 3.8 eV, which may represent the resonance at 2.2 eV for bare clusters; again, shifted by 1.6 eV. Surprisingly the $^2\Pi_u$ resonance found at 3.1 eV for the bare clusters, seems to be substantially quenched because no obvious resonance at about 4.7 eV (assuming the expected shift of 1.6 eV) is present in the $(CO_2)_4^-$ ion yield.

This study shows that for systems exhibiting several resonant features in a comparably small energy range (which is often the case in the sub-excitation regime, *i.e.* at electron energies below electronic excitation of the compound), the utilization of a high resolution electron beam is crucial. One should note that this tentative interpretation of the $(CO_2)_4^-$ data for HNDs demonstrates again that the relative abundance of the inelastic scattering features may change significantly in HNDs. Most pronounced seems to be the strong suppression of the $^2\Pi_u$ resonance, which is a rather unexpected result. However, also in the ion yield of $(CO_2)_nO^-$, where the $^2\Pi_u$ resonance abundantly contributes to a peak close to 4 eV for bare clusters, the contribution is substantially reduced in HNDs [313]. For these fragment cluster anions core excited resonances close to 13 eV dominate the ion yield.

# 10 Chemical reactions

HNDs, or 'flying nano-cryo-reactors' as they have been called [491], provide an extraordinarily unique medium for the study of many facets of chemistry. They provide an extremely cold superfluid bath of He atoms in which atoms and molecules, neutral or charged, as well as electrons, are free to move and to encounter each other and to undergo electron attachment, electron transfer or chemical bond redisposition as they interact with each other. Experiments with HNDs have demonstrated the occurrence of associative and dissociative electron attachment, bimolecular neutral reactions between selected neutral dopants and positive and negative ion-molecule/cluster reactions when doped or multiply doped droplets are exposed to low or high energy electrons.

The energies of reactants embedded in HNDs are extremely well defined. At 0.37 K, the reactants cool into their lowest quantum states and any electronically excited products will cool toward their lowest rovibrational states.

Of course the low temperature prevents the occurrence of exothermic chemical reactions that have even very small energy barriers. On the other hand, the helium matrix can stabilize reaction intermediates in local minima of potential energy surfaces when energy barriers are present. Collisional stabilization also can freeze bond cleavage of intermediate excited anions by dissociative electron attachment.

Varying the ratio of the neutral dopants or their relative beam positions of pickup, or delaying product detection after pickup, can be used to control and assess the extent of reaction. The size of the droplet (the number of helium atoms in the droplet) can be enhanced and this, together with the low temperature of the droplets, promotes the formation of neutral clusters within the droplet. $He^+$ ions can be produced by electron ionization at the surface of the droplet (after the electrons have spent about 1.5 eV to enter the droplet) and then, after resonant charge hopping from the surface, can cause ionization of embedded molecules or molecular cluster by electron transfer to $He^+$, and so initiate single or intra-cluster ion-molecule reactions. Electrons may also attach to metastable He* that can be produced by electron excitation and so form metastable $He^{*-}$ which has recently been found to be able to donate one or two electrons to the dopant to form mono or dianions [254] and indirectly singly or doubly-charged cations.

## 10.1 Neutral reactions

When atoms and molecules are added sequentially into a superfluid He droplet (> $10^3$ atoms per droplet) that is allowed to cool by evaporation of He atoms, any chemical reactions that may occur between them will proceed at a uniquely low temperature of 0.37 K. The depletion of the reactants and/or formation of products can be followed by electron-impact or femtosecond photoionization mass spectrometry or by chemiluminescent spectroscopy or laser spectroscopy (beam depletion).

The classic example reported first in 2000 is the highly exothermic chemiluminescent reaction of Ba atoms with N$_2$O, which produces electronically excited BaO* via reaction (1) proceeding in He droplets.

$$Ba + N_2O \rightarrow BaO^* + N_2 \qquad (14)$$

Emission from BaO* was observed in a rovibrational progression of sharp lines beyond 550 nm; indicating that the reaction occurs only inside the HNDs [491].

Cold reactions of alkali-metal clusters with water clusters embedded in HNDs have been investigated using femtosecond photoionization, as well as electron-impact ionization, for the detection of reaction products [309]. Weakly bound van der Waals complexes of the form Na$_k$(H$_2$O)$_m$ (k = 1, 2, 3) were the predominant products for the reaction of Na$_n$ clusters, with only minor amounts of Na hydroxide clusters of the type Na(NaOH)$_n$. In sharp contrast, Cs clusters were found to completely react with water clusters. Oxidative hydrolysis to form (CsOH)$_n$ appeared to occur when water was present in excess of Cs, although ionization led to dissociation products that had to be understood with high-level molecular structure calculations. Excess Cs led to the formation of compounds of the type Cs$_n$(CsOH)$_m$ for n > 1, Cs$_{n-1}$(CsH)$_m$, Cs$_{n-2}$(Cs$_2$O)$_m$, Cs(CsOH)CsH and others.

Later studies of reactions of Mg atoms and clusters with O$_2$ reported in 2010 by Krasnokutski and Huisken were able to produce the first rate constant measurement of a reaction proceeding within a HND [492]. The reaction products Mg$_2$O, Mg$_2$O$_2$, Mg$_3$O, Mg$_3$O$_2$ and Mg$_4$O$_2$ were identified with electron-impact mass spectrometry. The energy released in their formation was sufficient to eject them out of the He droplets. Chemiluminescence (CL) could be detected when more than one Mg atom was picked up per He droplet and electronically and vibrationally excited Mg$_n$O$_2$ (n ≥ 2) left the He droplets.

The experimental approach is illustrated in Figure 43. In rate measurements the point of oxygen pickup is translated along the He droplet beam so that the reactants are given different times to react before they are detected. The ion signal intensity of a reactant or product molecule is measured as a function of the position of the O$_2$ source. Measurements at two different flight times (1.17 and 1.56 ms) showed no difference in the mass spectra and so indicated the occurrence of a fast reaction complete in less than 1 ms. Measurements of the intensity of chemiluminescence along (and away from) the He droplet beam provided another kinetic method with better time resolution, albeit the reaction time and the excited state lifetime of the product (probably larger than the reaction time in this case) combine to determine the time between the O$_2$ pickup and the emission measurement. This method provided a first-order rate constant > 5×10$^4$ s$^{-1}$ for product formation in the oxidation of Mg clusters with O$_2$.

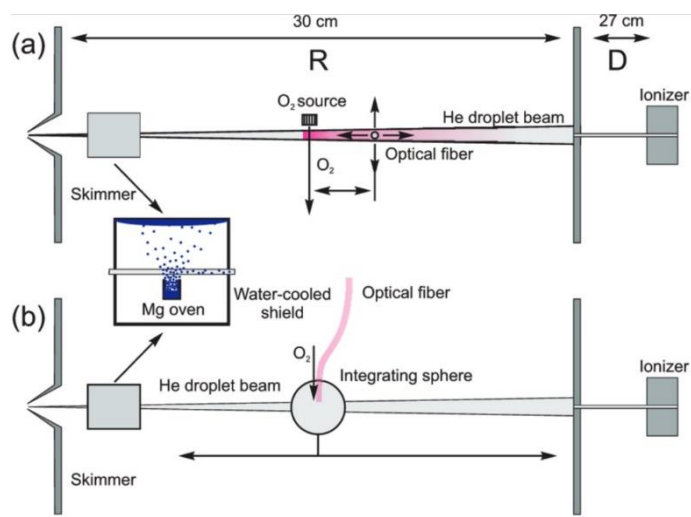

*Figure 43 Sketch of the experimental setup used for measuring reactions of Mg atoms and clusters with $O_2$ molecules in He droplets. The upper part (a) shows the configuration employed for the study of reaction kinetics by monitoring the chemiluminescence emitted by the reaction products as a function of position and time. The mass spectrometric measurements were carried out in configuration (b) depicted in the lower part of the figure. The integrating sphere served as pick-up source for $O_2$ and as light collector for the chemiluminescence. Reprinted with permission from Ref. [492]. Copyright 2010 American Chemical Society.*

A subsequent study, by the same authors, of the reaction of Si atoms and clusters with $O_2$ was directed in part to the development of another method to monitor reaction kinetics. The energy release during the reaction and the concomitant evaporation of helium provided a measure of the extent of reaction [493]. In essence the He droplet acts as a "nanocalorimeter". The depletion of the He droplet beam is measured in terms of a background pressure (in the detection chamber). The incorporation of Si and $O_2$ together was seen to cause a depletion considerably larger than the depletion normally caused by the incorporation of the reactants individually. The exothermicities of the bimolecular reaction to produce SiO and the termolecular reaction to form $SiO_2$ are 221.33 kJ mol$^{-1}$ (= 2.29 eV) and 615.89 kcal mol$^{-1}$ (= 6.38 eV), respectively. In the kinetic measurements, the reaction time was again varied by moving the position of $O_2$ pickup along the beam. A 30 mm change in this distance provided no change in the depletion of the He droplet beam, indicating a fast reaction with k > 5×10$^{-14}$ cm$^3$ molecule$^{-1}$ s$^{-1}$. In addition to SiO and $SiO_2$, mass spectra recorded ions of the products $SiO_3$, $Si_2O$, $Si_2O_2$ and $Si_2O_3$ and there was evidence for the presence of reactant $Si_n$ clusters with n up to 4. Separate experiments with $H_2O$ indicated no reactions with Si atoms and clusters of Si.

In recent studies by this group, in which HNDs were doped with SiO molecules, mass spectrometric measurements were able to track the oligomerization of SiO up to (SiO)$_n$ with n up to 10 [494]. Use of the calorimetric technique provided reaction energies for the first two steps that matched theoretical predictions based on the formation of cyclic (SiO)$_n$ clusters. The oligomerization was found to proceed in a barrierless fashion at the low temperature of HNDs. Furthermore, the incorporation of numerous SiO molecules into HNDs resulted in the formation of amorphous SiO grains that could be collected and characterized with various spectroscopic techniques. Grain formation was demonstrated with different dopant combinations, such as Si atoms and $H_2O$ molecules and with SiO, $O_2$ and $H_2O$ molecules.

Further studies also were performed by Krasnokutski and Huisken that explored the low-temperature reactivities of aluminum [495] and iron [496] atoms with $O_2$ and $H_2O$ and, in the case of Fe, also with $C_2H_2$. The Al atoms were shown to remain separated as atoms even at high doping concentrations and to react with $O_2$ to trap the intermediate adduct $AlO_2$ in its potential well. In contrast, no reaction of single Al atoms with $H_2O$ was observed, apparently due to an energy barrier predicted by quantum chemical calculations. Reactions were also observed with both $O_2$ and $H_2O$ clusters. Al depletion experiments were employed to monitor the reactions. Analogous experiments with Fe atoms showed only the formation of weakly bound adducts with $O_2$, $H_2O$ and $C_2H_2$ that were easily dissociated by electron transfer to $He^+$.

Most recently, with the development of a source of pure C atoms, studies have become feasible of chemical reactions of C atoms under HND conditions. Nanocalorimetry has been applied in a demonstration of the occurrence of the reaction of C atoms with $H_2$ by a barrierless insertion to form HCH [497] and high-resolution mass-spectrometric studies have explored the reaction of C atoms with $C_{60}$ [498]. The experiments with $C_{60}$ molecules have shown that C atoms can bridge bond onto $C_{60}$ without cage growth to form carbenes of the type $C_{60}C$: and higher members $C_{60}(C:)_n$ with n up to 6, and so dramatically transform the chemical reactivity of the fullerene. The C atoms also were observed to build bridges between fullerenes to form fullerene dimers, such as the known dumbbell $C_{60}=C=C_{60}$, for example.

## 10.2 Intra-droplet and intra-cluster ion-molecule reactions

### 10.2.1 Ionization initiated by electron and dissociative electron transfer to $He^+$

As indicated in Section 7.1, when $He^+$ is formed by electron ionization at the surface of a He droplet, the charge on the $He^+$ is propagated into the droplet by charge hopping until eventually an electron is transferred to the $He^+$ from the molecule or the molecular cluster embedded in the droplet. Molecular cations are formed, either single or clustered, or, when the molecules are fragile, the molecular cations can dissociate into fragment ions according to reactions (15) and (16).

$$He^+ + M_n \rightarrow M_n^+ + He \tag{15}$$

$$\rightarrow (M_n\text{-}X)^+ + X + He \tag{16}$$

The influence of energy transfer to the He matrix or intermolecular relaxation in a cluster on the extent of stabilization of short-lived intermediates (either parent cations or early dissociation products) can be assessed by comparisons with product-ion formation by electron impact ionization of the isolated molecules in the gas phase. For example, a recent He droplet study of the small diatomic molecules $O_2$, $N_2$, and CO was able to demonstrate a significant suppression of the formation of dissociative ion products according to reaction (15) and the reaction scheme depicted in Figure 25 compared to observations of the known gas-phase reactions of these molecules with $He^+$ at room temperature [499].

A recent study of three polyfluoroethers that were subjected to both ionization with 70 eV electrons in the gas phase and electron transfer to $He^+$ in cold He droplets provides a useful illustrative example of charge transfer reactions [500]. Mass spectra were recorded with perfluorodiglyme (PD, diethylene glycol dimethyl ether, $C_6F_{14}O_3$), perfluorotriglyme (PT, triethylene glycol dimethyl ether, $C_8F_{18}O_4$) and perfluorocrownether (PC, eicosafluoro-[15]-crown-5 ether, $C_{10}F_{20}O_5$). Pick-up conditions in the droplet experiments were adjusted to embed only one molecule of PT or PC and a few molecules of PT into the He droplet. Figure 44 provides results obtained with PC.

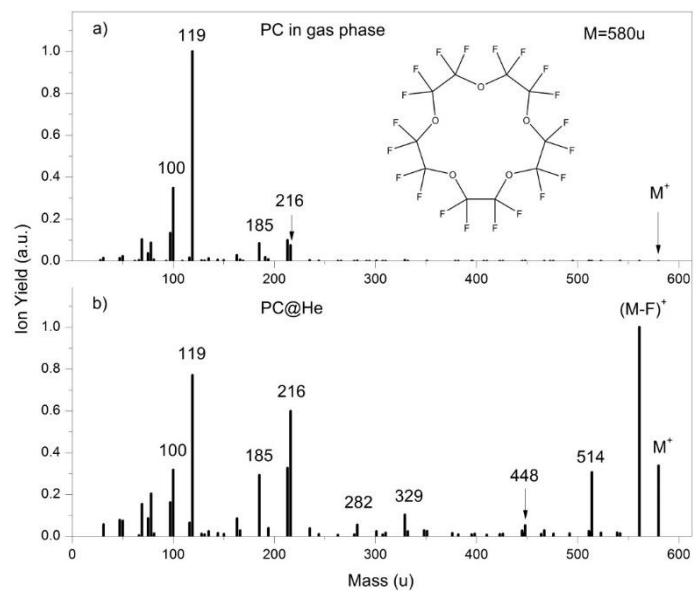

*Figure 44 Electron ionization mass spectrum of (a) isolated perfluorocrownether (PC) and (b) when the same molecule is embedded in ultracold HNDs. Reproduced with permission of John Wiley & Sons Ldt from Ref. [500].*

With embedded single molecules of PC and PT, the observations indicate enhanced formation of parent and higher-mass fragment cations. For embedded PD cluster ions one observes single $M_n^+$ as well as solvated cluster ions $M_nX^+$. The point has been made that the decomposition mechanisms for these large molecular ions is complex and various decomposition mechanisms may contribute such as: electronic decomposition along a repulsive potential energy surface (fs time scale, *e.g.* loss of F), vibrational predissociation (ps time scale) or metastable decomposition (μs to ms time scale). The slower fragmentations might be expected to be suppressed or "frozen" by HNDs, unlike fast decomposition processes.

Studies of electron impact ionization with the amino acid valine in the gas phase and in HNDs have extended the droplet comparison to the He$^+$ ionization of larger clusters, both in the absence and presence of co-embedded water [501]. Isolated valine cations were found to be extremely fragile in the gas phase, but less so in the clusters and even less in mixed valine/water subclusters present in HNDs [502].

10.2.2 Dynamics of ionization initiated by electron and dissociative electron transfer to He$^+$
The dynamics of ionization or dissociative ionization of a molecule embedded in a low temperature He environment has been scrutinized by Lewis *et al.* [162] for the dipolar molecules HCN and HC$_3$N and the quadrupolar molecule C$_2$H$_2$, as well as two isomeric van der Waals complexes of HCN(C$_2$H$_2$) and HCN(HC$_3$N) [503]. The efficiency of electron transfer to He$^+$ was measured as a function of droplet size (number of He atoms) and this dependence was analyzed in terms of the trajectory of the migrating He$^+$ in the presence of charge steering by the long-range electrostatic interactions between the He$^+$ ions and the neutral molecules. Both the trajectory of the migrating charge and the site of electron transfer from the embedded molecule were found to depend on its electrostatic nature and isomeric structure and lead to differences in ionization outcomes from purely gas-phase experiments. Boatwright *et al.* [504] have provided a larger overview of the droplet/gas-phase comparison by investigating He$^+$ reactions with H$_2$O, SO$_2$, CO$_2$, CH$_3$OH, C$_2$H$_5$OH, CH$_3$F and CH$_3$Cl that were allowed to form clusters, as well as CF$_2$Cl$_2$ and CH$_3$I that behaved as single molecules embedded in the droplets. The chemistry exhibited by many of the cluster ions and at least one of the single molecular ions in HND experiments was found to be very different from that observed for the same species in the gas phase.

### 10.2.3 "Self" and "mixed" ion-molecule reactions: experimental strategies

Experimental strategies for the initiation and investigation of both primary and secondary ion-molecule reactions proceeding at the very low temperature provided by HNDs were first delineated in a benchmark paper by Fárník and Toennies in 2005 [505]. Experiments were also described that demonstrate the stabilization of selected transient intermediate complexes by the rapid quenching of internal degrees of freedom within the He droplets. Earlier mass-spectrometric studies by the Toennies group of positive ions formed after electron impact of doped HNDs identified primary reactions between $He^+$ ions with embedded $SF_6$ [64] or cold clusters of Ar, Kr, Xe, $H_2O$, and $SF_6$ formed prior to ionization inside the droplet [65]. Fárník and Toennies extended such studies to $D_2$, $N_2$ and $CH_4$ [505]. The $He^+$ ionization processes were identified using "difference spectra" obtained from mass spectra of the He droplets with and without various amounts of dopant gas (see Figure 45). Primary electron transfer and dissociative electron transfer channels were seen and compared to gas-phase results. Furthermore, the mass spectra provided evidence for the occurrence of secondary self-reactions, such as (17), (18) and (19), which are well known from gas-phase studies:

$$D_2^+ + D_2 \rightarrow D_3^+ + D \qquad (17)$$

$$N_2^+ + N_2 + He \rightarrow N_4^+ + He \qquad (18)$$

$$CH_4^+ + CH_4 \rightarrow CH_5^+ + CH_3 \qquad (19)$$

The novel approach of adding a second dopant to investigate "mixed" reactions was applied to $N_2$ and $CH_4$, each mixed with varying amounts of $D_2$. Again, reactions were observed that have previously known gas-phase analogues, reactions (20), (21) and (22).

$$N_2^+ + D_2 \rightarrow N_2D^+ + D \qquad (20)$$

$$CH_4^+ + D_2 \rightarrow CH_4D^+ + D \qquad (21)$$

$$CH_3^+ + D_2 + He \rightarrow CH_3D_2^+ + He \qquad (22)$$

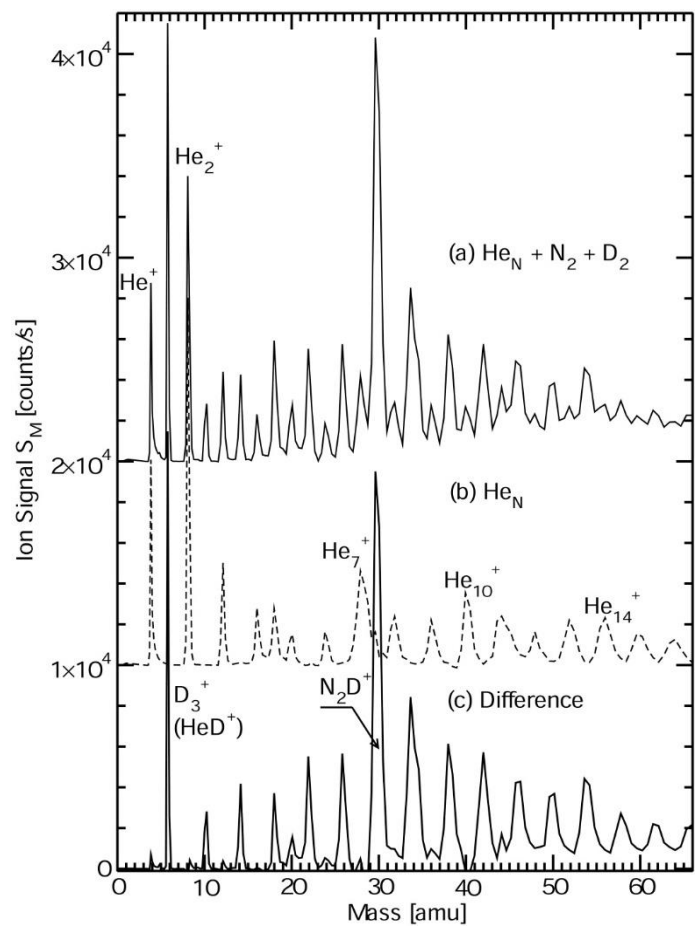

*Figure 45 The mass spectrum of He droplets containing $N_2$ and $D_2$ molecules (a) (thin line) embedded simultaneously in the scattering chamber at a total pressure $p(N_2 + D_2) = 6.9 \times 10^{-6}$ mbar and mixing ratio $D_2:N_2 = 3:1$. The spectrum (b) (dashed line) corresponds to the pure droplet. The trace (c) (heavy line) is the difference spectrum between (a) and (b). The strong $N_2D^+$ ion peak indicated is due to the secondary reaction between $N_2^+$ and $D_2$. The other peaks are attributed to $He_m \cdot D^+$ ions from the $He^+ + D_2$ reaction. Reprinted from Ref. [505], with the permission of AIP Publishing.*

More recently $N_2/CH_4$ mixtures embedded in HNDs have been studied with a view to possible C-N bond formation by ion-molecule reactions, as might occur in the atmosphere of Titan [506, 507]. Major ions of the type $CH_nN_2^+$ (n = 1 to 5) and minor ions of the type $CH_nN_4^+$ (n = 1 to 5) were followed as a function of increasing $CH_4$ content in the droplet. Their formation was interpreted in terms of the initial ionization of a methane molecule in methane clusters containing one or two nitrogen molecules. Electronic structure calculations suggested the possible formation of a C-N bond in $CH_3N_2^+$.

10.2.4 Intra-cluster reactions with $H_2$, $CH_4$

Dopant cluster formation can proceed with the combination of two or more dopant atoms or molecules and, when the dopant is molecular, $He^+$ ions can initiate cluster and intra-cluster ion-molecule reactions. For example, in He droplets ($10^6$ He atoms) containing clusters of $H_2$, in the presence of 70 eV electron bombardment, the $He^+$ produced by ionization can ionize the embedded $H_2$ cluster to produce $H_2^+$ and $H^+$, initially [399]. These two ions can then interact further with $H_2$ within the cluster to form $H_3^+$ according to reactions (23) and (24) below. High resolution mass spectrometric detection of the ions released showed formation of clusters of the type $He_nH^+$, $He_nH_2^+$ and $He_nH_3^+$ with n = 1 to 30 (and their deuterium analogs in experiments with $D_2$) and interesting "magic numbers" in their distributions. Kinetic energy release experiments indicated that the $He_2H_2^+$ species was metastable with a linear, centrosymmetric ion structure of $[He-H-H-He^+]$ [508].

$$\text{He}^+ + (\text{H}_2)_n \rightarrow [(\text{H}_2)_n^+ \rightarrow \text{H}_3^+\text{H}(\text{H}_2)_{n-2}^*] + \text{He} \rightarrow \text{H}_3^+(\text{H}_2)_{n-2} + \text{H} + \text{He} \quad (23)$$

$$\text{He}^+ + (\text{H}_2)_n \rightarrow [(\text{H}_2)_n^+ \rightarrow \text{H}^+\text{H}(\text{H}_2)_{n-1}^*] + \text{He} \rightarrow \text{H}_3^+(\text{H}_2)_{n-2} + \text{H} + \text{He} \quad (24)$$

With clusters of pure methane, high resolution mass spectra showed strong peaks for $(\text{CH}_4)_n^+$ and $\text{CH}_5^+(\text{CH}_4)_n$ ions, with weaker contributions from $[\text{C}_2\text{H}_x(\text{CH}_4)_n]^+$ ions with x = 2–7. This is very similar to observations in the gas phase [294]. Both the $\text{CH}_5^+(\text{CH}_4)_n$ and $[\text{C}_2\text{H}_x(\text{CH}_4)_n]^+$ ions can be attributed to $\text{He}^+$ ionization of the neutral cluster followed by intracluster ion-molecule reactions (25) and (26), with the latter involving C-C bond formation.

$$\text{He}^+ + (\text{CH}_4)_n \rightarrow [(\text{CH}_4)_n^+ \rightarrow \text{CH}_5^+\text{CH}_3(\text{CH}_4)_{n-2}]^* + \text{He} \rightarrow \text{CH}_5^+(\text{CH}_4)_{n-2} + \text{CH}_3 + \text{He} \quad (25)$$

$$\text{He}^+ + (\text{CH}_4)_n \rightarrow [(\text{CH}_4)_n^+ \rightarrow \text{C}_2\text{H}_7^+\text{H}(\text{CH}_4)_{n-2}]^* + \text{He} \rightarrow \text{C}_2\text{H}_7^+(\text{CH}_4)_{n-2} + \text{H} + \text{He} \quad (26)$$

Furthermore, the formation of cluster dications $(\text{CH}_4)_n^{2+}$ was also reported and proposed to occur when the initial ionization of $(\text{CH}_4)_n$ by $\text{He}^+$ is followed by Penning ionization with $\text{He}^*(2^3\text{S})$ metastable atoms (see section 7.3 and 7.4).

Ion-molecule reactions between clusters containing $\text{H}_2/\text{D}_2$ and $\text{O}_2$ have been investigated in superfluid HNDs initiated by electron-induced ionization at 70 eV by Renzler *et al.* [509]. Reaction products were identified mass spectrometrically and could be explained by primary reactions involving $\text{H}_3^+$ or $\text{H}_3^+(\text{H}_2)_n$ clusters and their deuterated analogues. Representative mass spectra are shown for deuterated species in Figure 46. Relative ion yields suggested that $\text{DO}_2^+$ produced from the reaction of $\text{D}_3^+$ with $\text{O}_2$, well known from room temperature gas-phase studies [510, 511], was diminished by a secondary reaction of $\text{DO}_2^+$ with $\text{D}_2$ to produce $\text{D}_2\text{O}_2^+$ + D. In contrast, $\text{DO}_4^+$, not known from previous gas-phase experiments, was the most abundant product of the proposed reaction of $\text{D}_3^+$ with the oxygen dimer $(\text{O}_2)_2$, in HNDs. $\text{D}_3\text{O}_x^+$ ions become the most abundant ions for large $(\text{O}_2)_n$ clusters and their formation was attributed to secondary association reactions between $\text{DO}_x^+$ and $\text{D}_2$.

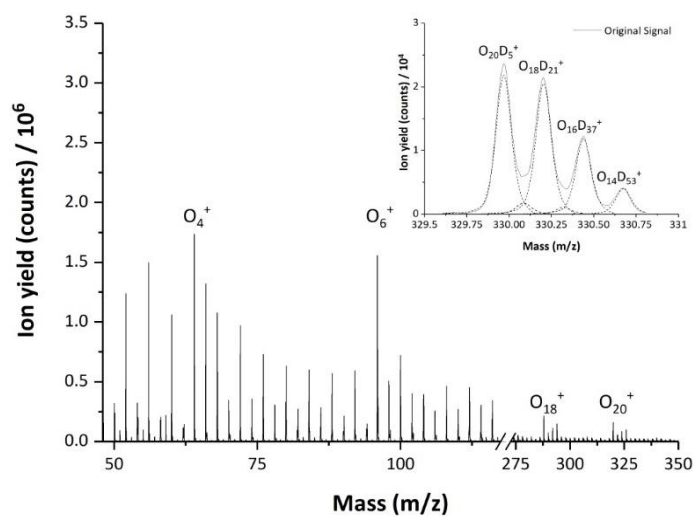

*Figure 46 Part of the mass spectrum from electron ionization of HNDs containing a $D_2/O_2$ mixture. In the main image peaks for the $O_6^+$ and $O_8^+$ ions are highlighted. Built upon the $O_6^+$ and $O_8^+$ peaks are additional peaks from ions with added D atoms. Those ions with an even number of D atoms have masses which notionally coincide with $He_n^+$ cluster ion peaks. However, the mass resolution is high enough to distinguish between these peaks. The high mass resolution is demonstrated in the inset. A deconvolution leads to the dashed lines in the inset, which show underlying contributions from ions with particular values of m/z. Reproduced from Ref. [512] with permission from the PCCP Owner Societies.*

## 10.2.5 Intra-cluster reactions of alcohols

High resolution mass spectrometric experiments with large He clusters (*ca.* $10^5$ He atoms) have been used to access the chemistry of very large clusters of methanol and ethanol initiated by $He^+$ [512]. The dominant products observed with methanol were protonated methanol cluster ions, $(CH_3OH)_nH^+$, containing up to 100 methanol molecules (the parent cluster ions were not seen) [512]. A representative mass spectrum is shown in Figure 47. The formation of these cluster ions can be attributed to the chemical sequence given in reaction (27) and is analogous to those given above for $H_2$ and $CH_4$, leading to the expulsion of $OCH_3$ in this case.

$$He^+ + (CH_3OH)_n \rightarrow [(CH_3OH)_n^+ \rightarrow CH_3OH_2^+OCH_3(CH_3OH)_{n-2}]^* + He \rightarrow CH_3OH_2^+(CH_3OH)_{n-2} + OCH_3 + He \quad (27)$$

Other significant product cluster ions that were seen, although with lower abundance, were formed by expulsion of $C_2H_6O$, presumed to be dimethyl ether. This chemical sequence, which leaves behind $H_2O$, is shown in reaction (28).

$$[(CH_3OH)_nH^+]^* \rightarrow H^+(H_2O)(CH_3OH)_{n-3} + CH_3OH + CH_3OCH_3 \quad (28)$$

The cluster ions $H^+(H_2O)(CH_3OH)_n$ became noticeably more prominent at n = 7 and peaked at n = 9. Reaction (28) was observed previously in multiphoton ionization experiments with clusters of methanol in the gas phase [512, 513].

Still other distributions of cluster ions with up to 5 water molecules, $H^+(H_2O)_m(CH_3OH)_n$ (m = 1–5), were observed in the droplet experiments with maxima shifted to higher values of n with increasing m. Apparently, and remarkably, reaction (28) can occur more than once for sufficiently large methanol clusters.

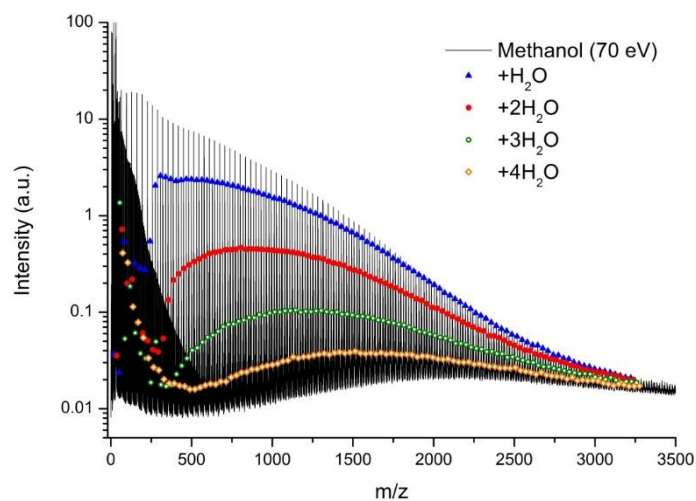

*Figure 47 Observed abundances of $H^+(H_2O)_m(CH_3OH)_n$ clusters for m = 1–4. Reproduced from Ref. [512] with permission from the PCCP Owner Societies.*

Ethanol was reported to behave in a similar way under similar operating conditions, although some differences were seen in the shapes, maxima and intensities of the cluster ion distributions.

10.2.6 Intra-cluster reactions with co-embedded dopants: $C_{60}/H_2O$ and $C_{60}/NH_3$

Ionization of doubly doped cold He droplets allows the exploration of the formation of novel mixed cluster ions. To this end, high-resolution experiments have been performed with droplets containing either $H_2O$ or $NH_3$ together with $C_{60}$ [514, 515]. The general features of the observed mixed cluster ions were seen to be quite similar. $C_{60}$ aggregates were detected with unprotonated, protonated and dehydrogenated water or ammonia (A = ammonia) with up to four $C_{60}$ molecules. Figure 48 shows the mass spectrum of HNDs doped with $C_{60}$ and $NH_3$.

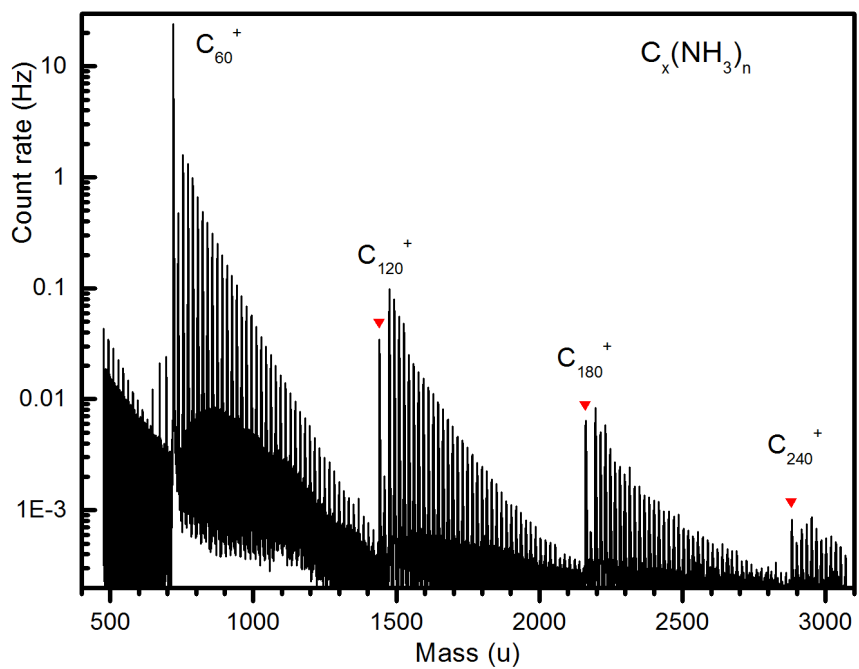

Figure 48 Mass spectrum of HNDs doped with $C_{60}$ and $NH_3$. Pure fullerene aggregates $C_x^+$ (x = 120, 180, 240) are marked by triangles. Reproduced from Ref. [515] with permission from the PCCP Owner Societies.

The unprotonated cluster signal $C_{60}(A)_n^+$ exceeds that of the protonated signal $C_{60}(A)_nH^+$ for small n. Magic numbers are seen for the $C_{60}(H_2O)_n^+$ and $C_{120}(H_2O)_n^+$ series at n = 4. $C_{60}$ appears to suppress the protonated signal more strongly than $C_{120}$ (also see Section 12.1).

Formation of the protonated cluster ions $C_{60}(A)_nH^+$ can proceed by an intra-cluster ion-molecule reaction after an electron is lost to the incoming $He^+$ ion. This is illustrated in equations (29) and (30) for water and ammonia, respectively.

$$He^+ + C_{60}(H_2O)_n \rightarrow [C_{60}(H_2O)_n^+ \rightarrow C_{60}OH(H_2O)_{n-2}H_3O^+]^* + He \rightarrow C_{60}(H_2O)_{n-1}H^+ + OH + He \quad (29)$$

$$He^+ + C_{60}(NH_3)_n \rightarrow [C_{60}(NH_3)_n^+ \rightarrow C_{60}NH_2(NH_3)_{n-2}NH_4^+]^* + He \rightarrow C_{60}(NH_3)_{n-1}H^+ + NH_2 + He \quad (30)$$

Formation of the dehydrogenated cluster ions is somewhat counterintuitive in that fragmentation of clusters ions embedded in He usually is suppressed. The intervention of doubly-charged $C_{60}$, or double ionization, has been proposed to lead to their formation according to equations (31) and (32) for water and ammonia, respectively.

$$He^+ + C_{60}(H_2O)_n \rightarrow [C_{60}^+(H_2O)_n^+ \rightarrow C_{60}^+OH(H_2O)_{n-2}H_3O^+]^* + He \rightarrow [C_{60}OH]^+ + (H_2O)_{n-2}H_3O^+ \quad (31)$$

$$He^+ + C_{60}(NH_3)_n \rightarrow [C_{60}^+(NH_3)_n^+ \rightarrow C_{60}^+NH_2(NH_3)_{n-2}NH_4^+]^* + He \rightarrow [C_{60}NH_2]^+ + (NH_3)_{n-2}NH_4^+ \quad (32)$$

### 10.2.7 Intra-cluster reactions with co-embedded Na and SF$_6$: self-assembly of salt nanocrystals

In a quest to make crystalline salts in small clusters, Na and SF$_6$ were co-embedded by injecting them separately into HNDs that were then exposed to energetic electrons [296]. The two reagents are likely to be directly in contact and frozen in the interior of the droplets, rather than being segregated. The major cations observed were (NaF)$_n$Na$^+$ and (NaF)$_n$(Na$_2$S)$_m$Na$^+$. Major negative ions included (NaF)$_n$F$^-$ and (NaF)$_n$S$^-$. Figure 49 provides abundance plots for these ions in which magic numbers are clearly visible. The magic numbers at n = 13, 22 and 37 have been attributed to the formation of structures consisting of one, two and four complete unit cells of sodium fluoride that are especially stable because they maximize the attractive Coulombic interaction between the constituent ions.

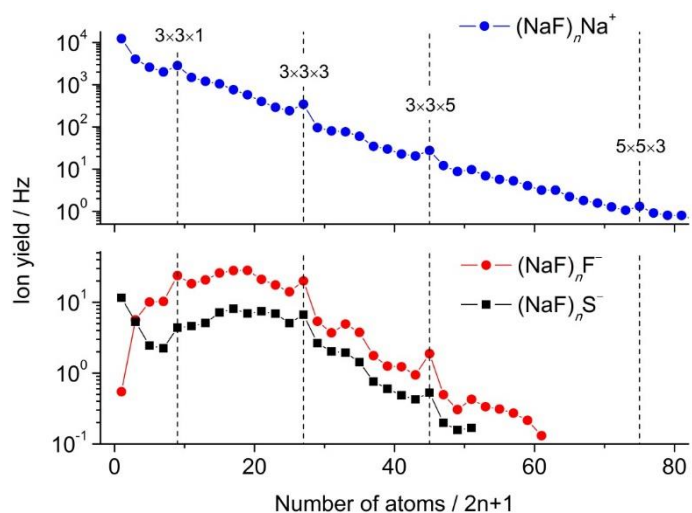

*Figure 49 Abundance plots for the major cations (top) and anions (bottom). The labels refer to the number of ions located along Cartesian coordinates, e.g. 3x3x3 refers to the cubic unit cell of sodium fluoride. Reproduced with permission of John Wiley & Sons Ldt from Ref. [296].*

Less certain are the details of the chemical reactions that lead to all of the cluster ions in Figure 49. The formation of NaF as part of ionic structures clearly is critical. Formation of Na$^+$, F$^-$ and S$^-$ is proposed to result from the impact of He*, He$^+$ or He*$^-$ with clusters of Na$_n$(SF$_6$)$_m$. The formation of NaF is less certain, but reasons are provided that favor electron initiated chemistry in which the various impinging energetic He species can activate the clusters and achieve NaF formation. The neutral reaction between Na and SF$_6$ has too large an activation energy and ejected products, either neutral or ionic, of possible highly exothermic reactions of Na$_n$ and (SF$_6$)$_m$ were not detected.

### 10.2.8 Reactions of He*$^-$ with fullerenes: formation of multiply-charged ions and ion/ion reactions with cations.

The very recent discovery of the presence of metastable, highly-mobile He*$^-$ anions in superfluid HNDs [176] has opened the door to novel investigations of their chemistry with both neutral molecules and with cations. Fullerene clusters of the type (C$_{60}$)$_n$ and (C$_{70}$)$_m$ have recently provided insight into both types of chemistry [176, 177].

Survey anion mass spectra (Figure 50) have shown the production, in the presence of He*$^-$, of both monoanions (C$_{60}$)$_n^-$ and, at much lower intensities, the dianions (C$_{60}$)$_n^{2-}$ for n ≥ 5. The formation of these anions has been attributed to the chemical reactions (33) and (34), respectively, where $N$ is the number of He atoms in the droplet.

$$\text{He}^{*-} + (C_{60})_n \rightarrow (C_{60})_n^- + \text{He} \qquad (33)$$

$$\text{He}^{*-} + (C_{60})_n@\text{He}_N \rightarrow \text{He}_2^+ + (C_{60})_n^{2-} + (N-1)\text{He} \qquad (34)$$

Formation of the dianions from reactions of He*$^-$ with monoanions was excluded because of Coulomb repulsion, and the onset of dianion formation by Eq. (34) at n = 5, also seen for $(C_{70})_n^{2-}$, was attributed to Coulomb explosion of the smaller anions, rather than an energy threshold.

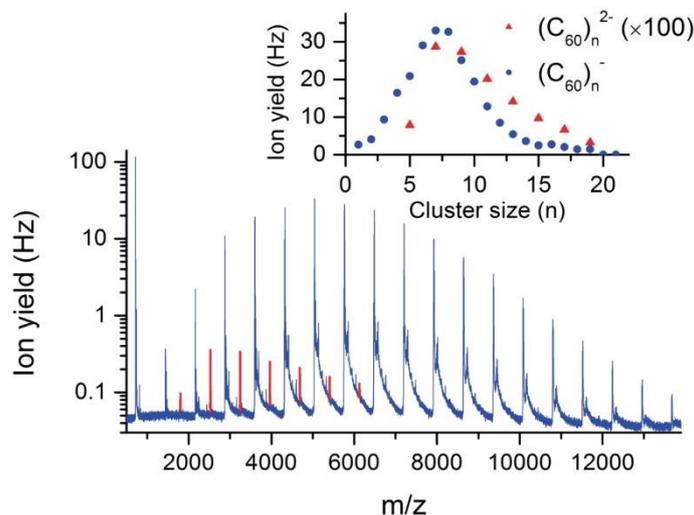

Figure 50 Anion survey mass spectrum. The dominant peaks arise from $(C_{60})_n^-$. At much lower abundance $(C_{60})_n^{2-}$ dianions can also be seen and are marked in red. The inset compares the observed size distributions of the monoanions and dianions. Reprinted from Ref. [177], with the permission of AIP Publishing.

The formation of multiply-charged fullerene cluster cations, $(C_{60})_n^{z+}$ with z up to 4, was also observed in these experiments. Details of this mechanism are discussed in chapter 6.4.4. Once these are formed in the droplet, strong Coulomb attraction will draw He*$^-$ toward them. Through a sequence of He*$^-$ production events, multiply-charged fullerene cluster cations with increasing charge can then be generated by reaction (35).

$$\text{He}^{*-} + (C_{60})_n^{z+} \rightarrow (C_{60})_n^{(z+1)+} + \text{He} + 2e^- \qquad (35)$$

Again, Coulomb repulsion would prevent the occurrence of electron transfer to He$^+$.

# 11 Spectroscopic investigations of highly reactive molecules, clusters and ions

## 11.1 Free radicals

HNDs provide a cold and inert environment in which to trap highly reactive molecules. Good examples are free radicals, which have been extensively studied in the gas phase with traditional supersonic molecular beam techniques. In contrast, attempts to investigate free radicals in HNDs were slow to begin. One of the classic routes to produce free radicals is photodissociation and therefore one might consider trapping a precursor in a HND and subjecting it to UV radiation to break the appropriate bond(s). This technique works in solid matrices, such as argon, since the fragments can separate, dissipate energy into the matrix, and then become trapped at different sites. However, this approach is ineffective for HNDs. In small droplets the fragments have too much kinetic energy and simply exit [516], while in large droplets they may remain but will inevitably recombine at some point because there are no distinct, rigid trapping sites [517]. Consequently, the only viable source of free radicals in HNDs is by generating them externally and then using a pick-up procedure.

A rather simple case is the pickup of NO. This was first added to HNDs by von Haeften *et al.* and the infrared (IR) spectrum was then recorded by a depletion technique [518]. Depletion spectroscopy relies on some of the energy absorbed by the dopant molecule making its way out into the surrounding matrix and causing evaporative loss of helium atoms. This shrinks the HND, a change that can be detected by electron ionization mass spectrometry downstream of the laser excitation region. The basic idea is that a loss of helium atoms decreases the electron ionization cross section of a HND, and hence fewer ions are registered by the mass spectrometer. Despite being inert and behaving as a superfluid, it is important to recognize that the surrounding helium can have an impact on the spectra of molecules. While the effect of the helium on molecular vibrations tends to be very small, this is not the case for molecular rotations and one often finds that rotational constants are dramatically reduced by the helium. This happens when rotation is sufficiently slow to allow some of the neighboring helium density to rotate with the molecule, delivering a higher effective moment of inertia. Hence the rotational structure, even if resolved, cannot be used to extract quantitative structural information for the molecule. The NO study by von Haeften *et al.* is a good example of this [518]. The rotationally-resolved spectrum gave a rotational constant which is 24% smaller than the gas phase value. Hyperfine structure was also recorded in this study and the $\Lambda$-doubling splitting in the $^2\Pi$ ground state was found to be substantially larger than the value for NO in the gas phase.

The first successful production of a small and highly reactive molecular free radical in HNDs was achieved by the group of Roger Miller. In this experiment pyrolysis was used to induce thermal dissociation of propargyl bromide, generating the propargyl radical ($HCCCH_2$) at temperatures in excess of 1000 K [519]. This 'hot' radical was then captured by a beam of HNDs and an IR depletion spectrum was recorded in the acetylenic CH stretching region near 3300 cm$^{-1}$. As found in earlier gas phase work, the molecule was found to adopt $C_{2v}$ symmetry in superfluid helium, making it a near prolate symmetric rotor.

The most extensive body of research on free radicals in HNDs has been generated by the group of Douberly. Much of this work achieved full or partial rotational resolution. Examples of molecular free radicals studied in this way include hydroxyl [520], allyl [521], vinyl [522] and ethyl [523]. An interesting variation on this theme is to generate a new radical inside the HND by combining a pyrolytically-generated radical with another molecule. This has been used by the Douberly group to make the methyl peroxy radical, $CH_3OO$, through the $CH_3 + O_2$ reaction [524]. This is a barrierless reaction with a modest exothermicity ($\sim$ 130 kJ mol$^{-1}$ (1.35 eV)), which makes it possible to generate and contain the products inside the relatively small HNDs (< $10^4$ helium atoms) preferred for depletion spectroscopy. All three CH stretching vibrations were identified in the IR spectrum, as shown in Figure 51, along with rotational contours of each band in higher resolution scans. Other peroxy radicals made in this way include HOOO [520] and $C_3H_3OO$ (propargyl peroxy) [525].

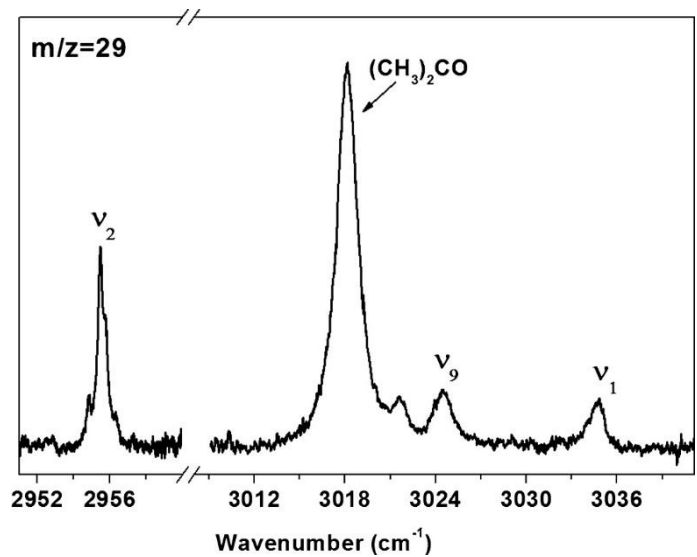

*Figure 51 IR survey spectrum for HNDs containing $O_2$ and the products form the pyrolysis of di-tert-butyl peroxide. This spectrum was recorded by detecting depletion at m/z 29 ($C_2H_5^+$). The most intense band is from acetone, a pyrolysis product. The three bands assigned to vibrations $\nu_1$, $\nu_2$ and $\nu_9$ are from the $CH_3OO$ radical. Reprinted with permission from Ref. [524]. Copyright 2012 American Chemical Society.*

## 11.2 Clusters and complexes

As well as individual molecules, HNDs afford the opportunity to systematically explore the aggregation of molecules into clusters and complexes. Indeed, the formation of clusters in some of the droplets is inevitable given the statistical nature of the pick-up process, unless the pressure of the dopant is particularly low and/or the droplets are very small. By increasing the partial pressure of the dopant in the pick-up zone it is possible to use techniques such as mass spectrometry or optical spectroscopy to follow the growth of clusters and to explore their properties. To illustrate this, some examples are shown in the next two sections, with the emphasis on observation by infrared (IR) spectroscopy.

### 11.2.1 Molecular clusters

Molecular clusters are entities which bridge the transition from isolated molecules to the bulk condensed phase. Through the study of small clusters one might hope to gain insight into various aspects of the bulk, such as identifying key structural moieties. If we take a very important practical example, the case of water, the investigation of small and medium sized water clusters can reveal new information about the structure and dynamics of the hydrogen bonding network and how it evolves with size. Water clusters, $(H_2O)_n$, have been subjected to many experimental studies in the gas phase, as well as numerous theoretical studies, but here we focus on contributions from HND experiments.

The first spectroscopic investigation of water clusters was carried out by Fröchtenicht *et al.* in the mid-1990s [526]. This group utilized infrared depletion spectroscopy and recorded spectra in the O-H stretching region. Specific ions were monitored by mass spectrometry in order to gain size-specific information on the absorbing neutral cluster. For example, any potential contribution from the $H_2O$ monomer is eliminated by monitoring $H_3O^+$, and similarly detection of $H_3O^+(H_2O)$ removes any possible contributions from the monomer and the dimer. By using this mass selective detection in combination with control of the dopant pressure, Fröchtenicht *et al.* could identify bands from the water dimer, trimer and tetramer [526]. In the case of the trimer, which adopts a cyclic structure in which each water molecule acts as both a single hydrogen bond donor and acceptor, it was shown for the first time that there are two IR absorption bands in the bonded OH stretching region, signifying an asymmetric ring structure. Furthermore, the surrounding helium was found to cause only small shifts in the IR absorption band positions when compared with the gas phase. These findings have been largely confirmed by subsequent studies of small water clusters [527, 528].

A few years after the work by Fröchtenicht *et al.*, a remarkable finding was reported by Nauta and Miller [95]. Instead of using a mass spectrometer, depletion events induced by IR absorption were registered via a bolometer. This provides a highly sensitive detection scheme which enabled Nauta and Miller to record high quality IR spectra extending up to the water hexamer. Most spectral features were unremarkable save for a single new band, whose dependence on the water pressure indicated that it arose from the hexamer. It was known at that time that the water trimer, tetramer and pentamer formed rings as their lowest energy structure. However, non-cyclic structures (prism and cage isomers) of the hexamer have lower energies than the cyclic isomer because the formation of double donor and double acceptor hydrogen bonding modes that become possible in non-cyclic structures tip the energy balance in the non-cyclic direction at this threshold size. Although this structural switch does indeed occur in gas phase experiments, in HNDs the cyclic water hexamer forms. This surprising finding represents a significant milestone in the study of water clusters since the cyclic water hexamer is the basic building block of hexagonal ice ($I_h$ phase). Subsequent work has suggested that the cyclic water clusters form sequentially by insertion of a water molecule into an existing ring [529]. This route appears to provide a very low energy insertion barrier and once the cluster is formed it can be rapidly cooled by the surrounding helium, explaining why the cyclic hexamer does not revert to a lower energy non-cyclic structure. Nevertheless, a second and lower frequency band near 3200 cm$^{-1}$ was assigned by Nauta and Miller to a non-cyclic structure of the hexamer. However, a recent theoretical study has suggested that this band is simply an overtone transition from the water bending mode and that it, too, probably derives from the cyclic water hexamer [530]. The water hexamer case is a beautiful example of the ability of HNDs to form and trap metastable structures.

A recent experiment provides a new twist in the water cluster story. As mentioned above, the cyclic structures are the lowest energy forms of small water clusters, but Douberly *et al.* have shown via IR spectroscopy that addition of a single neon atom to a HND *prior* to the addition of water molecules inhibits the formation of the cyclic water tetramer. Instead, significant quantities of an isomer are formed which consists of a cyclic trimer with a fourth water molecule hydrogen bonded to one of the ring water molecules [531]. It was suggested that formation of this so-called 3+1 structure becomes possible because of a simple steric effect by the weakly attached neon atom, which inhibits rearrangement to the cyclic structure. The rapid cooling in helium traps the system in the higher energy 3+1 well.

There have been several spectroscopic studies of other molecular clusters in HNDs. Some recent examples include the study of the rotation of individual methane molecules in methane clusters containing up to $4\times10^3$ CH$_4$ molecules [532], the investigation of ethane clusters [533], and the observation of large (NH$_3$)$_n$ clusters that show IR spectra consistent with a crystalline arrangement of the ammonia molecules [534]. The ability of the HND technique to access new structures is nicely illustrated in a recent study of (NO)$_n$ clusters by Hoshina *et al.* [535]. In addition to recording IR spectra of clusters larger than the dimer, evidence was presented for the first observation of the weakly bound *trans* isomer of (NO)$_2$.

As mentioned above, one of the opportunities offered by HNDs, and which sets them apart from gas phase experiments, is the ability to explore unusual metastable states of clusters. Another impressive illustration by Nauta and Miller was for clusters of the strongly dipolar HCN and HCCCN molecules [93, 94]. When molecules such as HCN are added sequentially to a HND, the first molecule entering the droplet will cool rapidly. The second molecule will also cool rapidly on entry and will be drawn towards the first by long range attractive forces. As the two molecules approach each other the dipole-dipole attraction will steer the second molecule into an orientation such that its electric dipole moment points in the same direction as that of the other molecule. In other words, a head-tail-head-tail chain can form that will extend as further HCN molecules are added. This dipolar chain is not the lowest energy structure of $(HCN)_n$ or $(HCCCN)_n$ but, as shown by IR spectroscopy, in the low temperature environment of a HND the clusters become trapped in these metastable configurations.

### 11.2.2 Molecular complexes

We make the distinction between clusters, which are composed of multiple atoms or molecules of the same type, and complexes, where there is more than one type of atom or molecule. In the same way that HNDs provide a means of systematically studying molecular clusters, molecular complexes are accessible via the addition of two or more distinct dopants.

An excellent illustration of the strengths and weaknesses of using HNDs to probe molecular complexes is provided by a series of recent studies of $HCl(H_2O)_n$. The ionic dissociation of HCl in water to form aqueous hydrochloric acid is standard fayre in basic chemistry. However, one might ask if this ionic dissociation can be achieved between a single HCl molecule and a small number of water molecules, and indeed how many water molecules are needed to trigger this event? The same questions could arise for other solute-solvent combinations but the $HCl/H_2O$ system is convenient because both molecules are easily obtained in the gas phase and can therefore readily be added to HNDs.

The simplest complex, $HCl(H_2O)$, was first observed using IR depletion spectroscopy in 2007 [536]. In this initial work a single band attributed to the H-Cl stretching vibration was reported. Subsequent work by Skvortsov *et al.* probed the OH stretching region and, through a combination of theoretical predictions and careful measurements of the dependence of the intensities of specific bands on the dopant partial pressures, features were assigned to $HCl(H_2O)_n$ for n = 1−3 [537]. At around the same time considerable excitement was generated by a report of bands that could be assigned to the separated ion-pair, $H_3O^+(H_2O)_3Cl^−$ [538]. This suggested that dissociation into solvent-separated ions could occur with as few as four water molecules, in agreement with a number of prior theoretical predictions. However, shortly afterwards the bands assigned to the dissociated $H_3O^+(H_2O)_3Cl^−$ complex, were reassigned to the undissociated $(HCl)_2(H_2O)_2$ complex [539]. One can appreciate some of the challenges in assigning bands by referring to the spectra in Figure 52. The upper scan is a depletion spectrum recorded by detecting $H_3O^+$ at *m/z* 19, while the scan below was collected at *m/z* 72. In the case of *m/z* 19, the mass selectivity reveals only that the neutral complex must contain at least one water molecule. Several ions could contribute to the signal at *m/z* 72, including $(H^{35}Cl)_2^+$, $(H_2O)_4^+$ and $He_{18}^+$. However, the absence of the lower frequency peak seen in Figure 52 (a) from the spectrum in (b) was taken as evidence that this peak arises from complexes containing $^{37}Cl$. This is useful information but is insufficient to make firm assignments of the bands. Other measurements were also employed, including careful study of how the band intensities depend on the pressure of both HCl and $H_2O$, as well as the effect of a strong electric field on the band intensities (vibrational Stark effect) [539].

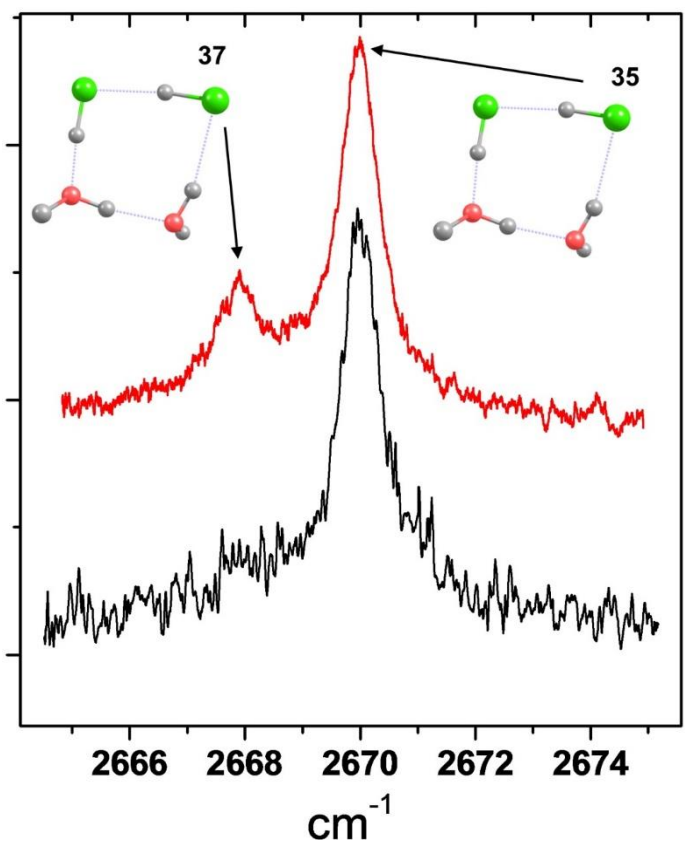

Figure 52 IR scans from doped HNDs containing HCl and $H_2O$. The Scan in red was measured by monitoring depletion of $H_3O^+$ (m/z 19) while the black one was recorded at m/z 72 (potentially containing contributions from several ions including $(H^{35}Cl)_2^+$). Reprinted with permission from Ref. [539]. Copyright 2010 American Chemical Society.

Unfortunately, clear observation of ionically dissociated $HCl(H_2O)_n$ complexes in HNDs remains elusive. Electric dipole moments for $DCl(H_2O)_n$ (n = 3–9) have, however, been measured via beam deflection [540]. While sharp IR bands are observed for relatively small complexes, broad underlying features become evident as the dopant partial pressures increase. Sources of this broadening could include contributions from multiple isomers from differently sized complexes, as well as broadening introduced by intramolecular vibrational redistribution, including interaction with excitation of the helium bath. Letzner *et al.* have presented evidence that the dissociated complexes with n ≥ 4 are significant contributors to a broad IR signal seen below ~ 2660 $cm^{-1}$ [541]. However, it seems unlikely that much significant information can be extracted from such unstructured spectral features.

Other types of solutes provide a potentially fertile ground for new HND experiments. For example, salts are vitally important solutes in solution chemistry and in principle HNDs provide a means of bringing a salt molecule into contact with solvent molecules such as water. However, unlike HCl, which is a gas, salts are involatile solids. Nevertheless, when heated to sufficiently high temperatures they evaporate molecularly and so molecules such as NaCl can be added to HNDs. Two recent studies have recorded IR spectra of small $NaCl(H_2O)_n$ and $NaCl(CH_3OH)_n$ complexes [542, 543]. Structures of the complexes were established in which each solvent molecule (for up to three solvent molecules) binds to the anion via an ionic hydrogen bond in a similar way to small $HCl(H_2O)_n$ complexes.

The rapid cooling and low temperature attained in HNDs also offers promise for the study of pre-reaction complexes. Although these can sometimes be made using conventional supersonic molecular beams techniques, the rapid quenching and low temperature in HNDs is particularly suited to reactive systems with very shallow entrance channel minima and/or very low barriers to reaction. A good example is provided by the reaction between the group 13 metal atoms M = Al, Ga and In, with HCN. Reactions to produce both MCN and MNC are known but by combining the reactants in a HND there is the prospect of observing the pre-reaction complexes. In work by Merritt *et al.* the complexes were formed by successive pick-up of metal atoms (produced from a high temperature oven) and HCN [544]. For Ga and In, the IR spectra showed that two distinct complexes were formed for each metal, both of which were linear. The metal atom can bind at either the H or N end of the HCN molecule. Binding at the H atom leads to a $^2\Sigma^+$ electronic state in which the unpaired electron on the metal atom resides in a p-orbital pointing at the H atom. At the nitrogen end, the metal binds in a $\pi$ orientation giving a $^2\Pi$ electronic state. These orientations of the 2p unpaired electron density on the metal atom are dictated by the electronic properties of the other atom. Specifically, electron-electron repulsion generated by the high electron density at the N atom favors a $\pi$ orientation for the approaching metal atom, whereas the relatively low density at the H atom favors a $\sigma$ orientation. In the case of Al, it proved impossible to isolate the Al-NCH complex, presumably because the barrier to reaction is too low for being quenched even in a HND, and thus reaction occurs. Liang and Douberly proved this to be true, observing the Al-NCH reaction product via CH stretch at 2620 cm$^{-1}$ [545].

Complexes between other radical atoms and molecules have also been studied in HNDs. Particularly interesting are reactions of halogen atoms with molecules, which often serve as prototypical reactions in the study of chemical kinetics and dynamics. HNDs have been successfully used to explore entrance channel complexes such as X-HF (X= Cl, Br and I) [546] and X-HCN [547]. Using a variety of evidence, including vibrational band positions, resolved rotational structure (where possible), and Stark effect measurements, linear complexes have been found for these systems. In a more recent example Moradi and Douberly recorded IR spectra of the Cl-HCl complex [548]. Although one might expect to see a linear hydrogen-bonded Cl⋯H-Cl complex, no such complex was observed. Instead, an L-shaped complex was found. Information such as this can provide a sensitive test of the entrance channel region of the Cl + HCl reaction potential energy surface.

The complexes formed between volatile organic compounds and the OH radical is another area where HND studies are proving beneficial. Such reactions are highly significant in tropospheric chemistry. One example is the reaction between OH and methanol, which has two possible outcomes:

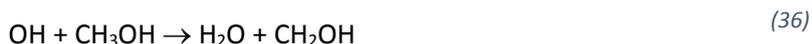
$$OH + CH_3OH \rightarrow H_2O + CH_2OH \qquad (36)$$

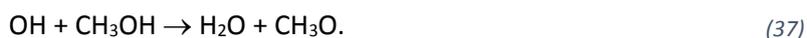
$$OH + CH_3OH \rightarrow H_2O + CH_3O. \qquad (37)$$

The dominant channel at room temperature is (36) but as the temperature is decreased below 200 K the reaction rate dramatically increases and reaction (37) becomes favored. The explanation seems to derive from formation of a pre-reaction complex, which once formed can facilitate channel (37) by quantum tunneling through the reaction barrier. It is therefore important to learn more about this complex and this has recently been characterized for the first time in HND experiments [549]. The OH radical was formed by pyrolysis of *tert*-butyl hydrogen peroxide, which generates acetone, hydroxyl and methyl radicals in equal measures. Figure 53 shows a low resolution IR spectrum which includes a band assigned to the OH(CH$_3$OH) radical. The equilibrium structure has the OH acting as a hydrogen bond donor to the O atom in the methanol. However, a more detailed analysis of the rotational contour and its dependence on the Stark effect shows that the complex formed is rather floppy, an observation which is significant to any model of the quantum tunneling process.

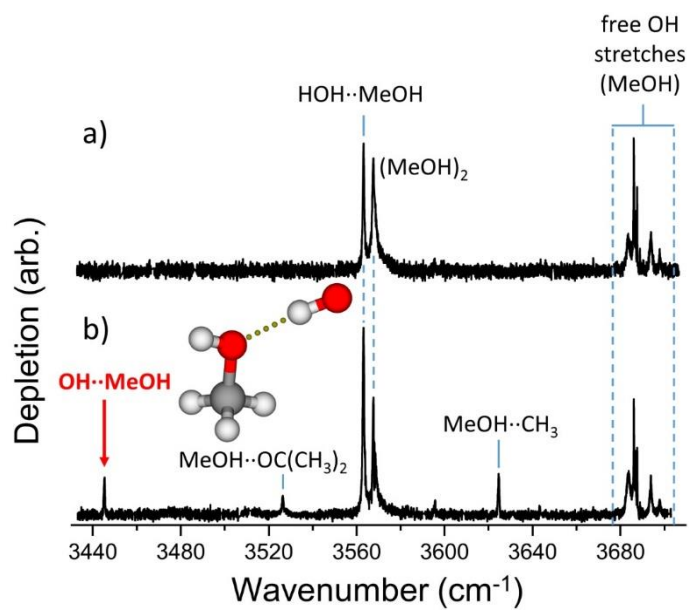

*Figure 53 IR spectrum recorded by pick-up of methanol and the pyrolysis products from tert-butyl hydrogen peroxide. Trace a) was obtained with the pyrolysis source at room temperature while trace b) was obtained at ca. 1000 °C. A single peak in trace b) is assigned to the OH-MeOH radical-molecule complex. Reprinted with permission from Ref. [549]. Copyright 2015 American Chemical Society.*

To close this section, we note that it is also possible to form radical + molecule complexes in HNDs which correspond to post-reaction (exit channel) complexes. Examples are the $CH_3 \cdots HF$ [550] and $CH_3 \cdots H_2O$ [551] complexes, which are exit channel complexes for the two hydrogen atom abstraction reactions shown below.

$$F + CH_4 \rightarrow CH_3 + HF \tag{38}$$

$$OH + CH_4 \rightarrow CH_3 + H_2O \tag{39}$$

### 11.2.3 Molecular ions

The spectroscopic investigation of molecular ions in HNDs has lagged well behind the study of neutral molecules. However, the challenge is similar to that of neutral free radicals in that it requires finding an efficient way to dope the droplets with ions and then record their spectra.

The first successful spectroscopic study of molecular ions in HNDs was only reported as recently as 2010. Instead of external generation of ions followed by pick-up, Smolarek *et al.* produced the ions *in situ* using resonance-enhanced multiphoton ionization (REMPI) [552]. This initial study focused on the aniline cation, which was produced in its electronic ground state through 1+1 REMPI via the $\tilde{A}^2 A_2$ state of the neutral molecule. IR spectra were then recorded by detecting ions ejected from the droplets after resonant IR laser excitation. It is important to emphasize that this is a new way of recording IR spectra and quite different from the depletion technique conventionally used for neutral molecules. As we have seen earlier, the depletion approach is a form of action spectroscopy which, instead of measuring light absorption directly, registers this absorption indirectly through some secondary phenomenon derived from the primary absorption event. In depletion spectroscopy this 'action' is the thermal loss of helium atoms after some or all of the energy absorbed by the dopant is dissipated into the surrounding helium. By way of contrast, the ejection of ions from the droplet is a non-thermal process. Of great practical significance is that the method developed by Smolarek *et al.* is effectively a zero background method, since no ions are ejected from the droplets when the IR is not resonant with any dopant transitions. This is very different from the depletion spectroscopy of neutral molecules and results in ion spectra showing high signal-to-noise ratios. In practice, some ions are ejected from the droplets during the initial photoionization process but these can be eliminated from the detection scheme through appropriate time-resolved gating of the ion detector.

Armed with this new technique, Smolarek *et al.* initially investigated IR spectra of the aniline cations, as well as the aniline dimer cation, in the NH stretching region near 3 μm. Subsequent work has extended the spectroscopy of these ions into the fingerprint region [553]. In the same study two other ions were considered, the styrene and 1,1-diphenylethylene cations. Rather than using REMPI, these ions were produced by non-resonant laser photoionization. This change of ionization scheme had no deleterious effects on the IR spectra, which remained of very high quality. One of the key conclusions drawn from the IR spectra of this small collection of molecular ions is that, just like the case of neutral dopant molecules, the helium induces only very small shifts of 1–2 cm$^{-1}$ of vibrational bands when compared with the gas phase. This may be considered surprising given that the interaction between an ion and the surrounding helium would be expected to be much stronger than for a neutral molecule. However, the key point is not the magnitude of this interaction but the fact that it shows little change as the vibrational state is changed. As a consequence, the available evidence suggests that HNDs will provide an excellent spectroscopic matrix for studying ions.

Drabbels and co-workers have shown that their IR technique can also be used to record electronic spectra of cations [554]. One would now expect a more severe effect of the helium on the electronic spectra of ions, because the distribution of the electron density may alter significantly upon electronic excitation. The change in electron distribution will alter the ion-induced dipole interactions with the helium atoms through a change in polarization of the ion. To accommodate the change in ion size the cavity usually has to expand upon electronic excitation because of Pauli repulsion, which would cause a blue shift in the electronic transition frequency if acting alone. However, the polarization effect acts in the opposite sense and so the size and direction of the actual band shifts will be dictated by the net result of these two effects. One might also anticipate substantial band broadening through matrix-enhanced dephasing processes such as intermolecular vibrational redistribution and internal conversion. However, from the observations made so far some of these concerns are not borne out in practice. Drabbels and co-workers have reported vibrationally-resolved electronic spectra of two cations, aniline [554] and 2,5-difluorophenol [555]. The electronic systems in both cases are shifted by a few tens of cm$^{-1}$ relative to the gas phase, which is significant but not huge. The bands are asymmetrically broadened by the excitation of phonon wings, *i.e.* collective excitations in the surrounding helium. Nevertheless, the vibrational structure is clear, as illustrated in Figure 54.

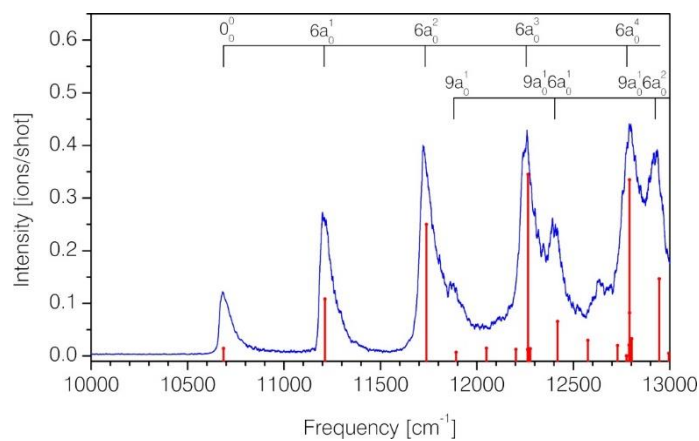

Figure 54 Vibrational structure from the $\tilde{A}^2 A_2 - \tilde{X}^2 B_1$ electronic transition of the aniline cation in HNDs. Also shown in red are predicted band positions from ab initio calculations. Reprinted with permission from Ref. [554]. Copyright 2011 American Chemical Society.

Instead of generating ions *in situ*, Bierau *et al.* have successfully introduced ions into HNDs from an external electrospray ion (ESI) source [80]. This novel approach makes it possible to add large molecules, including biomolecules, to the HNDs. The apparatus employed, which is illustrated in Figure 55, couples a stand]ard electrospray interface to a HND source and an ion trap. Ions produced by the electrospray process pass through two ion guides and are then mass selected by a quadrupole filter, before being turned 90°. The selected ions are then introduced into a hexapole ion trap and accumulated for several seconds until the space-charge limit is reached ($10^6$–$10^7$ ions per cm$^3$). The ion trap behaves essentially as a pick-up cell, but the HNDs come from a pulsed source and are only added once the trap is full. The train of droplet pulses is added over a period of tens of seconds and the droplets sweep a significant quantity of ions from the trap by utilizing their large kinetic energies to overcome the trapping potential barrier. The assumption is that at the low ion density in the trap each droplet will pick-up, at most, a single ion. In this preliminary ESI experiment both singly charged phenylalanine and multiply charged cytochrome C ions were added to the HNDs.

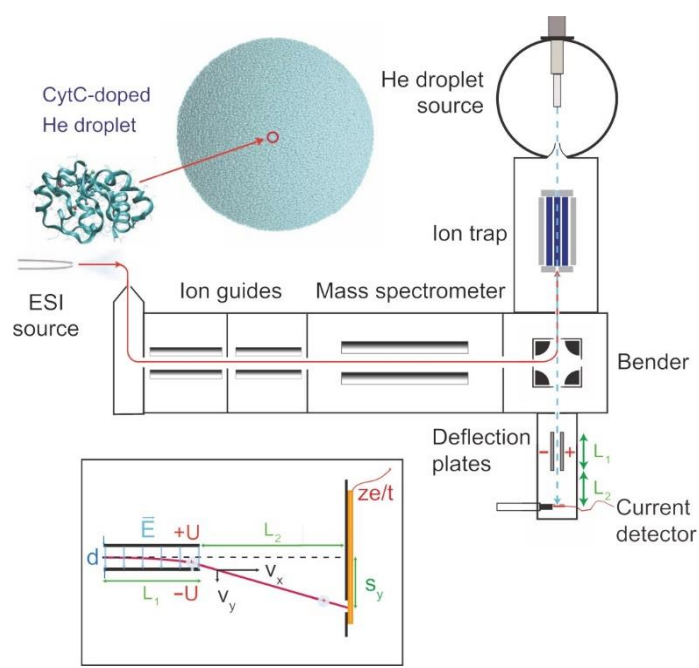

*Figure 55 Apparatus employed to trap ions generated from an electrospray ionization source in HNDs. Large charged biomolecules, such as cytochrome C in this case, can be added. Reprinted figure with permission from Ref. [80]. Copyright 2010 by the American Physical Society.*

In subsequent work this system has been coupled with a tunable laser to record optical spectra. The first example focused on the electronic spectrum of the singly charged cation of hemin, which is an iron-containing porphyrin with a molecular mass of 652 u [556]. The spectrum recorded between 370–395 nm consists of a broad absorption feature similar to that recorded previously in the gas phase. The primary source of the broadening is the very short lifetime of the excited electronic state, rather than broadening due to the helium matrix. Indeed, the HND absorption band was found to be narrower and only moderately shifted (*ca.* 3 nm) from the gas phase value. As in experiments by the Drabbels group, the HND spectrum was recorded by detecting ions ejected from the droplets as a function of the laser wavelength.

A drawback of the approach by Bierau *et al.* [80, 556] is the need to periodically replenish the contents of the ion trap. When the pulsed HND source is switched on the ESI source is effectively switched off and the quadrupole bender is grounded. The HNDs will then pick-up ions for several seconds before the trap becomes significantly depleted. At this stage the HND source is switched off and the ESI source effectively switched back on to accumulate more ions. Nevertheless, this is an inconvenience rather than a major weakness and the fact that the technique is a zero background technique means that it can deliver optical spectra of ions with high signal/noise levels. Another illustration of this is the recent recording of the IR spectrum of the protonated pentapeptide leu-enkephalin [557].

# 12 Adsorption on fullerenes and fullerene aggregates

Physisorption of gases on the surface of graphite has been studied for decades (for reviews see Steele [558, 559] and articles in the book edited by Bottani and Tascón [560]). The competition between adatom-adatom and adatom-substrate interactions combined with the corrugation of the substrate gives rise to rich phase diagrams which include transitions between different commensurate phases, commensurate-incommensurate phases, and solid-liquid phases. These can be identified by, for example, measurements of specific heat or neutron scattering. Many inorganic gases preferentially adsorb at hollow sites, *i.e.* at the centers of the carbon framework hexagons. The formation of commensurate phases at low temperature depends on a number of parameters, including the strength of the corrugation of the substrate, the size of the adsorbate molecule, and the ratio of molecule-molecule versus molecule-substrate interaction energies [561]. Relatively strong adatom-adatom interaction combined with weak corrugation will favor the formation of islands, whereas relatively weak adatom-adatom interaction combined with strong corrugation will favor adsorption at specific sites and, at appropriate coverage and temperature, the formation of commensurate phases.

Fullerenes, nanotubes and graphene make it possible to fine-tune parameters that affect the adsorbate layer. First, the curvature of nanotube or fullerene surfaces enhances the corrugation relative to that of graphene while it reduces the adsorption energy [562]. Second, the adsorption energy can be increased by stacking several graphene sheets [563]. Third, the spacing between molecules at adjacent hollow adsorption sites increases as the curvature of the (convex) surface is increased. It thus becomes possible to adsorb molecules on adjacent hexagonal sites, which is impossible over planar graphitic surfaces because the distance between these sites is only 2.46 Å, less than the nearest-neighbor distance in any van der Waals bound system. The densest commensurate phase that He, $H_2$, $CH_4$ and other adsorbates form on planar graphite or graphene is the $\sqrt{3}\times\sqrt{3}$ phase, in which 2/3 of all hexagon centers remain unoccupied. Over a curved convex surface the denser $1\times 1$ phase becomes accessible. The ability to vary the adsorption strength, the corrugation and the distance between adjacent registered sites, and to introduce additional corrugation on a different length scale has other implications, too. For example, for helium a decrease in corrugation over graphene sheets may lead to a transition from an ordered phase to a two-dimensional superfluid phase [563].

Furthermore, fullerene aggregates offer two other types of adsorption sites with much larger corrugation, namely sites in the groove between two adjacent fullerenes (analogous to groove sites between parallel nanotubes [564]), and at dimples between three adjacent $C_{60}$ (analogous to the preferred sites on the surface of a $C_{60}$ crystal [561]). The interplay between these corrugations leads to several interesting phenomena [565, 566].

Adsorption of inorganic and organic gases and vapors on fullerenes, mostly in the solid phase, has been reviewed by Suárez-García *et al.* [567]. The current section focuses on the physisorption of atoms and molecules on $C_{60}$, $C_{70}$ and their aggregates in the gas phase, with an emphasis on systems prepared in HNDs. These species are then ionized by collisions with electrons and mass-analyzed.

Ionization-induced dissociation leads to the appearance of anomalies in the ion abundance, which reflect the relative stabilities of the ions.

By analyzing these data in the light of molecular dynamics simulations, one can determine the nature and storage capacity of various adsorption sites.

Several topics related to the material discussed here will not be covered; we merely provide a few references: theoretical studies of $C_{60}$ complexed with just one or two molecules [568, 569], endohedral fullerenes [570, 571], gas-phase reactions of fullerene cations or dications with molecules [572, 573], adsorption of sulfur or phosphorous on $C_{60}$ cations and anions [574, 575], adsorption on polycyclic aromatic hydrocarbons such as coronene [415, 416, 576], graphene, nanotubes and films or crystals of $C_{60}$ [561, 567, 577, 578], fulleranes and other chemisorbed species [579-581].

## 12.1 Adsorption on $C_{60}$ and $C_{70}$

The Innsbruck group has analyzed mass spectra of HNDs doped with $C_{60}$ (or $C_{70}$) plus $H_2$, $N_2$, $O_2$, $H_2O$, $NH_3$, $CO_2$, $CH_4$, and $C_2H_4$. With proper adjustment of the droplet size and partial pressures in the pickup cells, electron ionization reveals cations containing one or a few fullerenes and up to several dozen small molecules. As an example, see the mass spectrum of $C_{60}$ doped with $NH_3$ in Figure 48 (Section 10.2.6), where one can easily identify $(C_{60})_m(NH_3)_n^+$ ions for m = 1 to 4 and n $\leq$ 42. The mass of $(NH_3)_{43}$ exceeds that of $C_{60}$, and therefore $(C_{60})_m^+(NH_3)_n$ ions with n > 42 are masked by ions containing an additional $C_{60}$. Even so, these ions may still be mass-resolved provided the mass resolution of the experiment is high enough. Another way to identify ions containing > 42 $NH_3$ molecules is to increase the $NH_3$ pressure in the pickup cell and/or reduce the $C_{60}$ vapor pressure to reduce the formation of $C_{60}$ oligomers.

Figure 48 (Section 10.2.6) displays several local anomalies in the ion yield of $(C_{60})_m^+(NH_3)_n$ *versus* n, most notably local minima at $(C_{60})_m^+NH_3$ for m = 1, 2, 3, and step-wise drops at $C_{60}^+(NH_3)_4$ and $(C_{60})_2^+(NH_3)_5$. Figure 56 displays the abundance of $C_{60}^+(NH_3)_n$ extracted from this mass spectrum after correcting for interfering ions and isotopologues (see 2.1 for details). For $NH_3$ and other hydrogen-containing molecules the ion abundance features groups of mass peaks that are separated by 1 u. They correspond to protonated, unprotonated, and dehydrogenated ions, as discussed in Section 10.2.6; they can be mass-resolved if the total ion mass is less than a few thousand u. In most cases local anomalies are identical for protonated, unprotonated, and dehydrogenated ion series if one specifies the number n of adsorbate molecules in terms of n for $NH_3$, O for $H_2O$, *etc*.

Figure 56 summarizes results for adsorption on $C_{60}^+$. Data for $C_{60}^+He_n$ are included; these ions are observed if the HNDs are large and no other gas is introduced into the pickup cell. The most striking feature in Figure 56 is a step or local maximum at n = 32 for helium, hydrogen (or deuterium), $CH_4$, $C_2H_4$, and $N_2$. This is remarkable given the fact that $C_{60}$ has 32 faces. $O_2$ shows just a hint of a maximum at n = 32 but $CO_2$, $H_2O$ and $NH_3$ do not. Experiments with $C_{70}$ reveal an anomaly at n = 37 for He, $H_2$ and $CH_4$; other systems were not investigated for $C_{70}$.

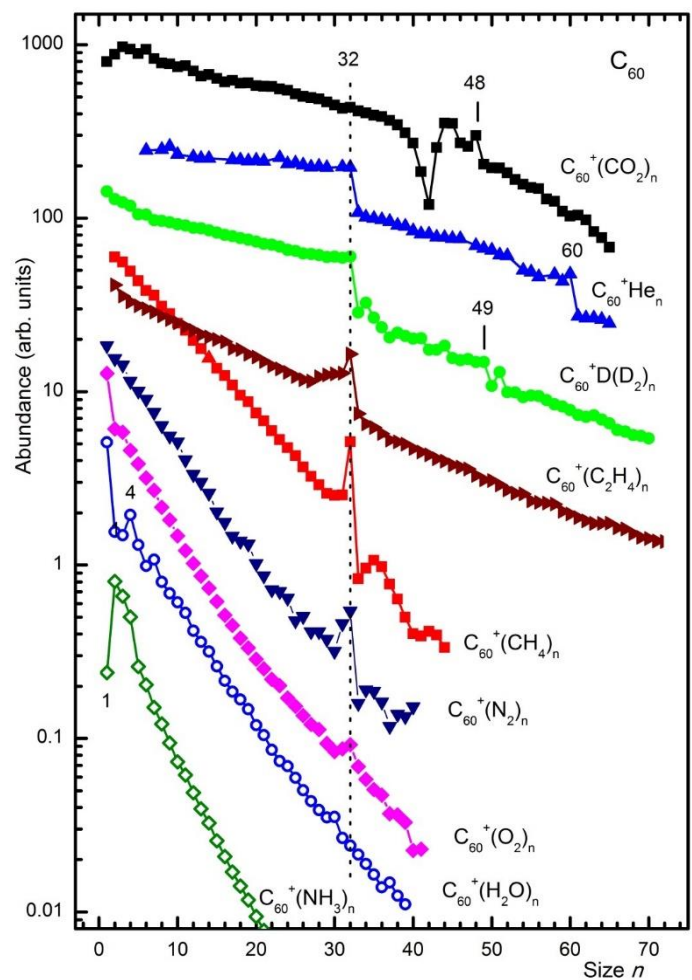

*Figure 56 Abundance distributions for adsorption of helium and molecules on $C_{60}^+$. For references see Table 5.*

$C_{60}^+He_n$ and $C_{60}^+D(D_2)_n$ show another step at 60 and 49, respectively. The steps shift to slightly larger values when $C_{60}$ is replaced by $C_{70}$. A very different pattern is observed for $C_{60}^+(CO_2)_n$. The distributions of $C_{60}^+(NH_3)_n$ and $C_{60}^+(H_2O)_n$ show small steps at n = 4 but they are largely featureless for larger n values.

Table 5 compiles all anomalies reported for adsorption on $C_{60}$ and $C_{70}$ cations and anions, and fullerene dimer cations. Note that local anomalies are judged relative to the envelope of the distribution, which depends on the droplet size, vapor pressure in the pickup cell, and degree of fragmentation upon ionization. We indicate the nature of the anomaly, *i.e.* a stepwise decrease, a local maximum, or local minimum, even though the distinction is not always unambiguous. Only those anomalies that are statistically significant (as judged by comparing mass spectra recorded under different experimental conditions) are included.

Table 5 Anomalies in the abundance distributions of $C_{60}^{+/-}A_n$ and $C_{70}^{+/-}A_n$ for various adsorbates A. Also shown are results for adsorption on $(C_{60})_2^+$ and fullerene anions. Values refer to local maxima (numbers in bold), minima (italic), and abrupt drops (underlined).

| Adsorbate | fullerene | n (fullerene monomer) | n (fullerene dimer) | Reference |
|---|---|---|---|---|
| He | $C_{60}^+$ | 32, 60 | | [375] |
| He | $C_{60}^-$ | 32, 60 | | [418] |
| He | $C_{70}^+$ | 37, 62 | | [375] |
| He | $C_{70}^-$ | 37, 65 | | [418] |
| $H_2$ or $D_2$ | $C_{60}^+$ | 32, 49 | 58 | [249, 582] |
| $H_2$ or $D_2$ | $C_{70}^+$ | 37, 51 | 66 | [249, 582] |
| $N_2$ | $C_{60}^+$ | **32** | *7* | [583] |
| $O_2$ | $C_{60}^+$ | **32** | *7* | [583] |
| $CO_2$ | $C_{60}^+$ | *42*, 48 | **8** | [584, 585] |
| $CO_2$ | $C_{60}^-$ | *31*, 33 | | [584, 585] |
| $CH_4$ | $C_{60}^+$ | **32** | 7, **56** | [372, 586] |
| $CH_4$ | $C_{70}^+$ | **37** | | [372, 586] |
| $C_2H_4$ | $C_{60}^+$ | **32** | 6, **28, 55** | [587] |
| $H_2O$ | $C_{60}^+$ | **4** | **4**, 7 | [514, 515, 588] |
| $NH_3$ | $C_{60}^+$ | 4 | **2, 5** | [515] |

What do these anomalies tell us? Growth of neutral clusters in HNDs is a statistical process. The size distribution of neutral aggregates will be featureless; anomalies in the abundance of $(C_{60})_m^+A_n$ arise from fragmentation upon ionization. Although fragmentation kinetics could possibly play a role, the dominant factor turns out to be local variations in the adsorption energies $D_n$ (also called dissociation, evaporation or separation energies), *i.e.* the difference between total energies $E_n$ of cluster ions of adjacent size in their most stable configurations,

$$D_n = -E_n + E_{n-1}. \quad (40)$$

Total energies $E_n$ are negative; $D_n$ values are positive. The exact relation between $D_n$ and the abundance $I_n$ depends on several factors, including the heat capacity of the clusters [589]. The relation simplifies for cluster ions whose heat capacity is much less than the classical value [375, 376]. In that case one can derive "experimental" adsorption energies $D_{exp,n}$ from the relation

$$D_{exp,n} = \frac{I_n}{I_{av}} D_{av} \quad (41)$$

where $I_{av}$ is the local average over $I_n$; it may be obtained by fitting a low-order polynomial to a limited region of $I_n$, or by a weighted average of $I_n$ over several adjacent sizes [371]. Similarly, $D_{av}$ is the local average over "true" $D_n$ values, for example calculated values $D_{th,n}$. If those are not known one may set $D_{av} = 1$, and equation (41) merely provides relative adsorption energies. Either way, the experimental quantity $I_n/I_{av}$ provides a measure of the shape and strength of local anomalies.

We illustrate the correlation for helium adsorbed on $C_{60}$ ions. This system comes closest to the ideal case of zero heat capacity, for two reasons. First, the vibrational temperature is only a few K, as estimated from the concept of the evaporative ensemble [350] (the computed dissociation energy of $C_{60}^+He_n$ [375] exceeds the evaporation energy of bulk helium by an order of magnitude, therefore the vibrational temperature of the former will be an order of magnitude higher than that of the latter). At this temperature the heat capacity of $C_{60}$ is negligible. Second, the adsorbate has no internal degrees of freedom.

Figure 57 a) displays the abundance distributions $I_n$ of $C_{60}^+He_n$ and $C_{60}^-He_n$ [375, 418]. Both distributions show stepwise decreases at n = 32 and 60. Other anomalies in the anion data probably arise from interference with contaminants. The top trace in Figure 57 b) represents $D_{av} \cdot I_n/I_{av}$ for $C_{60}^+He_n$ where $D_{av}$ is the envelope of computed dissociation energies $D_{th,n}$ shown in the middle trace [375]. The bottom trace represents dissociation energies computed by Calvo [417]. Not included in Figure 57 b) are dissociation energies computed by Shin and Kwon for neutral $C_{60}He_n$ [590]. They show an even stronger drop at n = 32.

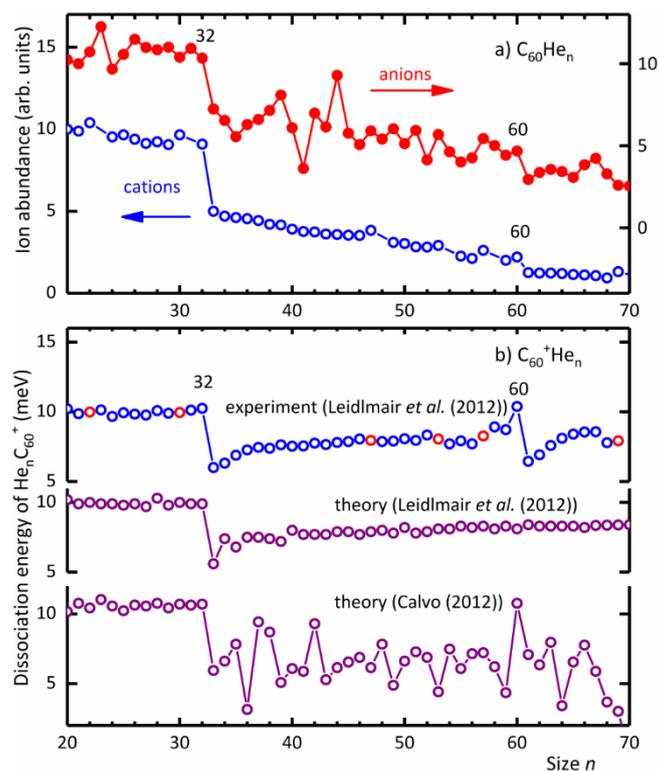

*Figure 57 Panel a: Ion abundances of $C_{60}^+He_n$ and $C_{60}^-He_n$ [375, 418]. The top trace in panel b represents dissociation energies of $C_{60}^+He_n$ derived from the ion abundance; the other traces represent computed dissociation energies [375, 417].*

All three traces in Figure 57 b) show a nearly constant dissociation energy up to n = 32, followed by an abrupt drop by about 45 % for $C_{60}^+He_{33}$. Also common to the top two traces is a gradual recovery after the minimum at n = 33 but the computed values (panel b), middle trace) miss the local maximum in the experimental data at n = 60. Values computed by Calvo show several anomalies above n = 32, including one at n = 60 [417].

The agreement between experiment and theory for n ≤ 33 is encouraging. Disagreements for larger sizes may arise from inadequate description of He-He and He-$C_{60}^+$ interactions and quantum effects. Another system that shows excellent agreement between experimental [370] and computed [377] dissociation energies is $Ar^+He_n$; it was discussed in section 8.1.

Having established a close correlation between local anomalies in the size dependence of dissociation energies and cluster ion abundances, we move on to discuss the physical origin of these anomalies. As mentioned in the Introduction to this section, on the hexagonal graphite lattice many inorganic gases preferentially adsorb at hollow sites [559]. $C_{60}$ has 12 pentagonal plus 20 hexagonal hollow sites, readily accounting for the anomaly observed at $C_{60}^+A_{32}$. The curvature of the substrate leads to increased separation between adsorbate molecules on adjacent hollow sites, making the commensurate 1×1 phase accessible for He, $H_2$, $CH_4$, $C_2H_4$ and $N_2$. $C_{70}$ has an additional 5 hexagonal sites and, in fact, all adsorbates studied so far (He, $H_2$, $CH_4$), show an anomaly at n = 37.

A computational study of $C_{60}^+He_n$ shows that at 4 K atoms at registered sites are localized [375]. A vacancy in the commensurate layer will not be visited by other atoms within 5 ns, the duration of the simulation. The simulation also shows that more than 32 He atoms can be accommodated into the first solvation shell; their average distance from the center of $C_{60}^+$ is only 6 % larger than for atoms at registered sites.

This raises two questions: First, how many atoms or molecules can be accommodated in the first solvation shell? Second, are there molecules that are too large to form the commensurate 1×1 phase on $C_{60}^+$? It appears that the steps in the ion abundance at $C_{60}^+He_{60}$ and $C_{70}^+He_{62}$ indicate closure of the first shell; the values are slightly smaller than obtained in calculations that did not include quantum effects [375]. Calvo applied path-integral simulations and concluded that the first layer accommodates about 72 atoms [417]. Interestingly though, he observed that 60 helium atoms form a particularly rigid, incommensurate layer with a high adsorption energy while for most other sizes the layer formed a homogeneous fluid.

Similarly, we assign anomalies at $C_{60}^+H(H_2)_{49}$ and $C_{70}^+H(H_2)_{51}$ to closure of the first solvation shell. Several theoretical studies of the fullerene-hydrogen system have been published but they considered either chemisorption (*i.e.* fulleranes) [581], hydrogen interaction with $C_{60}$-metal complexes [591] (also see section 12.3), or physisorption of no more than 2 $H_2$ [582].

On the other hand, $C_{60}^+(CH_4)_n$ shows no feature beyond the maximum in the abundance at n = 32 that might correspond to closure of a solvation shell. The prominent spike at n = 32 suggests that the size of $CH_4$ provides a near-perfect match to the distance between molecules adsorbed at hollow sites. Theoretical studies show, indeed, that closure of the first solvation shell coincides with completion of the commensurate 1×1 phase [372]. The radial distribution function calculated for $C_{60}^+(CH_4)_{500}$ exhibits a distinct peak at about 6.7 Å which comprises 32 $CH_4$ molecules; their exact distance from the center of $C_{60}^+$ depends on their orientation and adsorption site [372]. The histogram of calculated vertical desorption energies features a peak at about 200 meV comprising 32 molecules; the desorption energies of all other molecules is less than 130 meV.

So size matters, but it is not only size that matters. The abundance distribution of $C_{60}^+(C_2H_4)_n$ is similar to that of $C_{60}^+(CH_4)_n$, with a maximum at n = 32 but no distinct feature beyond, suggesting that the first solvation layer closes at n = 32. On graphite, though, the commensurate $\sqrt{3}\times\sqrt{3}$ phase has escaped detection in all but one experimental study [592]. At low temperature, isolated $C_2H_4$ prefers to lie flat on graphite [558, 559, 593]; in that orientation its projected size is too large to form the commensurate $\sqrt{3}\times\sqrt{3}$ phase. The orientation of the molecule will change as coverage increases, leading to an exceptionally rich phase diagram including 7 solid monolayer phases. First, the molecule tilts around its long (C=C) axis to reduce crowding in the adsorption layer. As the coverage approaches that of the $\sqrt{3}\times\sqrt{3}$ phase (coverage x = 1) an increasing number of molecules will have their long axis perpendicular to the substrate; all molecules will be in that orientation when the coverage exceeds x = 1.05 [558, 559].

Zöttl *et al.* have investigated neutral $C_{60}(C_2H_4)_n$ by simulated annealing [587]. Figure 58 displays the cumulative number Σ of molecules *versus* distance from the center of the nearest fullerene, equivalent to the integral over the radial distribution function. One sees a distinct plateau when the first adsorption layer is filled at n = 32 for the $C_{60}$ monomer; the flatness and width of the plateau indicates spatially distinct first and second layers.

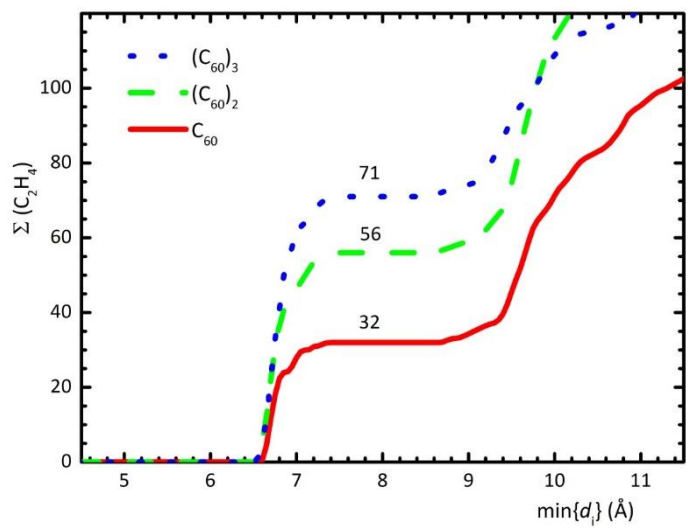

Figure 58 The cumulative number of $C_2H_4$ molecules versus their distance, in Å, from the nearest fullerene for $C_{60}$ monomers, dimers and trimers. Data taken from Ref. [587].

At low temperature all molecules in $C_{60}(C_2H_4)_{32}$ reside at hollow sites [587]. Only two molecules are oriented perpendicularly to the substrate; the other molecules have their C=C axis parallel to the substrate but with a significant tilt around that axis. On a curved surface this tilt increases the radial distance from the center and thus the distance between neighboring molecules. The tilt increases with increasing temperature but even at 110 K, the estimated vibrational temperature of $C_{60}^+(C_2H_4)_n$ in the experiment, 60 % above the melting temperature of solid phases on graphite, the molecules are locked in place for the 2 ps duration of the simulation. Presumably, the enhanced corrugation over the curved surface of $C_{60}$ significantly increases the melting temperature.

$CO_2$ is another highly anisotropic molecule [558, 559]. X-ray diffraction of $CO_2$ adsorbed on graphite shows a solid monolayer with herringbone structure that is incommensurate [594]; theory suggests the existence of a commensurate $\sqrt{3}\times\sqrt{3}$ phase that is marginally more stable than the incommensurate phase [559]. However, the central carbon atom would be over the midpoint of a C-C bond of the substrate; this arrangement is rather unfavorable for the curved fullerene surface.

It is not surprising, then, that $C_{60}^+(CO_2)_n$ displays no anomaly at n = 32 (see Figure 59). Instead one observes a deep minimum at n = 42 and maxima or steps at 45 and 48. Surprisingly, the abundance distribution of $C_{60}^-(CO_2)_n$ anions is very different, with a broad bump around n = 33. A molecular dynamics study predicts that the maximum number of molecules in the first solvation shell are 33 and 49 for the anion and cation, respectively, in good agreement with anomalies in the abundance distributions [585]. The large difference in adsorption capacity arises from the strong quadrupole moment of $CO_2$. Electrostatic interaction between the positive carbon center and $C_{60}^-$ favors a flat orientation for $CO_2$. On the other hand, the electrostatic interaction between the negative terminal O atoms in $CO_2$ and $C_{60}^+$ favors a tilted orientation, with $CO_2$ standing up by 40–60 degrees. Just before completion of the first shell there is a range of sizes (indicated as ROR in Figure 59) in which the adsorbate uniformly covers the fullerene but still accommodates additional molecules, at the price of strong orientational rearrangement. For the cation the computed ROR, from n = 41 to 49, agrees closely with the region in which abundance anomalies are observed.

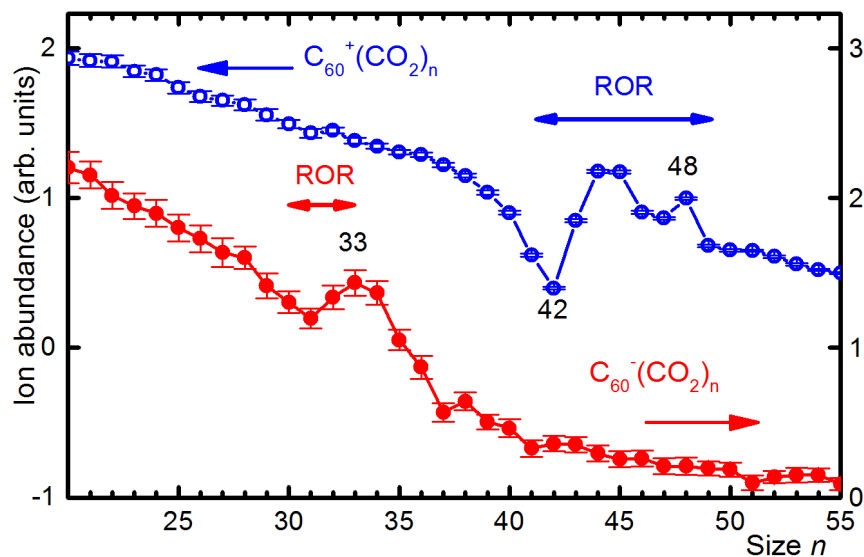

Figure 59 Abundance of $C_{60}^+(CO_2)_n$ and $C_{60}^-(CO_2)_n$. Note that the trace showing cations (left ordinate) is offset. ROR indicates computed regions of rearrangement. Data taken from Ref. [585].

Another class of anisotropic molecules is represented by the polar molecules $H_2O$, $D_2O$ and $NH_3$. Their abundance distributions are featureless beyond the step at n = 4 (Figure 56) [514, 515, 588]. Theoretical studies of these systems are demanding. Denifl *et al.* have reported ab-initio calculations of neutral and cationic $C_{60}$ complexes containing up to 6 $H_2O$ molecules [588]; Hernandez *et al.* have investigated neutral [595] and cationic [596] $C_{60}$ with up to 20 $H_2O$ molecules using empirical potentials. The studies agree that the interaction of $H_2O$ with $C_{60}$ is relatively weak; the water molecules tend to form compact clusters which become three-dimensional for $n \geq 8$ [596], in line with the strong heliophobicity of fullerenes. However, the structure does depend on the charge state of the fullerene. Mass-analyzed ion kinetic energy scans suggest that $C_{60}^+(H_2O)_n$ dissociates for n = 3, 4, and 6 by desorbing the complete water cluster rather than desorbing a single water molecule [588].

The enhanced stability of the water tetramer would explain the local maximum in the abundance at $C_{60}^+(H_2O)_4$ [588, 596]. Several other, larger $C_{60}^+$-water complexes are predicted to be particularly stable [596] but the abundance distribution in Figure 56 does not show any corresponding anomalies. At any rate, it is clear that water will not form the commensurate 1×1 phase.

The $C_{60}$-ammonia system shows an anomaly at n = 4 for stoichiometric ions, but for non-stoichiometric ions the anomalies are different [515]. Few calculations are available for this system but it is worth pointing out that the very low abundance of $C_{60}^+NH_3$ correlates with its predicted low (< 0.25 eV) stability [597]. A theoretical study of $C_{60}$ and $C_{60}^{5-}$ solvated in liquid ammonia concludes that the first solvation shell contains 44 and 41 molecules, respectively [598]. No significant anomalies are seen in that size range although it should be noted that $C_{60}$ complexes containing more than 42 $NH_3$ molecules are overwhelmed by the $(C_{60})_2^+(NH_3)_n$ series.

From the examples discussed so far it may appear that any atom or reasonably isotropic molecule that is no larger than $CH_4$ will favor the commensurate 1×1 phase. Preliminary data suggest that the abundance distribution $C_{60}^+Ar_n$ does indeed feature an anomaly at n = 32. However, only xenon and possibly krypton form the commensurate $\sqrt{3}\times\sqrt{3}$ phase on graphite [558, 599]; early observations of this phase for the argon-graphite systems have not been confirmed [600]. A numerical study of neutral $C_{60}$ complexed with neon, argon, krypton and xenon reports enhanced stability at n = 32 for xenon only. Argon and krypton form commensurate, energetically favorable structures at n = 44 [601].

## 12.2 Adsorption on fullerene aggregates

The abundance of cationic $C_{60}$ and $C_{70}$ dimers complexed with various molecules is shown in Figure 60. $N_2$, $O_2$ and $CH_4$ show steps at n = 7, $C_2H_4$ at 6, and $CO_2$ at 8. Theoretical studies of $CH_4$ [372, 586], $C_2H_4$ [587], and $CO_2$ [584] reveal that this feature has a common origin; it reflects the number of molecules in the groove region. These molecules are more strongly bound than other molecules in the first layer because they are in close proximity with two fullerenes.

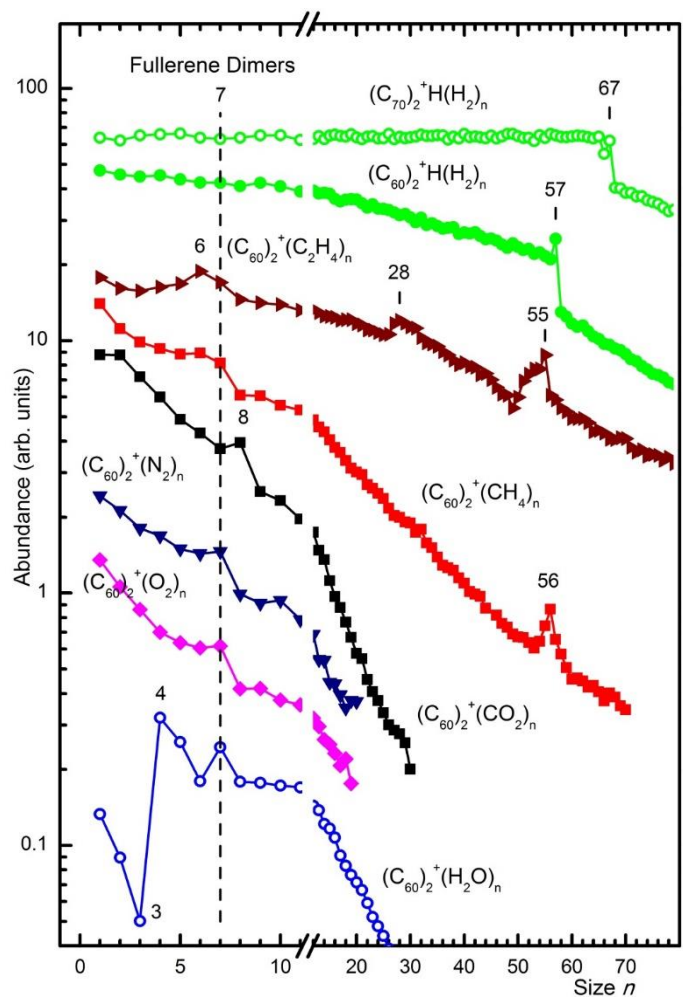

*Figure 60 Abundance of $(C_{60})_2^+$ and $(C_{70})_2^+$ complexed with various molecules. Several systems show an anomaly near n = 7. An anomaly near n = 56 is another prominent feature common to several adsorbates on $(C_{60})_2^+$. For references see Table 5.*

One way to count the number of molecules in the groove region is by plotting the cumulative sum $\Sigma$ of molecules that lie in the midplane of the dimer *versus* the distance from the dimer axis. The plateau in the bottom trace in Figure 61 indicates that 6 $C_2H_4$ molecules lie in the groove region, in agreement with experiment [587]. The inset in Figure 61 illustrates the structure of the $C_{60}$ dimer with all groove sites being filled. The variation in the number of groove sites, from 6 to 8, is rather small because the larger the adsorbate molecules the further away they are located from the dimer axis, and the more space there is in the groove.

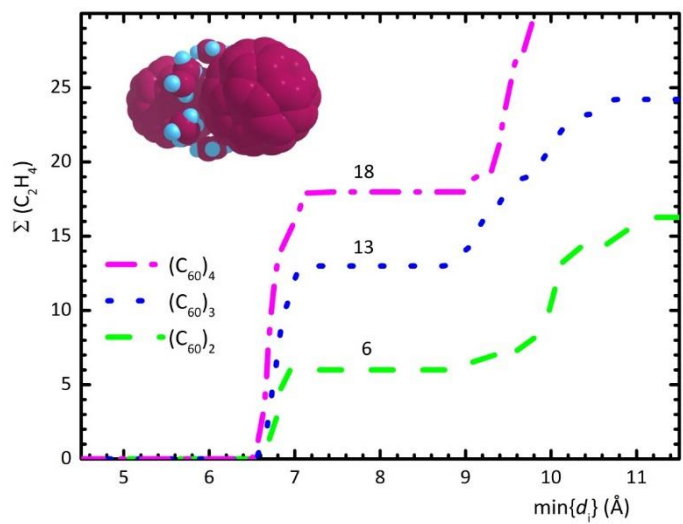

*Figure 61 The cumulative number of $C_2H_4$ molecules that lie in the midplane between any pair of fullerenes for the $C_{60}$ dimer, trimer and tetramer (bottom to top trace), versus their distance from the corresponding dimer axis. The inset shows the structure of $(C_{60})_2^+(C_2H_4)_6$. Data taken from Ref. [587].*

$(C_{60})_2^+(H_2O)_n$, too, shows a maximum at n = 7. Preliminary theoretical work, however, indicates that this feature is of different origin; water clusters adsorbed on the $C_{60}$ dimer keep their compact structure [602].

Another prominent feature in Figure 60 is the local maximum in the abundance of $(C_{60})_2^+(C_2H_4)_{55}$. This is just one unit below the number of molecules predicted for the first solvation shell of $(C_{60})_2^+$ (see Figure 58). For $CH_4$ Figure 60 reveals an anomaly at n = 56; the calculated spatial distribution predicts an anomaly at 58 [372]. Why does the much smaller $H_2$ molecule show an anomaly at nearly the same value, n = 57? Because the $C_{60}$ dimer offers approximately 56 hollow adsorption sites, independent of the nature of the adsorbate. 56 because two separate $C_{60}$ molecules offer 64 sites, minus 2 sites that are blocked when the fullerenes are in close proximity, minus 6 because each adsorbate molecule in a groove site will block two hollow sites (the structure of the $C_{60}$ dimer has been the subject of several theoretical studies, see [587] and references therein). A nice confirmation for this simple model comes from the anomaly observed for $(C_{70})_2^+(H_2)_{67}$: each $C_{70}$ contributes an additional 5 hexagonal adsorption sites.

Let us finally consider $C_{60}$ aggregates larger than the dimer. Experimental data are summarized in Figure 62. $CO_2$, $CH_4$ and $C_2H_4$ adsorbed on $(C_{60})_3^+$ exhibit a small step at n = 2. Calculated vertical desorption energies and spatial distributions reveal that the anomaly is due to enhanced binding at the two dimple sites of the trimer, above and below the molecular plane [372, 587].

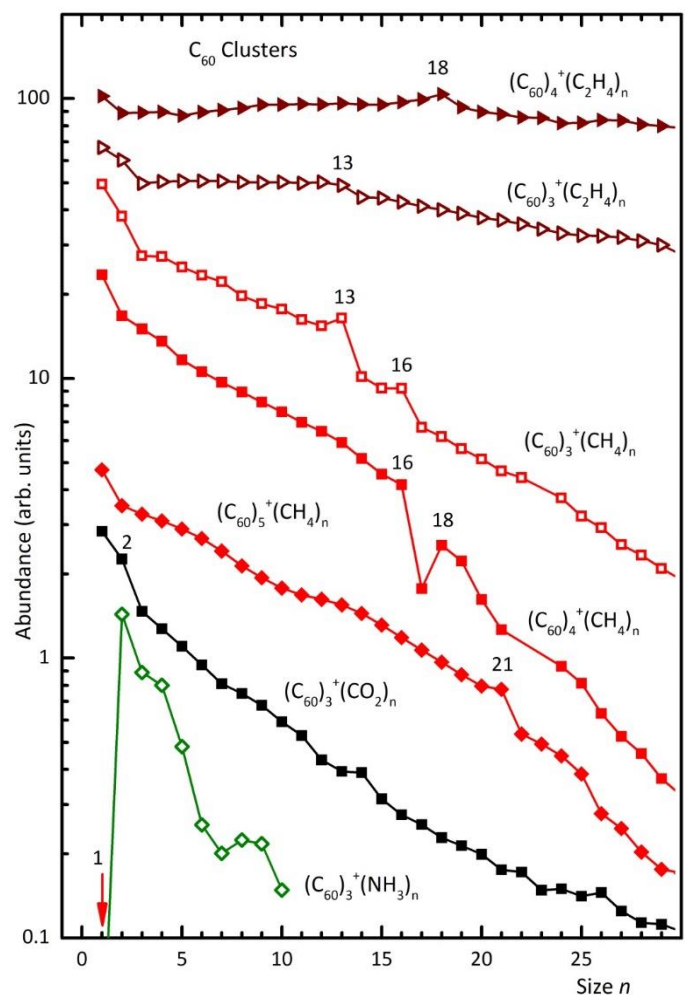

Figure 62 Abundance of $C_{60}$ aggregates complexed with $C_2H_4$, $CH_4$, $CO_2$, and $NH_3$. The abundance of $(C_{60})_3{}^+NH_3$, indicated by the arrow, is off-scale. For references see Table 5.

The $C_{60}$ trimer shows another anomaly at n = 13 for $CH_4$ and $C_2H_4$, and another one at n = 16 for $CH_4$. n = 13 agrees with the number of molecules in groove sites calculated for $C_2H_4$ (see Figure 61) and $CH_4$ [372]. The origin of the anomaly at $(C_{60})_3{}^+(CH_4)_{16}$ remains unclear.

$(C_{60})_4{}^+$ offers 4 dimple sites which correlates with a minor step in the abundance distribution at $(C_{60})_4{}^+(CH_4)_4$. The calculated number of groove sites for this system is 17, in good agreement with the prominent anomaly at n = 16 (Figure 62). For the $(C_{60})_4$-ethylene system the computed number of groove sites (18, see Figure 61) agrees perfectly with an anomaly in the ion abundance (Figure 62).

Several other significant anomalies have been observed for adsorption of $CH_4$ and $C_2H_4$, including completion of the solvation shell for $(C_{60})_3{}^+(CH_4)_n$ at n = 70 (predicted for n = 71, see Figure 58). One other feature, not shown here, is worth mentioning: anomalies observed for $(C_{60})_m{}^+(C_2H_4)_n$ (m = 2, 3) indicate that solvation shells are filled successively, one fullerene at a time [587].

## 12.3 Fullerene-metal complexes

Adsorption on nanotubes, fullerenes and other porous carbonaceous materials shows promise for high-density storage of $H_2$ and other hydrogen-rich molecules [603, 604]. However, the adsorption energy of $H_2$ over planar graphite and on isolated $C_{60}$ is only 0.05 to 0.06 eV (see [582] and references therein), too low for efficient storage at or near room temperature. One strategy that promises increased adsorption energies involves doping with alkali or earth alkaline metals. Theoretical studies suggest that storage capacities approaching 10 weight % may be attained [605, 606] but experiments, usually performed on solid samples, show less impressive values [607, 608].

Gas-phase studies of fullerene-metal systems, employing HNDs [418] or conventional gas aggregation sources for cluster growth [609], reveal that alkali atoms A bind ionically to $C_{60}$, resulting in particularly stable $C_{60}A_6$ building blocks although other particularly stable motifs exist as well [610-612]. Theoretical studies suggest that each alkali atom bound to $C_{60}$ can adsorb approximately 6 $H_2$ [605, 613].

HNDs offer an attractive way to synthesize fullerene-metal-hydrogen complexes in the gas phase. Kaiser *et al.* have reported mass spectra of droplets co-doped with $C_{60}$, Cs and $H_2$ [614]. Although cesium has no potential for hydrogen storage in light-weight materials, its large mass and the fact that it is monisotopic help in identifying the amount of hydrogen being adsorbed on the metal-fullerene complex. Mass spectra show that the first 10 $H_2$ molecules adsorbed on $C_{60}Cs^+$ are particularly abundant while density functional theory calculations suggest that $C_{60}Cs^+$ offers 13 energetically enhanced adsorption sites where six of them fill the groove between Cs and $C_{60}$ and 7 are located at the cesium. Furthermore, mass spectra suggest that completion of the commensurate $H_2$ layer is shifted from 32 molecules for bare $C_{60}^+$ to 42 molecules for $C_{60}Cs^+$.

# 13 Implications for extraterrestrial chemistry

The results of many of the HND chemistry experiments have provided useful insights into fundamental aspects of chemistry as it might occur in extraterrestrial environments, including interstellar and circumstellar environments as well as the atmosphere of Titan.

For example, the reactivity measurements of iron atoms that showed the low-temperature formation of only weakly bound adducts with acetylene and with water molecules have suggested only weak bonding of iron atoms at the surface of carbonaceous or icy interstellar dust grains [496]. Also, the observation of He droplet stabilization of intermediate $SiO_2$ in the reaction of Si with $O_2$ may translate to such stabilization occurring on cold interstellar dust grains [493].

Apparently, adamantane may play a role as a building block in dense interstellar clouds that contain microdiamonds [615, 616], While aggregation of adamantane, a building block of diamondoids, was readily observed in ionized form in He droplets at near 0 K, no adamantoid or microdiamond formation was evident in the presence of charged or excited helium species [282].The packing of adamantane into various "magic number" cluster structures with preferred arrangements of the adamantane molecules was observed instead.

The results of the SiO oligomerization experiments support the hypothesis that interstellar silicates can be formed in low-temperature regions of the interstellar medium by accretion through barrierless reactions [494]. Gaseous SiO is one of the main precursors of silicates in evolved stars [617]. While pure $SiO_x$ grains have not been observed in the ISM so far, the presence of Mg-Fe silicates has been proven [618].

Very recent droplet studies with embedded C atoms are providing some of the first experimental results for chemistry of pure C atoms at extremely low temperatures, without significant perturbations by multi-carbon species. The laboratory measurements have predicted a barrierless interstellar reaction between C atoms and $H_2$ to form the carbene HCH molecule and a pathway to single and multiple carbon atom hydrocarbon cations in mixtures of C and $H_2$ exposed to ionizing radiation [497]. The measurements with doped $C_{60}$ suggest that reactions of this molecule with C atoms can derivatize and activate the surface of this known interstellar and circumstellar molecule with bridging C atoms that provide carbene character in the formation of $C_{60}(C:)_n$ species [498]. The latter appear to react at 4 K with interstellar molecules such as $H_2$ and $H_2O$ and even further with $C_{60}$ molecules to form dumbbell structures, possibly extended as in $C_{60}=(C=C_{60})_n$. These results also have implications for the derivatization and activation by C atoms of graphitic molecules or particles in general that may be found in space.

A proposal has been made that the chemistry initiated in He droplets doped with $N_2$ and $CH_4$ and exposed to 70 eV electrons is akin to the chemistry initiated in Titan's nitrogen/methane atmosphere by electrons produced by cosmic rays [506]. Various heterogeneous $C_xH_yN_z^+$ cations were observed to be formed in the doped He droplets. Calculations of corresponding structures and binding energies have demonstrated the formation of chemical bonds, such as the CN bond in $CH_3N_2^+$. This important result draws attention to the possibility of the synthesis of true chemical bonds in chemical environments exposed to energetic electrons, such as the atmosphere of Titan.

The observation of a strong signal of $HO_4^+/DO_4^+$ with HNDs containing $H_2/D_2$ and $O_2$, and exposed to 70 eV electrons, has identified a possible signature ion for interstellar $O_2$, at least in those environments in which oxygen clusters may be formed. Surprisingly high (several %) concentrations of molecular oxygen have recently been reported to be present in the nucleus of a comet [619] and the infrared spectrum of $HO_4^+$ predicted recently from ab initio calculation [620] should improve the chances of the interstellar detection of this ion.

We have seen throughout this review that the superfluid He environment within HNDs is very favorable for the formation of He adducts of ions generated within these HNDs. Indeed, He adducts with a large variety of cations have now been reported to be formed in these environments and these may provide some measure of the possible formation of similar He adducts of ions in extraterrestrial environments, somewhere in the (known) universe where He is the second most abundant element and the temperature is low. One obvious example is the addition of He to the atomic, diatomic, and triatomic cations of hydrogen that are well established interstellar ions (H is by far the most abundant element in the universe). Indeed, the high-resolution mass spectra of HND doped with molecular hydrogen or deuterium have revealed that copious amounts of helium can be bound to $H^+$, $H_2^+$, $H_3^+$ and larger hydrogen-cluster ions at 0.37 K [399]. All conceivable $He_nH_x^+$ stoichiometries were identified below the mass limit of $\approx$ 120 u set by the resolution of the experiments. Observed anomalies in the ion yields of $He_nH_x^+$ for x = 1, 2 or 3 and n $\leq$ 30 have revealed particularly stable cluster ions. For example, the $He_nH_3^+$ series exhibited a pronounced anomaly at n = 12. Also, the dihydrogen ion $H_2^+$ was seen to retain helium with much greater efficiency than hydrogen-cluster ions, quite contrary to findings reported for other diatomic dopant molecules.

The earlier experimental study by Toennies and co-workers [505], in which the HNDs were doped with deuterium and subsequently ionized by electron impact, reported the observation of $He_nD^+$ ions with n up to 19 and drew attention to He solvation observed mass-spectrometrically for atomic noble-gas cations. Mass spectra of suitably doped HNDs have now demonstrated the He solvation of $Ne^+$, $Ar^+$, $Kr^+$ and $Xe^+$ and many other atomic, molecular, and cluster ions that are present in the interstellar medium [399]. As seen throughout this review, the mass spectrometric observation of He solvation of reactant and product cations at the low temperature in HNDs is quite commonplace. Even $C_{60}^+$, recently confirmed to be present in interstellar environments [621], has been found to become extensively solvated with He in HNDs [375] (and see Section 12.1). Observed mass spectra suggest that commensurate layers form when all carbon hexagons and pentagons are occupied by one He each, but that the solvation shell does not close until 60 He atoms are adsorbed on $C_{60}^+$, or 62 on $C_{70}$.

The adsorption of noble gases other than He, as well as of small molecules such as $H_2$, $N_2$, $O_2$, $H_2O$, $NH_3$, $CO_2$ and $CH_4$, on cations is of known importance in the chemistry of extraterrestrial environments and can be investigated in HNDs (see Section 12). Much attention has been given to the adsorption of these molecules by the fullerene cations $C_{60}^+$, $C_{70}^+$ and even cationic aggregates of $C_{60}$, as described in Sections 12.1 and 12.2. For example, more than several dozen of each of these molecules have been observed to adsorb to $C_{60}^+$ at 0.37 K in He droplets environments. To what extent the adsorptions of atoms and molecules by cations in He droplet environments are mimicked by adsorptions in extraterrestrial environments is not yet known, but it is clear from these studies that the physisorption of atoms and molecules by cations at low temperatures can proceed readily and extensively. An interesting case in point is the adsorption of $H_2$ by $C_{60}^+$, both known and important interstellar species. Copious amounts of molecular hydrogen are known to be able to physisorb $C_{60}^+$ in ordered layers at low temperatures [249]. Complexes of this type have been proposed to form in dense interstellar clouds (where thermal or photo desorption is diminished) if they contain fullerenes, but not so in the higher temperature diffuse clouds that are the major sources of the diffuse interstellar bands. In the case of He attachment to $C_{60}^+$, shifts and isomeric broadenings in the electronic excitation spectra have been observed for up to 100 helium atoms, despite the weak interaction of helium to the ionic molecule [622-624]. These line-shifts and line-broadenings observed even for helium tagging makes and extrapolation to bare spectra challenging.

# 14 Acknowledgments

This work was partially supported by FWF (P24443, P26635, I978, P30355, P31149). AME and SFY would like to take this opportunity to thank EPSRC (EP/J021342/1) and the Leverhulme Trust (RPG-2016-308, RPG-2016-272) for financial support. This study was partially supported by a grant of the Tyrolean Science Foundation.

# 15 Table of figures